\documentclass[a4paper,fleqn,usenatbib]{mnras}

\usepackage[T1]{fontenc}
\usepackage{ae,aecompl}
\usepackage{array}

\usepackage{graphicx}	
\usepackage{amsmath}	
\usepackage{amssymb}	
\usepackage{multirow}

\newcommand{\angstrom}{\mbox{\normalfont\AA}}

\usepackage{newtxtext,newtxmath}

\title[The origin of Ly$\alpha$ glow around high-z quasars]{AGN-driven outflows and the formation of Ly$\alpha$ nebulae around high-z quasars}

\author[Costa et al.]{
Tiago Costa$^{1}$\thanks{E-mail: tcosta@mpa-garching.mpg.de}, Fabrizio Arrigoni Battaia$^1$, Emanuele P. Farina$^2$, Laura C. Keating$^3$ \newauthor  Joakim Rosdahl$^4$ and Taysun Kimm$^5$
\\
$^{1}$Max-Planck-Institut f\"ur Astrophysik, Karl-Schwarzschild-Stra{\ss}e 1, D-85748 Garching b. M\"unchen, Germany \\
$^{2}$Gemini Observatory, NSF's NOIRLab, 670 N A'ohoku Place, Hilo, Hawaii 96720, USA \\
$^{3}$Leibniz-Institut f\"ur Astrophysik Potsdam, An der Sternwarte 16, D-14482 Potsdam, Germany \\
$^{4}$CRAL, Universit\'e de Lyon I, CNRS UMR 5574, ENS-Lyon, 9 Avenue Charles Andr\'e, 69561, Saint-Genis-Laval, France \\
$^{5}$Department of Astronomy, Yonsei University, 50 Yonsei-ro, Seodaemun-gu, Seoul 03722, Republic of Korea
}
\date{Submitted 2022}

\pubyear{2022}

\begin{document}
\label{firstpage}
\pagerange{\pageref{firstpage}--\pageref{lastpage}}
\maketitle
\begin{abstract}
The detection of Ly$\alpha$ nebulae around $z\gtrsim 6$ quasars provides evidence for extended gas reservoirs around the first rapidly growing supermassive black holes.
Observations of $z > 6$ quasars can be explained by cosmological models provided that the black holes by which they are powered evolve in rare, massive dark matter haloes. Whether these theoretical models also explain the observed extended Ly$\alpha$ emission remains an open question.
We post-process a suite of cosmological, radiation-hydrodynamic simulations targeting a quasar host halo at $z>6$ with the Ly$\alpha$ radiative transfer code {\sc Rascas}. 
A combination of recombination radiation from photo-ionised hydrogen and emission from collisionally excited gas powers Ly$\alpha$ nebulae with a surface brightness profile in close agreement with observations.
We also find that, even on its own, resonant scattering of the Ly$\alpha$ line associated to the quasar's broad line region can also generate Ly$\alpha$ emission on $\sim 100 \, \rm kpc$ scales, resulting in comparable agreement with observed surface brightness profiles. 
Even if powered by a broad quasar Ly$\alpha$ line, Ly$\alpha$ nebulae can have narrow line-widths $\lesssim 1000 \, \rm km \, s^{-1}$, consistent with observational constraints.
Even if there is no quasar, we find that halo gas cooling produces a faint, extended Ly$\alpha$ glow. However, to light-up extended Ly$\alpha$ nebulae with properties in line with observations, our simulations unambiguously require quasar-powered outflows to clear out the galactic nucleus and allow the Ly$\alpha$ flux to escape and still remain resonant with halo gas.
The close match between observations and simulations with quasar outflows suggests that AGN feedback already operates before $z \, = \, 6$ and confirms that high-$z$ quasars reside in massive haloes tracing overdensities.
\end{abstract}




\section{Introduction}
\label{sec:Introduction}

Luminous quasars have now been detected out to $z \, \gtrsim \, 7.5$, suggesting that supermassive black holes with masses $\gtrsim 10^9 \, \rm M_\odot$ have already assembled by the time the Universe is $\approx 680 \, \rm Myr$ old \citep{Banados:18, Yang:20, Wang:21}. These observations challenge theoretical models of galaxy evolution, which have to explain how such rapid black hole growth can take place at $z > 6$. Even if seed black holes produced at $z > 20$ are massive ($\sim 10^5 \, \rm M_\odot$), black hole growth has to proceed close to the Eddington rate for a Hubble time \citep[see e.g.][]{Wang:21}. Alternatively, supermassive black holes must undergo sustained super-Eddington accretion to reach masses $\sim 10^9 \, \rm M_\odot$ by $z \,  =  \, 7.5$.
In either growth scenario, the cosmological sites of $z > 6$ quasars have to ensure abundant gas inflow onto the quasar host galaxy and, once in the central galaxy, efficient transport towards the sphere of influence of the central black hole.

In order to explain masses of $\gtrsim 10^9 \, \rm M_\odot$ at $z > 6$, galaxy evolution models based on $\Lambda$CDM cosmology require the black holes powering bright quasars to grow in high-$\sigma$ peaks ($\sigma \gtrsim 4$) of the cosmic density field \citep[e.g.][]{Efstathiou:88, Volonteri:06}. These regions collapse prematurely, assembling rare dark matter haloes with virial masses $M_{\rm vir} \gtrsim 10^{12} \, \rm M_\odot$ by $z > 6$. Frequent merging \citep{Li:07} and smooth inflow of cold gas filaments into the central few kpc \citep{Sijacki:09,DiMatteo:12, Costa:14} allow these haloes to host rapid black hole growth to $\sim 10^9 \, \rm M_\odot$.  

Cosmological, hydrodynamic simulations following black hole growth in massive haloes with $M_{\rm vir} \gtrsim 10^{12} \, \rm M_\odot$ at $z > 6$ predict that quasars should trace gas overdensities and lie at the intersection of an extended network of cool $T \sim 10^4 \, K$ gas streams \citep[][]{DiMatteo:12,Dubois:12,Costa:14}. According to these simulations, gas streams should be flowing in towards the quasar host galaxy from multiple directions, colliding, cancelling angular momentum and sinking into the galactic nucleus.
This prediction remains observationally untested.

Another prediction, shared by virtually every model \citep{Dubois:13, Costa:14, Costa:15, Costa:18, Curtis:16, Barai:18, Ni:18, Lupi:21} is that active galactic nucleus (AGN) feedback \citep[see][for a review]{Fabian:12} should power large-scale outflows from the host galaxies of $z > 6$ quasars. The detection of broad absorption line features in $z > 6$ quasars \citep[e.g.][]{Mazzucchelli:17,Meyer:19,Schindler:20,Wang:21, Yang:21} points to the presence of small-scale winds with speeds up to $\sim 0.1 c$ in the nuclei of quasar host galaxies. According to detailed models \citep[e.g.][]{Costa:20}, such winds can drive out nuclear interstellar medium and power large-scale galactic outflows. Observational evidence for large-scale outflows at $z > 6$, however, remains anecdotal. \citet{Maiolino:12} and \citet{Cicone:15}, for instance, report the detection of a $30 \rm kpc$ scale outflow traced by \lbrack C\ensuremath{\,\textsc{ii}}\rbrack\,158\,$\mu$m emission in a quasar at $z \, = \,  6.4$ \citep[cf.][]{Meyer:22}. Targeting a larger sample of 17 $z > 6$ quasars, \citet{Novak:20}, however, find no evidence of ubiquitous large-scale outflows, at least as traced by \lbrack C\ensuremath{\,\textsc{ii}}\rbrack\,158\,$\mu$m emission, though \citet{Stanley:19} report potential outflow signatures using stacked spectra. 

For about a decade, observational evidence for the presence of extended Ly$\alpha$ nebulae around $z > 6$ quasars has been mounting \citep{Goto:09,Willott:11,Farina:17,Drake:19, Momose:19}.
In their REQUIEM survey, \citet{Farina:19} perform a comprehensive search for Ly$\alpha$ nebulae around $z > 5.7$ quasars. Out of a sample of 31 quasars in the redshift range $z \, = \, 5.7 \--  6.6$,  \citet{Farina:19} report the detection of 12 extended Ly$\alpha$ nebulae. These nebulae display a range of morphologies: at times approximately spherical and centred on the quasar, but often displaying strong asymmetries with the quasar lying on the outskirts of the emitting gas.
Proper diameters range from $15\, \rm kpc$ to $45 \, \rm kpc$ and total luminosities have a span of $L_{\rm Ly \alpha} \, = \, 10^{43} \-- 2 \times 10^{44} \, \rm erg \, s^{-1}$. When correcting for redshift dimming and scaling distances by the virial radius, \citet{Farina:19} find little redshift evolution in the median Ly$\alpha$ surface brightness (SB) profile down to $z \approx 3$, when Ly $\alpha$ nebulae are ubiquitously detected around bright quasars \citep{Borisova:16, Fumagalli:16,Husemann:18, Arrigoni-Battaia:19,Fossati:21}. A drop in surface brightness has, however, been reported at lower redshift \citep{Cai:19, O'Sullivan:20}.

In addition to mapping the cool gas reservoirs surrounding quasar host galaxies, Ly$\alpha$ nebulae encode a wealth of information about the dynamics of their circum-galactic media (CGM). The redistribution of Ly$\alpha$ photons in frequency space induced by scattering \citep[see e.g.][]{Dijkstra:14} is influenced by the temperature, density and velocity of neutral hydrogen in the CGM. The shape and width of the observed spectral line thus constrains the importance of large-scale rotation \citep[e.g.][]{Martin:15, Prescott:15}, inflows \citep{Villar-Martin:07,Humphrey:07,Martin:14} and outflows \citep{Gronke:16, Yang:16}. The observed morphology of Ly$\alpha$ nebulae may also contain information about the geometry of the quasar's light-cone, making it possible to pin-down the orientation of obscuring gas and dust on circum-nuclear scales \citep{DenBrok:20}.

Extended Ly$\alpha$ emission has, in fact, long been predicted to surround $z > 6$ quasars. \citet{Haiman:01} propose that, as cool gas is photo-ionised by the central quasar recombines, it should produce a Ly$\alpha$ ``fuzz'' on scales $\sim 10 \, \rm kpc$. Other origin scenarios include direct emission from collisionally excited gas \citep{Haiman:00, Furlanetto:05, Dijkstra:09, Faucher-Giguere:10, Goerdt:10} or scattering of Ly$\alpha$ photons produced in the interstellar medium embedded within galaxies \citep{Hayes:11, Humphrey:13, Beck:16}. These origin scenarios remain heavily disputed.

While multiple cosmological simulations have by now succeeded in reproducing the estimated masses of supermassive black holes at $z > 6$, no attempt has been made to predict their associated extended Ly$\alpha$ emission with such simulations. Studies have begun to employ cosmological simulations to pin down the origin of extended Ly$\alpha$ emission, but these have mostly concentrated on massive haloes at $z \approx 2$ \citep{Rosdahl:12,Cantalupo:14, Gronke:17} or lower mass haloes \citep{Smith:19,Mitchell:21,Byrohl:21}, in all cases without an on-the-fly treatment of quasar radiation.

This paper proposes a theoretical explanation for observations of extended Ly$\alpha$ nebulae around $z > 6$ quasars, though we argue we expect our results to equally apply to $z \approx 3$ quasars. 
In Section~\ref{sec:Simulations}, we describe the cosmological simulations employed in our study along with the Ly$\alpha$ radiative transfer technique that we adopt to generate mock datacubes from our simulations. In Section~\ref{sec:Results}, we present mock Ly$\alpha$ maps, surface brightness profiles, spectral line profiles and compare with available data. We discuss the broader implications of our results in Section~\ref{sec:discussion}. In Section~\ref{sec:Conclusions}, we summarise our main conclusions. Our reference observational sample, the REQUIEM Survey \citep{Farina:19}, assumes a $\Lambda$CDM cosmology with a Hubble constant of $H_{\rm 0} \, = \, 70 \, \rm km \, s^{-1} \, Mpc^{-1}$, a matter density parameter $\Omega_{\rm m} \, = \, 0.3$ and a dark energy density parameter $\Omega_{\rm \Lambda} \, = \, 0.7$, close to the cosmological parameters adopted in our simulations (Section~\ref{sec:Simulations}).

\section{ Simulations }
\label{sec:Simulations}

In this section, we describe the numerical simulations performed and analysed in this study. 
We start by describing our cosmological, radiation-hydrodynamic simulations (Section~\ref{sec:cRHD}). Section~\ref{sec:LyRT} outlines the radiative transfer code that is applied in post-processing to our radiation-hydrodynamic simulations in order to model Ly$ \alpha$ photon transport in quasar environments.

\subsection{Cosmological, radiation-hydrodynamic, ``zoom-in'' simulations}
\label{sec:cRHD}

The Ly$\alpha$ emissivity depends sensitively on the ionisation state of hydrogen. The non-equilibrium ionisation states are strongly influenced by the ionising fluxes of both young stellar populations and AGN (especially in the environment of a bright $z = 6$ quasar), but also by hydrodynamic processes, including gravitational accretion shocks and galactic feedback. Realistic modelling of Ly$\alpha$ emission in quasar environments can thus best be achieved via cosmological, radiation-hydrodynamic simulations.

\subsubsection{A quasar host halo at $z \, = \, 6$}
We use a set of cosmological, radiation-hydrodynamic (RHD), ``zoom-in'' simulations targeting a massive halo with $M_{\rm vir} \, = \, 2.4 \times 10^{12} \, \rm M_\odot$ at $z \, = \, 6$. These simulations comprise a small spherical volume with a radius $\approx 5 R_{\rm  vir} \approx 300 \, \rm kpc$, where the virial radius is $R_{\rm  vir} \approx 60 \, \rm kpc$ at $z \, = \, 6$, followed at high-resolution. In order to save computational cost and at the same time model the tidal torque field operating on the target halo, the remainder of the cosmological box is also followed, but only at coarse resolution.
The only selection criterion for our target halo required it to be among the most massive haloes found at $z \, = \, 6$ within a large cosmological box with a comoving side length of $500 \, \rm h^{-1} \, Mpc$ \citep{Costa:18, Costa:19}. As shown in \citet{Costa:14}, such haloes represent the likely hosts of bright $z > 6$ quasars. As a consequence of their low number density \citep[e.g.][]{Springel:05}, simulating them requires following an unusually large cosmological box.
Our simulations adopt a concordance cosmology with $H_{\rm 0} \, \approx \, 69.3 \, \rm km \, s^{-1} \, Mpc^{-1}$, $\Omega_{\rm m} \, \approx \, 0.3$, $\Omega_{\rm \Lambda} \, \approx \, 0.7$ in line with the cosmological parameters assumed in the REQUIEM Survey \citep[see][]{Farina:19}, and a baryonic density parameter $\Omega_{\rm b} \, = \, 0.048$ \citep{Planck:16}.

The simulations are performed with {\sc Ramses-RT} \citep{Rosdahl:13,Rosdahl:15}, the coupled radiation-hydrodynamic (RHD) extension of the Eulerian, adaptive mesh refinement code {\sc Ramses} \citep{Teyssier:02}, solving for coupled gas hydrodynamics and radiative transfer of stellar and AGN radiation.
In order to solve for radiative transport, {\sc Ramses-RT} takes the first two angular moments of the radiative transfer equation, obtaining a system of conservation laws which is closed with the M1 closure for the Eddington tensor \citep{Levermore:84}. Radiation is advected between neighbouring cells using an explicit, first-order Godunov solver. In order to avoid prohibitively shot time-steps, {\sc Ramses-RT} adopts the reduced speed of light approximation, valid when the speed of light exceeds other characteristic speeds (e.g. outflow speed, ionisation front speed). We adopt a global reduced speed of light of $0.03c$, which is more rapid than fastest outflows driven in the simulation $\approx 3000 \, \rm km \, s^{-1}$ \citep[see also][for convergence tests]{Costa:18a,Costa:18}. Note that the reduced speed of light approximation is only employed in order to advect the radiation field. Physical processes impacting gas dynamics, such as radiation pressure, are treated using the full speed of light \citep[see][]{Rosdahl:15}.

In order to increase the numerical resolution, {\sc Ramses-RT} employs adaptive refinement. A gas cell is refined if it satisfies $\left( \rho_{\rm dm} + \Omega_{\rm dm} / \Omega_{\rm b} \rho_{\rm \star} + \Omega_{\rm dm} / \Omega_{\rm b} \rho_{\rm gas} \right) \Delta x^3 > 8 \rho_{\rm dm}$, where $\rho_{\rm dm}$, $\rho_{\rm \star}$, $\rho_{\rm gas}$ are, respectively, the dark matter, stellar and gas density within the cell, $\Omega_{\rm dm} / \Omega_{\rm b} \, = \, \Omega_{\rm m} / \Omega_{\rm b} - 1$, and $\Delta x$ is the cell width. The minimum cell size in our fiducial simulations is $\Delta x_{\rm min} \, = \, 80 \, \rm pc$. About $15 \%$ of cells located within the virial radius at $z \, = \, 6$ are refined down to the minimum cell size allowed in the simulations.

The simulations also track the N-body dynamics of stars and dark
matter using a particle-mesh method and cloud-in-cell interpolation. Stars and dark matter are modelled with particles of minimum mass $m_{\rm \star} \, \approx \, 10^4 \, \rm M_\odot$
and $m_{\rm DM} \, = \,  3 \times 10^6 \, \rm M_\odot$, respectively. Due to the mass resolution afforded by our cosmological simulations, star particles should be viewed as sampling full stellar populations (see Section~\ref{sec:cooling}).

In the light of the results presented in this paper, it is worth emphasising that our simulations are in no way tuned to yield a realistic treatment of the CGM. The CGM is notoriously under-resolved even in ``zoom-in'' cosmological simulations \citep[see e.g. the resolution studies in][]{vandeVoort:19,Hummels:19}. The typical cell size in the CGM of our target halo is $\approx 350 \, \rm pc$. \citet{Bennett:20} explore enhancing the resolution in shock fronts (resolving structures as small as $\approx 20 \, \rm pc$) within a $z \, = \, 6$ halo, more in line with the halo targeted by our RHD simulations, finding a significant enhancement in the HI covering fractions even at $R > R_{\rm vir}$. As resolution increases, we should expect the CGM to comprise more numerous, smaller, but denser cloudlets than captured in current ``zoom-in'' simulations. Whether these cloudlets really fragment all the way to form a fine ``mist'' containing a large number of small clouds \citep{McCourt:18} or whether they coagulate to assemble larger clouds \citep{Gronke:20}, however, is a fundamental question that remains unanswered.

\subsubsection{Cooling, star formation and supernova feedback}
\label{sec:cooling}
Our simulations track the non-equilibrium ionisation states of both hydrogen (H) and helium (He), which are coupled to the radiative fluxes followed in the RHD simulations (see Section~\ref{sec:agnradiation}), following Bremsstrahlung, collisional excitation, collisional ionisation, Compton cooling off the cosmic microwave background, di-electronic recombination and photo-ionisation.
The cooling contribution from metals is computed at $T \geq 10^4 \, \rm K$ using {\sc CLOUDY} \citep{Ferland:98}, assuming photo-ionisation equilibrium with the UV background of \citet{Haardt:96}. For $T < 10^4 \, \rm K$ metal-line cooling is modelled using fine-structure cooling rates from \citet{Rosen:95}, allowing cooling down to a density-independent temperature of $T_{\rm 0} = 10 \, \rm K$.

Star formation rate is modelled assuming a Schmidt law with a variable star formation efficiency, closely following the turbulent star formation implementation described in \citet{Kimm:17}. Star formation can occur in a cell if:
\begin{itemize}
    \item the density is a local maximum and is locally growing with time, i.e. $\vec{\nabla} \cdot (\rho_{\rm gas} \vec{v}) < 0$,
    \item the local hydrogen number density satisfies $n_{\rm H} > 1 \, \rm cm^{-3}$ and the local overdensity exceeds the cosmological background mean by more than a factor $200$,
    \item the local gas temperature is $T < 2 \times 10^4 \, \rm K$,
    \item the turbulent Jeans length is unresolved. The turbulent Jeans length $\lambda_{\rm J}$ is defined following \citet{Federrath:12} as
    \begin{equation}
        \lambda_{\rm J} \, = \, \frac{\pi \sigma_{\rm gas}^2 + \left(36 \pi c_{\rm s}^2 G \Delta x^2 \rho_{\rm gas} + \pi^2\sigma_{\rm gas}^4 \right)^{1/2}}{6 G \rho \Delta x},
    \end{equation}
    where $G$ is the gravitational constant, $c_{\rm s}$ is the local speed of sound and $\sigma_{\rm gas}$ is the local gas turbulent velocity dispersion, obtained by computing the norm
    \begin{equation}
    \| \vec{\nabla} \vec{v} \|^2 \, = \, \mathrm{tr} \left( \vec{\nabla} \vec{v} \right) \left( \vec{\nabla} \vec{v} \right)^\mathrm{T}
    \label{Eq:norm}
    \end{equation}
    of the velocity gradient tensor using the six nearest neighbour cells. Eq.~\ref{Eq:norm} is computed after subtracting the symmetric divergence field and rotational velocity components from the local velocity $\vec{v}$.
\end{itemize}

Star particles are assumed to fully sample stellar populations.
Assuming a \citet{Kroupa:01} initial mass function (IMF), a fraction $\eta_{\rm SN} \, = \, 0.2$ of the initial stellar mass is returned by supernova (SN) explosions in a single event when their age exceeds $10 \, \rm Myr$. We assume an average supernova progenitor mass of $4 \, \rm M_\odot$, boosting the supernova rate above the value expected for a \citet{Kroupa:01} IMF by a factor $5$, giving a supernova rate of $0.05 \, \rm M_\odot^{-1}$. Like in \citet{Rosdahl:18}, this choice is made to (i) account for complementary feedback processes not followed explicitly by our simulations (e.g. cosmic rays, stellar winds, proto-stellar jets) and (ii) to ensure that supernova feedback is strong enough to reduce the stellar to halo mass ratio of central and satellite galaxies in our ``zoom-in" region to values broadly consistent with abundance matching predictions \citep[highly uncertain at $z > 6$, e.g.][]{Behroozi:19}. Our decision is also driven by results from \citet{Costa:14}, where strong supernova feedback is shown to be necessary in order to reconcile the number of observed satellite galaxies around $z \approx 6$ quasars with existing observational constraints.

Individual Type II SN explosions are assumed to inject an energy $E_{\rm SN} \, = \, 10^{51} \, \rm erg$ into their neighbouring cells.
Momentum is injected radially with a magnitude that depends on whether the Sedov-Taylor phase is resolved, as described in \citet{Kimm:14} and \citet{Kimm:15}. The aim of this model is to recover the correct terminal momentum associated to the snowplough phase of a supernova remnant, even if the cooling radius is unresolved. For each cell neighbouring a SN explosion site, the dimensionless parameter $\chi \, = \, \Delta M_{\rm swept}/ \Delta M_{\rm  ej}$ is computed, where $\Delta M_{\rm swept}$ is the sum of the original mass in the cell, and the share of mass it receives both from the SN ejecta $\Delta M_{\rm  ej}$ and from the SN host cell. The dimensionless parameter $\chi$ is then compared against a threshold value $\chi_{\rm tr}$ \citep[Eq. 20 in][]{Kimm:17} defining the transition between adiabatic and snowplough phases. If $\chi \geq \chi_{\rm tr}$, the adiabatic expansion phase of the SN remnant is poorly or not resolved. In this case, the radial momentum imparted to ambient gas is set to the value
\begin{equation}
    p_{\rm SN} \, = \, p_{\rm SN, snow} e^{-\Delta x / r_{\rm S}} + p_{\rm SN+PH} \left(1 - e^{-\Delta x / r_{\rm S}} \right) \, ,
\end{equation}
where the first term encapsulates the radial momentum associated to the SN snowplough phase and the second term accounts for the momentum boost obtained if the ambient gas through which the SN remnant propagates is pre-ionised \citep[][]{Geen:15}. The exponential factors ensure the \citet{Geen:15} correction is introduced only if the Str{\"o}mgren radius $r_{\rm S}$ is poorly resolved, i.e. if $r_{\rm S} \lesssim \Delta x$. The functional forms employed for $p_{\rm SN, snow}$ and $p_{\rm SN + PH}$ are, respectively,
\begin{equation}
    p_\mathrm{SN, snow} \, = \, 3 \times 10^5 \mathrm{km \, s^{-1} M_\odot} n_{\rm H}^{-2/17} \left( \frac{E_\mathrm{SN}}{10^{51} \, \mathrm{erg}} \right)^{16/17} Z^\prime{}^{-0.14}
  \end{equation} 
 where the $Z^\prime \, = \, \max{\left[Z/0.02, 0.01\right]}$ is the gas metallicity, and
\begin{equation}
    p_{\rm SN+PH} \, = \, 4.2 \times 10^5 \mathrm{km \, s^{-1} M_\odot} \left( \frac{E_{\rm SN}}{10^{51} \, \mathrm{erg}} \right)^{16/17} Z^\prime{}^{-0.14}.
\end{equation}

\noindent If $\chi < \chi_{\rm tr}$, then momentum is injected radially with a magnitude 
\begin{equation}
    p_{\rm SN} \, = \, \left( 2 \chi  f_{\rm e}(\chi) M_{\rm ej} E_{\rm SN}\right)^{1/2} \, , 
\end{equation}
where the function $f_{\rm e} (\chi) \, = \, 1 - \frac{\chi - 1}{3 (\chi_{\rm tr} - 1)}$ modulates the injected energy, smoothly connecting both high and low $\chi$ limits.

Besides energy, SN are assumed to inject metals with a yield of $0.075$. Gas phase metallicity is treated as a passive scalar and, in our simulations, is transported to the halo via outflows launched by SN and AGN. Gas is initiated with a homogeneous metallicity floor, which is used to compensate for the lack of molecular hydrogen cooling channels in our simulations. We adopt a metallicity floor of $0.00032 \, \rm Z_\odot$, callibrated such that the first stars form at $z \approx 15$.

\subsubsection{Stellar and AGN radiative feedback}
\label{sec:agnradiation}

\begin{table*}
    \centering
\begin{tabular}{ w{c}{3cm} w{c}{1cm} w{c}{1cm} w{c}{1cm} w{c}{1cm} w{c}{1cm} w{c}{1cm} w{c}{1cm} w{c}{1cm} w{c}{2cm}}
 \hline
 Photon Group & $\epsilon_{\rm 0}$ & $\epsilon_{\rm 1}$ & $\langle \epsilon \rangle$ & $\sigma_{\rm HI}$ & $\sigma_{\rm HeI}$ & $\sigma_{\rm HeII}$ & $\kappa_{\rm abs}^0$ & $\kappa_{\rm scat}^0$ & AGN contribution \\
   & $\rm [eV]$ & $\rm [eV]$ & $\rm [eV]$ & $\rm [cm^2]$ & $\rm [cm^2]$ & $\rm [cm^2]$ & $\rm [cm^2 \, g^{-1}]$ & $\rm [cm^2  \, g^{-1}]$ &  \\
  
 \hline
infrared   & 0.1   & 1        & 0.6  & 0 & 0 & 0 & 0 & 10 & 0.38  \\
optical    & 1     & 13.6     & 5.2  & 0 & 0 & 0 & $10^3$ & 0 & 0.46  \\ 
UVI        & 13.6  & 24.59    & 17.9 & $3.2 \times 10^{-18}$ & 0 & 0 & $10^3$ & 0 & 0.09 \\
UVII       & 24.59 & 54.42    & 33.0 & $5.9 \times 10^{-19}$ & $4.6 \times 10^{-18}$ & 0 & $10^3$ & 0 & 0.05    \\
UVIII      & 54.42 & $\infty$ & 73.3 & $5.9 \times 10^{-20}$ & $9.3 \times 10^{-19}$ & $8\times 10^{-19}$ & $10^3$ & 0 & 0.02 \\
 \hline
\end{tabular}
    \caption{Photon groups (first column), the lower and upper energies defining their energy interval (second and third columns), mean photon group energies (fourth column), ionisation cross-sections to HI, HeI and HeII (respectively, the fifth, sixth and seventh columns), the normalisation of the absorption and scattering opacities to dust (eighth and ninth columns) and energy fraction contributed by the quasar for each group (last column).}
    \label{tab:radiation}
\end{table*}

In order to inject radiation fluxes from stellar populations and AGN, we sample their frequency space with five ``photon groups''. 
The radiation frequency ranges, characteristic energies, ionisation cross-sections and dust opacities associated to each group are given in Table~\ref{tab:radiation}. These groups include three UV frequency bins, bounded by the ionisation potentials of HI, HeI and HeII. Besides photo-ionisation and photo-heating, photons in these UV groups interact with ambient gas via radiation pressure from photo-ionisation and from dust (see below). 
The other two photon groups include optical and infrared photons. These are not sufficiently energetic to ionise hydrogen or helium, but can still interact with ambient gas via radiation pressure on dust.
Dust is assumed to be mixed with gas in proportion to the local metallicity, with absorption and scattering opacities $\kappa_{\rm abs}^0$ and $\kappa_{\rm scat}^0$.  
The pseudo-dust number density is assumed to follow $n_{\rm dust} \, = \, \left( Z/\mathrm{Z_\odot} \right) n_{\rm H}$ following \citet{Rosdahl:15b}.
If absorbed by dust, the flux of any given photon group is then added to the infrared group, where the only interaction with gas occurs via multi-scattering radiation pressure \citep{Rosdahl:15, Costa:18a}.

The luminosity of stellar particles is set based on their age, mass and metallicity using the spectral energy 
models of \citet{Bruzual:03}, following the procedure described in Appendix D of \citet{Rosdahl:18}.
The quasar spectral energy distribution is modelled using the unobscured composite spectrum given in \citet{Hopkins:07}. We also experimented using the harder, unobscured spectrum of \citet{Sazonov:04}, finding no significant difference in our results. We model a single quasar by placing a black hole particle of mass $\sim 10^9 \, \rm M_\odot$ at the potential minimum of the most massive galaxy at some redshift $z_{\rm QSO}$.
Note that the original aim of the simulations presented in this paper was to conduct controlled experiments on the efficiency of AGN radiative feedback in the spirit of earlier simulations by \citet{Costa:18}, and to remove the sensitivity of our results on highly uncertain black hole growth models (see Section~\ref{sec:openquestions}).
After seeding a single black hole, we explore varying the AGN light-curve and the quasar bolometric luminosity. AGN radiation is `switched-on' at $z_{\rm QSO} \, = \, 6.5$ in most of our simulations, but we also explore switching it on at $z_{\rm QSO} \, = \, 7.7$ in order to test AGN feedback in the most distant quasars observed to date.

We adopt two types of light-curve: (i) the AGN radiates constantly at a fixed bolometric luminosity $L_{\rm bol}$, (ii) the AGN switches on and off, following a square-wave lightcurve with a specific period $\tau_{\rm cycle}$ and quasar lifetime $\tau_{\rm QSO}$.
We assume a duty cycle $\approx 90 \%$ \citep[consistent with the cosmological simulations][]{Costa:14} and a quasar lifetime of $1 \, \rm Myr$ \citep[consistent with the observational constraints of high-$z$ quasar lifetimes of][]{Khrykin:21}, such that $\tau_{\rm cycle} \, = \, 1.1 \, \rm Myr$.
We sample bolometric luminosities ranging from $L_{\rm bol} \, = \, 10^{47} \, \rm erg \, s^{-1}$ to $L_{\rm bol} \, = \, 5 \times 10^{47} \, \rm erg \, s^{-1}$, encompassing the typical range of observed $z \, = \, 6$ quasar luminosities. We do not probe fainter AGN luminosities, because these do not generate sufficient momentum flux to launch large-scale outflows in quasar host galaxies at $z > 6$ \citep{Costa:18}. 
We name our simulations according to the bolometric luminosity of the quasar. For instance, in \texttt{Quasar-L3e47}, the characteristic bolometric luminosity of the quasar is $3 \times 10^{47} \, \rm erg \, s^{-1}$, while in \texttt{Quasar-L5e47}, the bolometric luminosity is $5 \times 10^{47} \, \rm erg \, s^{-1}$. In one of our simulations (\texttt{noQuasar}), no quasar radiation is injected and there is thus no AGN feedback. This simulation, which still follows feedback from supernovae and stellar radiation, allows us to control for the impact of AGN feedback.
In other simulations, denoted e.g. \texttt{Quasar-L3e47-continuous}, AGN radiation is injected continuously at a constant rate.
All simulations account for photo-ionisation, photo-heating and radiation pressure on dust by stellar populations and (if present) from a quasar.

In order to model the ionising flux from sources external to our ``zoom-in'' region, we adopt the spatially homogeneous and time-evolving UV background of \citet{Haardt:12}.
We apply a self-shielding correction that damps the ionising background in cells with $n_{\rm H} > 0.01 \, \rm cm^{-3}$ \citep[see e.g.][]{Rosdahl:12}.

\subsection{Ly$\alpha$ radiative transfer}
\label{sec:LyRT}

Ly$\alpha$ is a resonant line and its absorption is followed by re-emission on a timescale of $\sim 10^{-9} \, \rm s$ \citep[see e.g.][]{Dijkstra:14}. When propagating through HI gas, Ly$\alpha$ photon transport can thus be treated as a scattering process. 
With every absorption event, the frequency at which Ly$\alpha$ photons are re-emitted is shifted due to both the temperature and velocity of the ambient HI gas. The emerging Ly$\alpha$ spectrum is thus shaped both by the properties of the Ly$\alpha$ sources and the medium through which the Ly$\alpha$ flux travels.
Tracing the spectral and spatial diffusion of Ly$\alpha$ photons in arbitrarily complex media, such as the interstellar medium or the CGM, requires detailed radiative transfer calculations.

\subsubsection{{\sc RASCAS}}
\label{sec:rascas}
We perform Ly$\alpha$ radiative transfer in post-processing using the publicly available, massively-parallel code {\sc Rascas} \citep{Michel-Dansac:20}. 
{\sc Rascas} employs a Monte Carlo technique in order to evolve the spatial and spectral diffusion of resonant line photons.

The photon distribution is sampled with a discrete number of photon `packets' $N_{\rm MC}$. We use $N_{\rm MC} \, = \, (1 \-- 5) \times 10^6$. The number of photon packets generated by any given source is proportional to its real number photon emission rate $\dot{N}_{\rm Ly\alpha}$. Thus, if the real, total photon production rate in the entire simulation domain is $\dot{N}_{\rm Ly\alpha}^{\rm tot}$, then the probability that a photon packet is emitted from a source is $\dot{N}_{\rm Ly\alpha} / \dot{N}_{\rm Ly\alpha}^{\rm tot}$.

The emission frequency of each photon packet is calculated in the reference frame of its parent cell. A photon packet's frequency is randomly drawn assuming a Gaussian line profile with a width $\Delta v_{\rm D}$, set by the thermal broadening caused by random motions of the constituent hydrogen atoms.
We compute $\Delta v_{\rm D} \, = \, \nu_{\rm 0} \left( 2 k_{\rm B} T/m_{\rm p} \right)^{1/2} c^{-1}$, where $c$ is the speed of light in vacuum, $k_{\rm B}$ is Boltzmann’s constant, $m_{\rm p}$ the proton mass, $T$ is the gas temperature and $\nu_{\rm 0} \, = \, 2.47 \times 10^{15} \, \rm s^{-1}$ is the frequency corresponding to Ly$\alpha$ line resonance. The frequency of any given photon packet is then shifted to an external frame according to the source's velocity.

Photon packets are initialised with random orientations -- Ly$\alpha$ sources are assumed to be isotropic.
In scattering events, however, the outgoing direction $\vec{k}_{\rm out}$ of a photon packet is related to its incoming direction $\vec{k}_{\rm in}$  through a phase function $P\left(|\vec{k}_{\rm in} \cdot \vec{k}_{\rm out}| \right)$ that depends on the photon's frequency in the scatterer's frame $\nu_{\rm scat, \, in} \, = \, \nu_{\rm in} \left(1 - \vec{k}_{\rm in} \cdot \vec{v}_{\rm scat}/c \right)$, where $\nu_{\rm in}$ is the incoming photon's rest-frame frequency and $\vec{v}_{\rm scat}$ is the scatterer's velocity. If $|\nu_{\rm scat, \, in} - \nu_{\rm scatt, 0}| \geq 0.2 \Delta \nu_{\rm D}$, where $\nu_{\rm scatt, 0}$ is Ly$\alpha$ line resonance frequency in the scatterer's frame, then
\begin{equation}
P\left(|\vec{k}_{\rm in} \cdot \vec{k}_{\rm out}| \right) \, = \, \frac{11}{24} + \frac{3}{24}|\vec{k}_{\rm in} \cdot \vec{k}_{\rm out}|^2 \, .
\label{Eq:phasefunc1}
\end{equation}

\noindent Otherwise, if $|\nu_{\rm scat, \, in} - \nu_{\rm 0}| < 0.2 \Delta \nu_{\rm D}$, then
\begin{equation}
P\left(|\vec{k}_{\rm in} \cdot \vec{k}_{\rm out}| \right) \, = \, \frac{3 \left( 1 + |\vec{k}_{\rm in} \cdot \vec{k}_{\rm out}|^2 \right)}{8} \, . 
\label{Eq:phasefunc2}
\end{equation}

\noindent Eqs.~\ref{Eq:phasefunc1} and~\ref{Eq:phasefunc2} are then inverted to compute $\vec{k}_{\rm out}$ \citep[see][]{Michel-Dansac:20}.

In order to model Ly$\alpha$ sources, {\sc Rascas} follows (i) recombination radiation from photo-ionised gas, (ii) collisional excitation and subsequent Ly$\alpha$ cooling, and (iii) stars and AGN. In the following, we outline how each of these processes are modelled in {\sc Rascas}.

\subsubsection{Recombination radiation}
In this scenario, Ly$\alpha$ radiation is generated as a result of a recombination cascade from photo-ionised gas.
Following \citet{Cantalupo:08}, the number of Ly$\alpha$ emitted per unit time in a given cell is calculated in {\sc Rascas} as
\begin{equation}
    \dot{N}_{\rm Ly\alpha , rec} \, = \, n_{\rm e} n_{\rm p} \epsilon_{\rm Ly\alpha}^{\rm B} (T) \alpha^{\rm B}(T) \left( \Delta x \right)^3 \, ,
\end{equation}
where $n_{\rm e}$ and $n_{\rm p}$ are the non-equilibrium free electron and proton number densities output by {\sc Ramses-RT}, $\alpha^{\rm B}(T)$ is the case B recombination coefficient \citep[set using the fit from][]{Hui:97} and $\epsilon_{\rm Ly\alpha}^{\rm B} (T)$ is the number of Ly$\alpha$ photons produced per recombination event \citep[equation 2 in][]{Cantalupo:08}. The latter is a weak function of temperature, varying between 0.68 and 0.61 for $10^4 \, \mathrm{K} < T < 10^{4.7} \, \mathrm{K}$, the typical temperature range of gas photo-ionised by young stars and AGN.

Recombination radiation has been proposed as the chief source of giant Ly$\alpha$ nebulae \citep[e.g.][]{Cantalupo:14}. Reconciling the surface brightness levels of observed Ly$\alpha$ nebulae at $z \, \approx \, 2 \-- 3$ with a recombination radiation origin alone is, however, only possible if ionised hydrogen reaches ISM-like densities $\gtrsim 1 \, \rm cm^{-3}$ \citep{Arrigoni-Battaia:15} at scales $\sim (10 \-- 100) \, \rm kpc$, comparable or beyond the virial radii of their host haloes. These high densities translate to clumping factors $C \, = \, \langle n^2 \rangle / \langle n \rangle^2\sim 1000$, consistent with a picture in which the CGM is pervaded by a fog-like distribution of ionised, low volume-filling cloudlets \citep{McCourt:18}.

\subsubsection{Collisional excitation}
In a different scenario, Ly$\alpha$ nebulae are generated by collisionally excited hydrogen. As the gas de-excites and cools, it generates Ly$\alpha$ photons.
The number of Ly$\alpha$ photons emitted per unit time in a given cell is computed as 
 \begin{equation}
    \dot{N}_\mathrm{Ly\alpha, col} \, = \, n_{\rm e} n_{\rm HI} C_{\rm Ly\alpha}(T) \left( \Delta x \right)^3 \, ,
\end{equation}
where $C_{\rm Ly\alpha}(T)$ is the rate of collisional excitations from level $1$s to level $2$p, evaluated using the fit of \citet{Goerdt:10}. This rate is a very strong function of temperature, increasing by $\sim 3$ orders of magnitude between the temperature range $10^4 \, \mathrm{K} < T < 3 \times 10^4 \, \mathrm{K}$. Above $T \approx 5 \times 10^4 \, \rm K$, collisional excitation is inefficient, as most hydrogen gas becomes ionised. Thus, while collisional excitation can be a very efficient source of Ly$\alpha$ photons, it only operates over a narrow temperature range.

\subsubsection{Scattering from the broad line region}
\label{sec:scatteringBLR}
An extreme case involves generating Ly$\alpha$ nebulae via scattering off neutral hydrogen starting from a point source.
Quasar spectra typically include a prominent, broad Ly$\alpha$ line. A potential origin scenario for giant Ly$\alpha$ nebulae is the direct transport of Ly$\alpha$ photons from the broad line region (BLR) to scales of $10 \-- 100 \, \rm kpc$ via scattering.

The challenge associated to this scenario is that it requires Ly$\alpha$ photons to either be in resonance with HI in the halo or to scatter efficiently in the wing of the Ly$\alpha$ line.
Both conditions can be difficult to satisfy. If Ly$\alpha$ photons are resonantly trapped in the central regions of the halo, where HI gas might be most abundant, escape likely occurs with a single fly-out if the photons are scattered into the wings of the Ly$\alpha$ line. To inflate a large nebula, the photons have to scatter in the wings of the line, which is possible only if optical depths remain high throughout the halo. 

We test this extreme scenario and model BLR Ly\--$\alpha$ emission by assigning a Gaussian line with width $\sigma_{\rm BLR}$ to a point source positioned at the location and at the rest-frame velocity of the black hole particle.
We use $\sigma_{\rm BLR} \, = \, 1000 \-- 1500 \, \rm km \, s^{-1}$. These values respectively correspond to full-widths-at-half-maximum of $\rm FWHM_{\rm BLR} \, \approx \, 2400 \-- 3500 \, \rm km \, s^{-1}$, consistent with the widths of broad line region emission lines in $z \approx 6$ quasars \citep[e.g.][]{Mazzucchelli:17, Reed:19}.
The Ly$\alpha$ luminosity of the source is parametrised as $f_{\rm Ly \alpha} L_{\rm bol}$, where $L_{\rm bol}$ is the quasar bolometric luminosity adopted in the parent radiation-hydrodynamic simulation.
We select $f_{\rm Ly \alpha}$, which is the fractional quasar luminosity associated to the Ly$\alpha$ line, based on observational constraints.
This fraction can vary from object to object, but is typically $f_{\rm Ly \alpha} \approx 0.001 - 0.1$ \citep{Lusso:15}. At $z \approx 2$, for instance, \citet{Cantalupo:14} find $f_{\rm Ly \alpha} \approx 0.006$. In line with values typical for $z \approx 6$ quasars \citep{Koptelova:17}, we adopt $f_{\rm Ly \alpha} \approx 0.005$ as our fiducial value. In our various experiments, we, however, test varying this parameter within the range $0.001 \leq f_{\rm Ly \alpha} \leq 0.01$.

Ly$\alpha$ emission from the broad line region at scales $\lesssim 1 \, \rm pc$ likely itself consists of reprocessed ionising flux from the AGN \citep[e.g.][]{Osterbrock:06}.
Adding a point source to model BLR Ly$\alpha$ emission is justified because the scales associated to the BLR cannot be directly resolved in our cosmological simulations. When we consider surface brightness profiles, nebula luminosities and spectra in Section~\ref{sec:Results}, the question arises whether extended emission resulting from BLR scattering can be simply added to that of resolved recombination radiation and collisional excitation without double-counting the Ly$\alpha$ luminosity. We take a conservative approach in which we consider these different emission scenarios independently as well as together in order to bracket all possible scenarios.

\subsubsection{Dust absorption}
\label{sec:dustabsorption}
We model Ly$\alpha$ absorption by dust following \citet{Laursen:09}, computing the dust number density $n_{\rm dust}$ as

\begin{equation}
    n_{\rm dust} \, = \, \frac{Z}{Z_{\rm 0}} \left( n_{\rm HI} + f_{\rm ion} n_{\rm HII} \right) ,
\end{equation}
\label{eq:dustdensity}
 where $f_{\rm ion} \, = \, 0.01$, $n_{\rm HI}$ is the neutral hydrogen number density, and $n_{\rm HII}$ is the ionised hydrogen number density. 

The dust absorption cross-section $\sigma_{\rm dust}$ is shown in \citet{Laursen:09} to be largely frequency-independent around Ly$\alpha$ resonance and is thus set to a constant value. We explore both ``Large Magellanic Cloud'' (LMC) and ``Small Magellanic Cloud'' (SMC) models introduced in \citet{Laursen:09}, which are incorporated into {\sc Rascas} and adopted in previous studies \citep[e.g.][]{Gronke:17}.
For the SMC case, which we adopt as our fiducial model, $\kappa_{\rm abs} \, = \, \sigma_{\rm dust} / m_{\rm p} \approx 960 \, \rm cm^{2} \, g^{-1}$, while $\kappa_{\rm abs} \approx 840 \, \rm cm^{2} \, g^{-1}$ in the LMC case, both close (respectively within $4\%$ and $15\%$) to the fiducial value for the dust absorption opacity adopted in our radiation-hydrodynamic simulations. We explicitly verified that both models result in almost indistinguishable Ly$\alpha$ escape fractions and surface brightness profiles. 

When a Ly$\alpha$ photon interacts with a dust grain, there is some probability that, instead of being absorbed, the photon is scattered. This probability is set by the dust albedo $a_{\rm dust}$, which we set to $a_{\rm dust} \, = \, 0.32$ following \citet{Li:01}. After dust scattering, the outgoing photon direction is set by the phase function
\begin{equation}
P\left(|\vec{k}_{\rm in} \cdot \vec{k}_{\rm out}| \right) \, = \, \frac{1}{2} \frac{1 - g^2}{\left(1 + g^2 - 2g|\vec{k}_{\rm in} \cdot \vec{k}_{\rm out}|\right)^{3/2}} \, ,
\end{equation}
where the asymmetry parameter $g$ is set to $g \, = \, 0.73$ following \citet{Li:01}.

\subsubsection{Data-cube construction}
We use the ``peeling algorithm'' to collect the Ly$\alpha$ flux in a data-cube with $N \times N$ spatial pixels and $N_{\rm \lambda}$ spectral bins of width $\Delta \lambda_{\rm obs}$. The peeling algorithm loops over every photon packet, treating each scattering event as a point source, and adding the flux contribution to each bin of the data-cube. Each photon packet contributes with a luminosity $L_{\rm Ly \alpha} / N_{\rm ph}$, where $L_{\rm Ly \alpha}$ is the total luminosity. The probability that a photon packet escapes into the line-of-sight of an observer at luminosity distance $D_{\rm L}$ and into a wavelength interval $\Delta \lambda_{\rm obs}$ is $P\left(|\vec{k}_{\rm in} \cdot \vec{k}_{\rm out}| \right) e^{-\tau_{\rm esc}(\lambda)}$, where $P\left(|\vec{k}_{\rm in} \cdot \vec{k}_{\rm out}| \right)$ is the phase function given in Eqs.~\ref{Eq:phasefunc1} and~\ref{Eq:phasefunc2} and $\tau_{\rm esc} (\lambda)$ is the optical depth towards the edge of the computational domain.
The spectral flux density $F^{\rm \lambda}_{\rm Ly\alpha, pix}$, defined as the amount of energy $\Delta E$ received in a pixel per unit time $\Delta t$, per unit area $\Delta A$, per observed wavelength interval $\Delta \lambda_{\rm obs}$ is 
\begin{eqnarray}
F^{\rm \lambda}_{\rm Ly\alpha, pix} & \, = \, & \frac{\Delta E}{\Delta t \Delta A \Delta \lambda_{\rm obs}} \\ \nonumber & \, = \, & \frac{L_{\rm \lambda}/N_{\rm ph}}{4 \pi D_{\rm L}^2 (1+z)}  \sum{P\left(|\vec{k}_{\rm in} \cdot \vec{k}_{\rm out}| \right) e^{-\tau_{\rm esc}(\lambda)}} \, ,
\end{eqnarray}
where $L_{\rm \lambda} \, = \, L_{\rm Ly \alpha} / \left[\Delta \lambda_{\rm obs} (1+z)^{-1} \right]$, and the sum is performed over all photon packets and all scattering events. Integrating over wavelength gives the Ly$\alpha$ flux $F_{\rm Ly\alpha, pix} \, = \, \sum_{\rm \lambda}{F^{\rm \lambda}_{\rm Ly\alpha}} \Delta \lambda_{\rm obs}$ per pixel.

If a pixel subtends a solid angle $\Delta \Omega_{\rm pix} \sim (\Delta \theta_{\rm pix})^2$, where $\Delta \theta$ is the pixel size in arcsec, we define the surface brightness $SB_{\rm Ly\alpha, pix}$ as
\begin{equation}
SB_{\rm Ly\alpha , \, \rm pix} \, =  \,  \frac{L_{\rm Ly \alpha}/N_{\rm ph}}{4 \pi D_{\rm L}^2 (\Delta \theta_{\rm pix})^2} \sum{P\left(|\vec{k}_{\rm in} \cdot \vec{k}_{\rm out}| \right) e^{-\tau_{\rm esc}(\lambda)}} \, .
\end{equation}
\noindent Since $\Delta \theta_{\rm pix} \approx l / D_{\rm A}$, where $l$ is the physical scale probed by a pixel and $D_{\rm A}$ is the angular diameter distance, and $D_{\rm A} \, = \, D_{\rm L} (1 + z)^{-2}$, note that the surface brightness scales as $SB_{\rm Ly\alpha , \, \rm pix} \propto (1+z)^{-4}$.

We use $N \times N \,=\, 250 \times 250$ pixels over a field of view $\approx 43 \, \rm arcsec$ centred on the position of the quasar. This field of view is chosen such that only the high-resolution region of our simulations is taken into account and is much larger than the sizes of Ly$\alpha$ nebulae in the sample of \citet{Farina:19}. Each pixel thus has a size $\approx 0.17 \, \rm arcsec$, comparable to the resolution of $\approx 0.2 \, \rm arcsec$ achieved by MUSE. Due to seeing, note that the resolution obtained in the observations of \citet{Farina:19} is somewhat lower $\approx 0.5 \, \rm arcsec$. We adopt $200$ spectral bins, covering a (rest-frame) wavelength range from $1205 \, \angstrom$ to $1225 \, \angstrom$, giving $\Delta \lambda_{\rm obs} \approx 0.7 \, \angstrom$ at $z \, = \, 6.2$. We choose a higher spectral resolution  than in \citet{Farina:19}, where $\Delta \lambda_{\rm obs} \approx 2.6 \, \angstrom$, in order to quantify how the spectral line properties are affected by resolution effects.

\subsubsection{IGM absorption}
\label{sec:igmabsorption}
Assuming that the neutral fraction drops rapidly at $z < 6.5$ and that quasars produce large proximity zones, we neglect absorption by the intergalactic medium (IGM). In this approximation, the formation of nebulae is likely more severely impacted by the quasar age. If very young \citep{Eilers:17}, light-travel time could restrict the sizes of Ly$\alpha$ nebulae around $z \approx 6$ quasars.

The approximation that IGM absorption can be safely neglected also breaks down in the environments of $z \approx 7.5$ quasars, when the neutral fraction is much higher and quasar proximity zones tend to be small $\sim 1 \-- 2 \, \rm Mpc$ \citep[e.g.][]{Banados:18, Wang:20}.
Modelling IGM absorption self-consistently is, however, not possible with our simulations, since (i) our ``zoom-in'' region is much smaller than the quasar proximity zones of even $z \, = \, 7.5$ quasars and (ii) much of IGM absorption occurs in the low-resolution region. We here choose to gauge the maximum effect of IGM absorption at $z \, = \, 7.5$ with a simple analytic model. We assume hydrogen is fully ionised within a region of radius $R_{\rm p}$, which we vary from $0.5 \, \rm Mpc$ to $3 \, \rm Mpc$, in line with the proximity zone sizes of $z \, = \, 7.5$ quasars. Beyond $R_{\rm p}$, hydrogen is assumed to be neutral.  
This simple setup does not take into account any residual neutral hydrogen that may lie within the ionized bubble and cause additional absorption even within the quasar proximity zone. We further neglect any peculiar velocity of the IGM gas in this calculation. The resulting Ly$\alpha$ optical depth along the line of sight is then calculated using the analytic approximation for a Voigt profile presented in \citet{TepperGarcia:06}, resulting in a normalised 1D Ly$\alpha$ forest spectrum that can be used to attenuate the emission from $z \, = \, 7.5$ Ly$\alpha$ haloes (Section~\ref{sec:lyalpha7}).

\section{Results}
\label{sec:Results}
In this section, we present the results of our Ly$\alpha$ radiative transfer computations. We start with a general overview of the properties of the simulated halo and its large-scale environment (Section~\ref{sec:overview}), before addressing the Ly$\alpha$ emission properties of the simulated system (Section~\ref{sec:Ly-escape}).
We first concentrate our analysis on simulation \texttt{Quasar-L3e47} at $z \, = \, 6.2$. This redshift allows enough time for AGN feedback to operate in our simulated halo (recall that this is only switched on at $z \, = \, 6.5$). In Section~\ref{sec:qsoluminosity}, we also present results at higher and lower redshift.

\begin{figure*}
    \centering
    \includegraphics[width=0.975\textwidth,trim={0 2cm 0 1cm},clip]{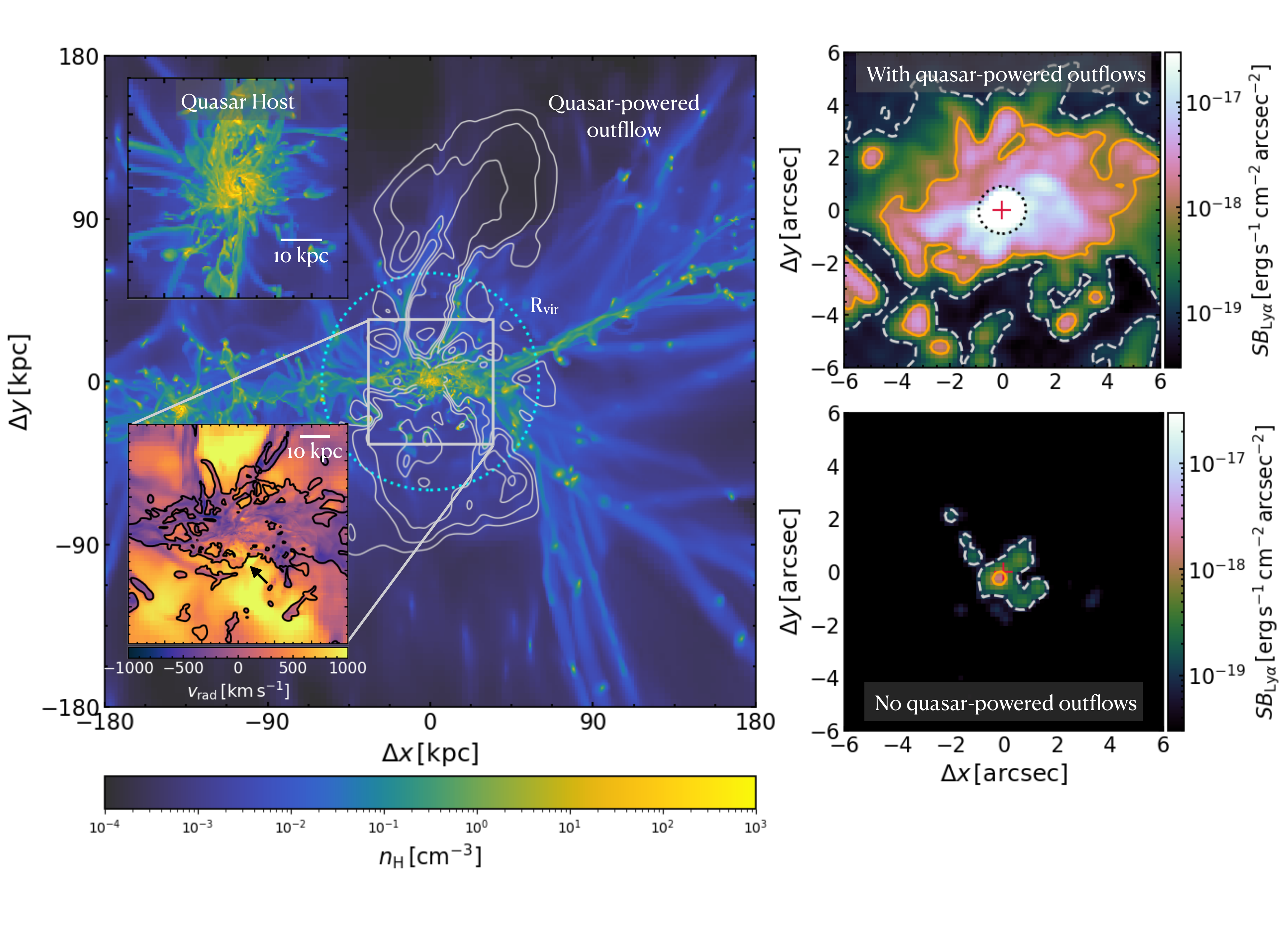}
    \caption{{\bf Left:} Gas density projected along a cubic volume of width of $360 \, \rm kpc$ centred on the quasar host galaxy at $z \, = \, 6.2$. Radial velocity contours (levels $100$, $300$ and $500 \, \rm km \, s^{-1}$) mark the quasar-driven outflow. The quasar host galaxy lies at the intersection of a network of cold, dense gas inflows. The outflow propagates into the voids beyond the virial radius (blue, dotted circle). Inset plots show the gas density around the host galaxy (top left) for a face-on configuration, while in the bottom left we show the gas radial velocity in the central $12 \, \mathrm{arcsec} \approx 70 \, \rm kpc$, showing large-scale outflows (orange regions) and inflows (purple regions). The contours trace dense gas (thick: $n_{\rm H} > 0.1 \, \rm cm^{-3}$), a phase which is mostly associated to inflowing gas.; {\bf Right:} Smoothed Ly$\alpha$ surface brightness (SB) map in the central regions of the quasar host halo for the broad line region scattering scenario. Contours give surface brightness levels of $10^{-18} \, \rm erg \,s^{-1} \, cm^{-2} \, \rm arcsec^{-2}$ (orange, solid) and $10^{-19} \, \rm erg \,s^{-1} \, cm^{-2} \, \rm arcsec^{-2}$ (white, dashed). In the simulation with quasar feedback (top), scattering from the BLR produces an extended Ly$\alpha$ nebula. Extended emission, however, vanishes completely if AGN feedback is absent (bottom).}
    \label{fig:intro}
\end{figure*}

\subsection{Overview}
\label{sec:overview}
The cosmological density field surrounding the massive halo targeted by our simulations is shown on the left-hand panel of Figure~\ref{fig:intro}.
The quasar host galaxy lies at the intersection of a network consisting of multiple gas filaments that extend well beyond the virial radius (dotted, blue circle).
These filaments stream inward towards the host galaxy and collide with one another, creating a circum-galactic ``cloud'' of dense ($n_{\rm H} \gtrsim 1 \, \rm cm^{-3}$) gas.

AGN radiation pressure on dust, in turn, gives rise to gas outflows \citep[see][]{Costa:18}. These also extend out to very large scales $> 100 \, \rm kpc$. These outflows are spatially anti-correlated with the large-scale filaments. Outflows take paths of least resistance, first breaking through the minor axis of the host galaxy and then venting into cosmic voids, even if feedback is isotropic at the scale of injection \citep{Costa:14, Costa:18}. 

The host galaxy is shown in the top-left inset plot. A disc, shown face-on, has a radius $\approx 5 \, \mathrm{kpc} \approx 0.9 \, \rm arcsec$ and connects to the larger scale CGM via the various infalling streams. At the very centre of the disc, we also see a gas cavity. This cavity results from the gas expulsion from the galactic nucleus caused by the momentum transfer associated to radiation pressure.
If there is no AGN feedback, as in \texttt{noQuasar}, this gap does not exist, consisting instead of large amounts of neutral gas.

The gas dynamics in the central region ($12 \, \mathrm{arcsec} \approx 70 \, \rm kpc$) is illustrated more clearly in the bottom-left inset panel of Figure~\ref{fig:intro}, where a radial velocity map is shown together with gas density contours. Dense gas with $n_{\rm H} \gtrsim 0.1 \, \rm cm^{-3}$ is concentrated in a flattened cloud measuring about $8 \-- 10 \, \mathrm{arcsec} \, \approx \, 45 \-- 55 \, \rm kpc$ across. Gas in this cloud is mostly inflowing, streaming inward at speeds as high as $\approx 700 \, \rm km \, s^{-1}$, higher than the virial velocity of the halo $V_{\rm vir} \, \approx \, 450 \, \rm km \, s^{-1}$. Flowing perpendicularly to this gas plane is the bipolar quasar-powered outflow (orange regions). At scales $\gtrsim 2 \, \mathrm{arcsec} \, \approx \, 12 \, \rm kpc$, outflows are mostly composed of low-density gas with $n_{\rm H} < 0.1 \, \rm cm^{-3}$. However, closer to the quasar host galaxy, outflowing gas (see arrow) can reach very high densities ($n_{\rm H} \gtrsim 10 \, \rm cm^{-3}$), despite high speeds $\gtrsim 800 \, \rm km \, s^{-1}$. The central few $\rm arcsec$ are thus characterised by a complex interaction between colliding, inflowing streams and the propagation of AGN-powered outflows. As a result, the circum-galactic cloud is associated to significant velocity dispersion; for the scales shown on the bottom-left inset panel of Figure~\ref{fig:intro} and excluding the host galaxy (approximately the central $5 \, \rm kpc$), the density-weighted gas velocity dispersion is $\approx 590 \, \rm km \, s^{-1}$.

In the top panel on the right-hand side of Figure~\ref{fig:intro}, we show a Ly$\alpha$ surface brightness map for the central $12 \, \mathrm{arcsec} \approx 70 \, \rm kpc$. In order to set the stage of our key findings, we here show the extended emission resulting by considering BLR scattering alone (here shown for the case where $\mathrm{FWHM}_{\rm BLR} \, \approx 2400 \, \rm km \, s^{-1}$), the most extreme scenario outlined in Section~\ref{sec:rascas}, in \texttt{Quasar-L3e47}. Surface brightness maps for other Ly$\alpha$ sources are shown in Section~\ref{sec:Ly-escape}. This map is smoothed with a Gaussian kernel of FWHM $0.5 \, \rm arcsec$ in order to mimic the effect of seeing in the observations of \citet{Farina:19}.
We can see that an extended Ly$\alpha$ nebula surrounds the central quasar.
Comparing with the radial velocity map shown on the left-hand panel, plotted on the same scale, we see that the Ly$\alpha$ nebula traces mostly inflowing dense gas, lying perpendicularly to the large-scale outflow. 

We also see that much emission traces the quasar host itself (central $\sim 1 \, \rm arcsec$). In order to reveal extended Ly$\alpha$ emission, \citet{Farina:19} first model the point-spread function (PSF) and subtract it from their data-cubes, effectively removing unresolved contributions from the quasar and likely the unresolved host galaxy. In the Ly$\alpha$ surface brightness maps of Figure~\ref{fig:intro}, the PSF associated to the quasar point source can be seen. We mark its location with a black dotted circle. As discussed in Section~\ref{sec:luminosities}, removing this component can have an impact on the reported nebula luminosity.

Finally, the bottom panel on the right-hand side of Figure~\ref{fig:intro} shows the Ly$\alpha$ surface brightness map obtained for BLR scattering at $z \, = \, 6.2$ in \texttt{noQuasar}. 
We here thus test whether a point source of Ly$\alpha$ photons is able to produce an extended nebula assuming the gas configuration that arises if AGN feedback is neglected. 
In order to illustrate a best-case scenario, we also neglect dust absorption\footnote{Note that dust absorption is included in the top panel. This is always included in our radiative transfer calculations, unless stated otherwise.} in this case.
We see that the nebula vanishes almost entirely. If dust absorption is taken into account, the nebula becomes even dimmer than shown in the bottom panel of Figure~\ref{fig:intro}.
As we show in Section~\ref{sec:whyQSO}, without AGN feedback, our simulations cannot reproduce the observations of \citet{Farina:19} with BLR scattering and, as we show later, with any of the Ly$\alpha$ mechanisms considered.

\subsection{A strong diversity in nebula morphology}

\subsubsection{Shapes and spatial offsets between Ly$\alpha$ emission and quasar position}

\begin{figure}
    \centering
    \includegraphics[width=0.45\textwidth,trim={0 1cm 0 0},clip]{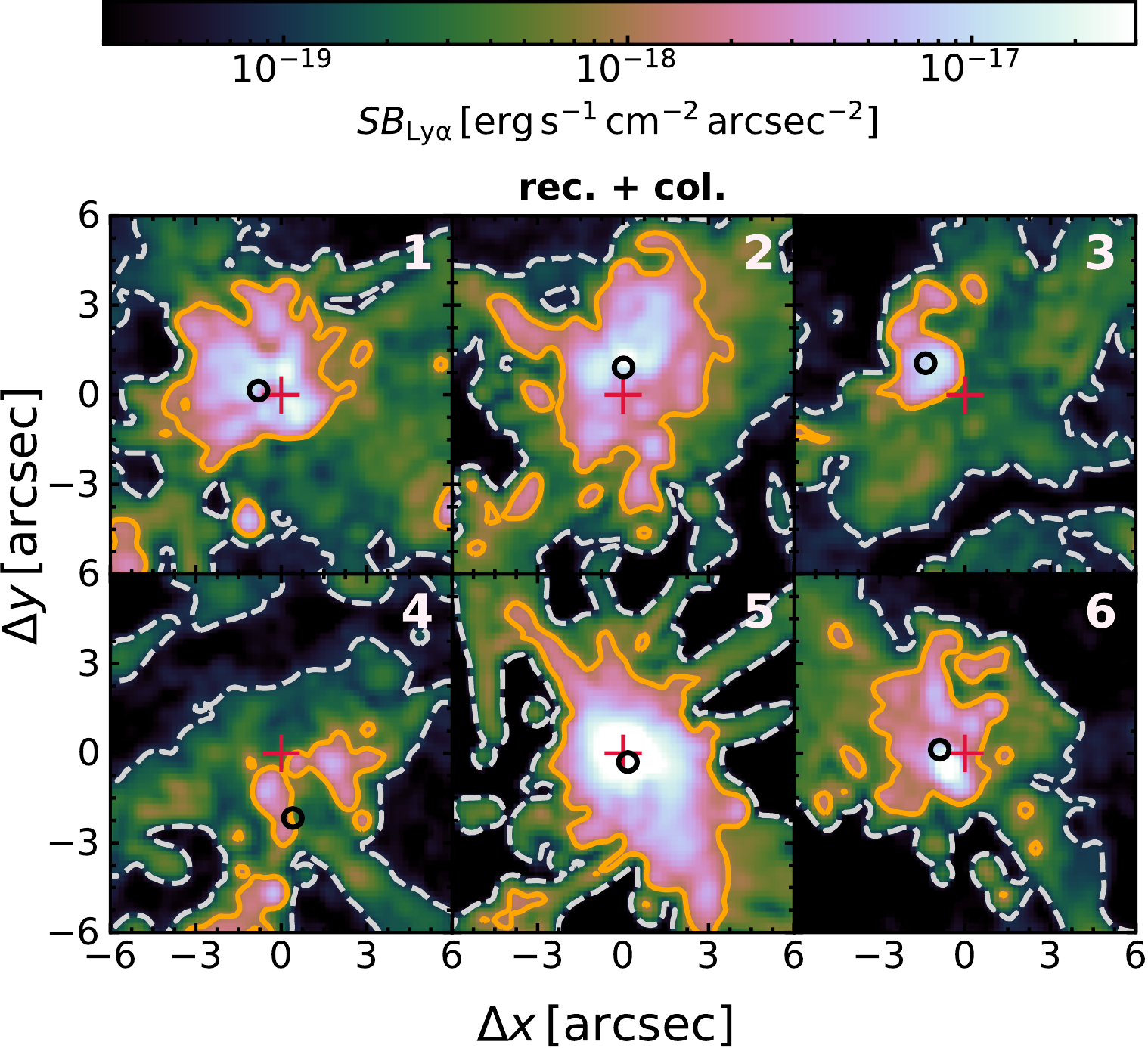}
    \includegraphics[width=0.45\textwidth,trim={0 1cm 0 2cm},clip]{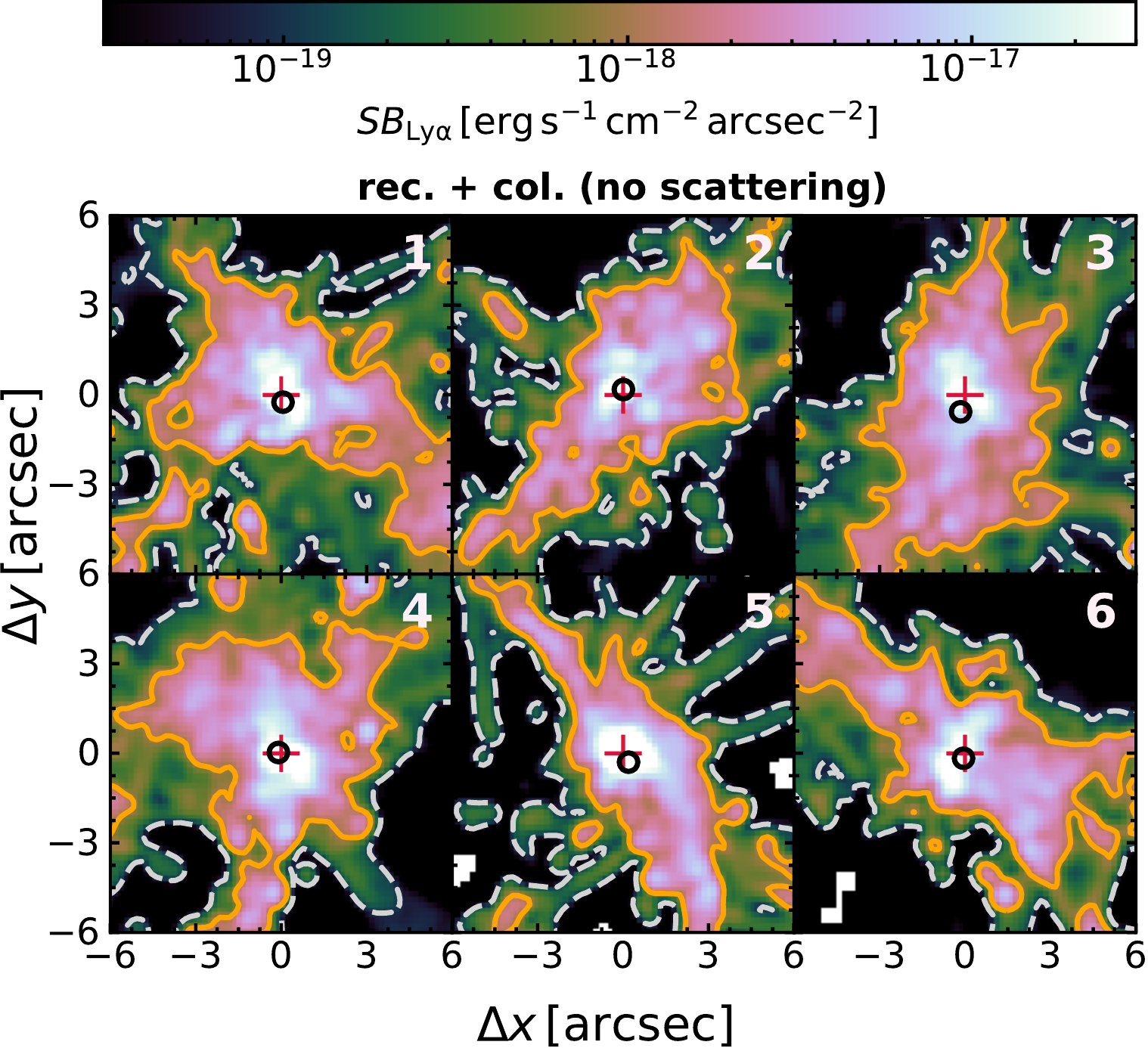}
    \includegraphics[width=0.45\textwidth,trim={0 0 0 2cm},clip]{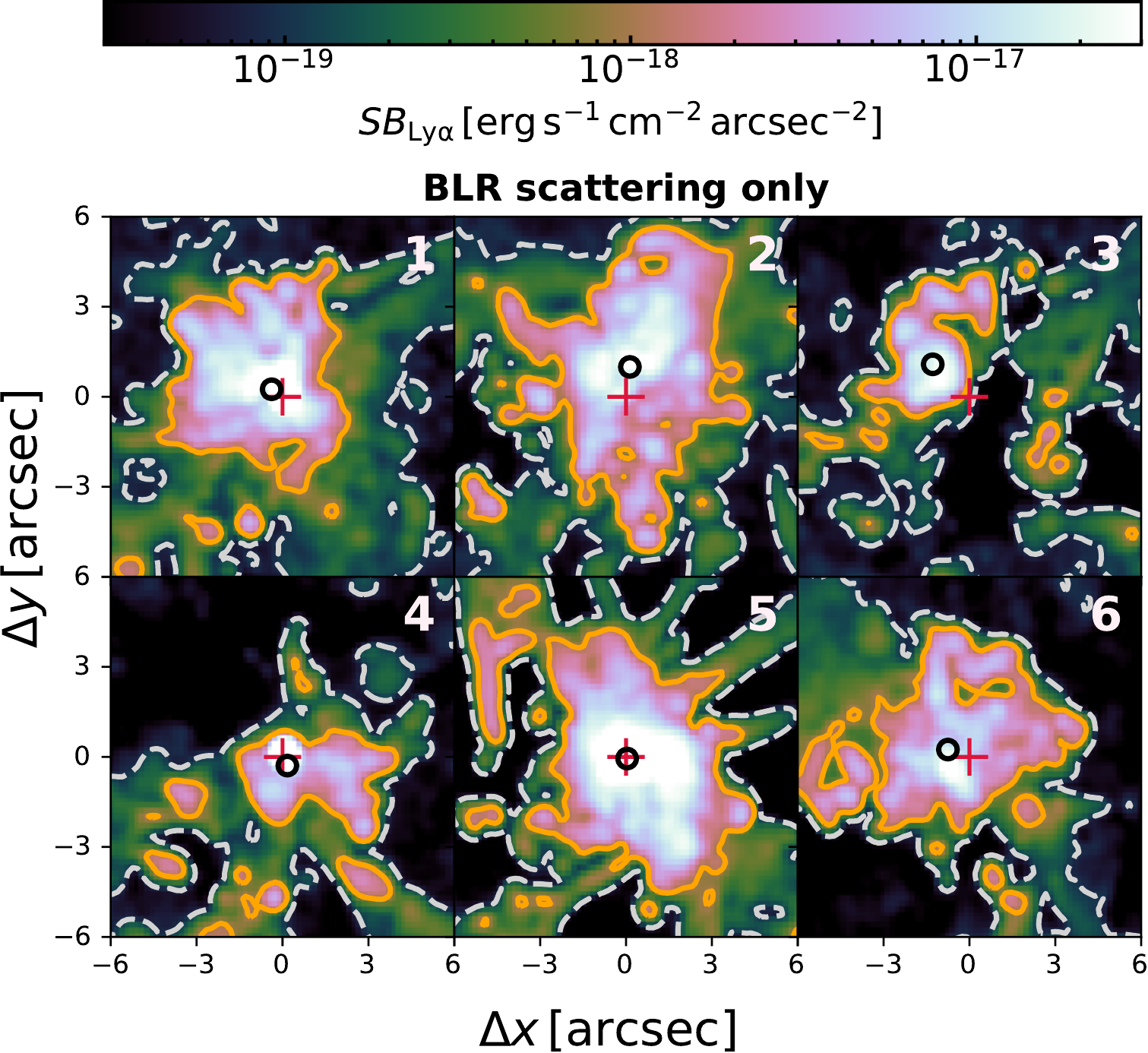}	
    \caption{Each sub-figure (group of six panels) gives Ly$\alpha$ surface brightness maps for different Ly$\alpha$ sources and transport properties. Different panels, numbered in the top right for convenience, show the nebula as seen from a different line-of-sight. Different sub-figures adopt the same lines-of-sight, i.e. panel 1 in the top sub-figure adopts the same line-of-sight as panel 1 in the central and bottom sub-figures.
    The surface brightness maps are smoothed with a Gaussian kernel with FWHM $0.5 \, \rm arcsec$ in order to match the resolution of \citet{Farina:19}. The position of the quasar is marked with a red $+$ sign and the SB-weighted centroid of the image with a black circle. In the first sub-figure, we consider recombination radiation and collisional excitation as sources and account for resonant scattering. In the second sub-figure, we neglect resonant scattering and show the intrinsic Ly$\alpha$ surface brightness. In the third sub-figure, we show the Ly$\alpha$ nebulae produced via pure scattering from the broad line region. Varying the viewing angle introduces striking diversity in the morphology of the nebula.}
    \label{fig_nebulae_los}
\end{figure}

Each sub-figure in Figure~\ref{fig_nebulae_los} shows smoothed Ly$\alpha$ surface brightness maps for \texttt{Quasar-L3e47} obtained for six random lines-of-sight at $z \, = \, 6.2$. Note that the variation of surface brightness profiles with redshift is discussed in Section~\ref{sec:qsoluminosity}.
The three sub-figures respectively illustrate (1) maps generated considering recombination radiation and collisional excitation (accounting for resonant scattering) and no BLR emission, (2) maps accounting for recombination radiation and collisional excitation but neglecting both resonant scattering off HI gas and BLR emission, and (3) maps generated for a pure scattering scenario involving only a broad Ly$\alpha$ line at the position of the quasar as a source (neglecting recombination radiation and collisional excitation).

Even though only one halo is investigated, we see a broad variety in nebula morphology. When viewed through some lines-of-sight, the nebula has an approximately spheroidal geometry (e.g. panels 2 and 5) and is centred on the position of the quasar (red plus sign). When looked at through other lines-of-sight, nebulae can acquire a more irregular geometry (e.g. panels 3 and 4). In such cases, there tend to be significant spatial offsets between the quasar position and the surface brightness-weighted centroid of the nebula, shown with a black circle, and nebulae appear lop-sided. Panels 3 and 4 give examples where nebula asymmetry is particularly strong; the quasar lies on the rim (or even outside) of the $10^{-18} \, \rm erg \,s^{-1} \, cm^{-2} \, \rm arcsec^{-2}$ isophote. For collisional excitation and recombination radiation (first sub-figure), the offset between the quasar position and the nebula centroid for these lines-of-sight is particularly large, approximately $ 2 \-- 3 \, \rm arcsec \, \approx \, 10 \--18 \, \rm kpc$.

For an asymmetric density distribution, the intrinsic Ly$\alpha$ surface brightness should also appear asymmetric.  
We, however, find that the large asymmetries and spatial offsets between the quasar position and the nebula centroids we see in Figure~\ref{fig_nebulae_los}, tend to become smaller if scattering is ignored, as shown by comparing panels 3 and 4 between the first two sub-figures of Figure~\ref{fig_nebulae_los}. Scattering can thus transform regular, spheroidal Ly$\alpha$ nebulae into lop-sided nebulae, such as those shown in panels 3 and 4 in the first sub-figure of Figure~\ref{fig_nebulae_los}. An explanation of this mechanism is provided in Appendix~\ref{appendix:a}. The detailed impact of scattering, however, depends on the line-of-sight. Panel 5, for instance, shows an example where scattering produces a more spheroidal nebula than obtained without scattering.

The morphological diversity we see here mirrors that of nebulae observed around $z \, = \, 6$ quasars \citep{Drake:19,Farina:19}, where nebulae are often seen to be lop-sided on scales of a few $\rm arcsec$ and down to a surface brightness of $\sim 10^{-18} \, \rm erg \, s^{-1} \, \rm cm^{-2} \, arcsec^{-2}$. This agreement provides a first hint that resonant scattering reconciles the properties of simulated nebulae with those detected around $z > 6$ quasars.  

The third sub-figure of Figure~\ref{fig_nebulae_los} leaves little doubt that resonant scattering operates effectively in our simulated halo. These panels illustrate a pure scattering scenario (as in Figure~\ref{fig:intro}) where the only Ly$\alpha$ source is the quasar broad line (here $\rm FWHM_{\rm BLR} \, \approx 3500 \, \rm km \, s^{-1}$) itself. The associated Ly$\alpha$ flux is nevertheless clearly able to inflate a spatially extended nebula for every line-of-sight. Qualitatively these nebulae resemble those generated via collisional excitation and recombination radiation (first sub-figure), to a great part sharing their morphology and spatial extent.
In Section~\ref{sec:SBprofile}, we perform a more quantitative analysis and present surface brightness profiles, where we further strengthen our argument that resonant scattering plays a central role in setting the properties of observed Ly$\alpha$ nebulae. 

\subsubsection{Ly$\alpha$ escape}
\label{sec:Ly-escape}

Here we show that, besides affecting nebular morphology, scattering also introduces variations in the Ly$\alpha$ nebula luminosity.
We construct a Cartesian coordinate system with a z-axis aligned with the quasar host galaxy's angular momentum vector, which we evaluate by measuring the angular momentum within the disc radius ($\approx 5 \, \rm kpc$).
Using a {\sc HEALPix} tessellation \citep{Gorski:05}, we decompose a spherical surface centred on the position of the quasar into 768 pixels of identical solid angle $\Omega_{\rm pix}$. We then compute the distribution of the \emph{final directions} of escaping photons, i.e. those that are not absorbed by dust at any point, on this surface. We select only photons produced within $R_{\rm vir} \, \approx \, 60 \, \rm kpc$, sufficient to encompass the scale of our simulated Ly$\alpha$ nebulae.
Two photon packets escaping along the same direction would fall on the same pixel, even if they escape from very different locations.
We can think of this procedure as a projection of the escaping Ly$\alpha$ flux onto a distant spherical surface with radius $R \gg R_{\rm vir}$. 
We evaluate the number of escaping photons per pixel $dn_{\rm esc}/d\Omega_{\rm pix}$ on this surface, and normalise it by the total number $N_{\rm tot}$ of photons (absorbed and escaping) generated within $R_{\rm vir}$. We consider two quantities:
\begin{enumerate}
    \item We first define an escape probability $p_{\rm esc} (\Omega) \, = \, \left( dn_{\rm esc}/d\Omega_{\rm pix}\right)   \left[N_{\rm tot} / (4 \pi) \right]^{-1}$. The term $\left[N_{\rm tot} / (4 \pi) \right]$ gives the photon number distribution expected if photon trajectories are isotropic, as is the case at emission. As defined, the escape probability is shaped both by (i) dust absorption (which can reduce $n_{\rm esc}$) and (ii) photon scattering (which can cause the escaping flux's direction to deviate from isotropy). In particular, it is possible for $p_{\rm esc} (\Omega)$ to exceed unity if dust absorption is inefficient and if scattering deflects photons into a preferred direction, enhancing their distribution above the value expected in the isotropic case.
    \item We define $f_{\rm esc} (\Omega) \, = \, \left( dn_{\rm esc}/d\Omega_{\rm pix}\right) \left( dN_{\rm tot}/d\Omega_{\rm pix}\right)^{-1}$ as the escape fraction. The escape fraction is subtly different from the escape probability: it only quantifies the efficiency of dust absorption along each pixel and is not sensitive to the redistribution of photons in solid angle. Thus $f_{\rm esc} (\Omega) \leq 1$.  
\end{enumerate}

\begin{figure}
    \centering
    \includegraphics[width=0.45\textwidth]{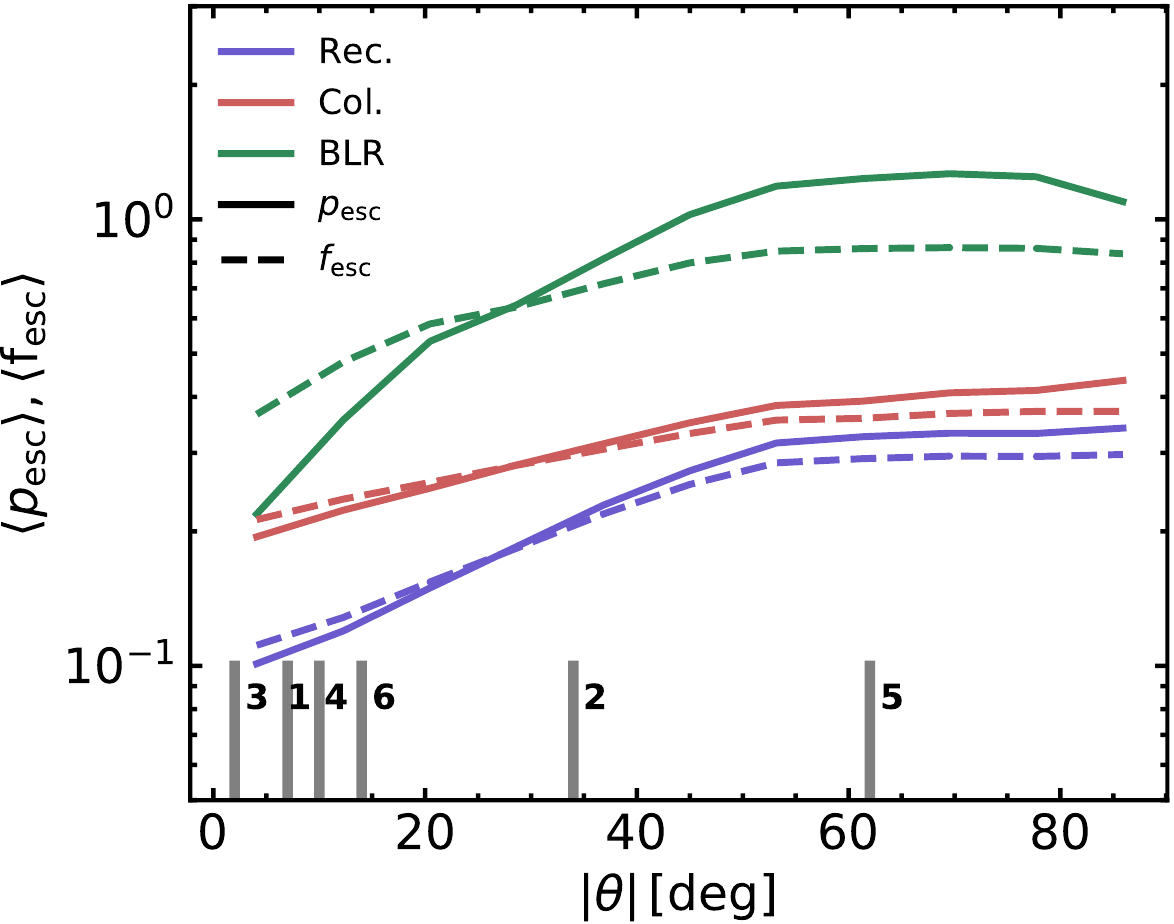}
    \caption{Ly$\alpha$ escape probability (solid curves) and escape fraction (dashed curves) as a function of elevation angle from the disc plane, for different emission processes. Resonant scattering introduces anisotropy in the Ly$\alpha$ escape, which is funnelled into the polar directions. Ly$\alpha$ nebulae are thus brightest when the quasar host disc is seen face-on. The orientation angles of the six random lines-of-sight adopted throughout this paper are marked with vertical grey lines.}
    \label{fig_escape}
\end{figure}

We then define $\theta$ as the angle between the disc plane and its angular momentum vector with $\theta \, = \, 0^\circ$ corresponding to directions along the disc plane and $|\theta| \, = \, 90^\circ$ corresponding to directions aligned with the disc poles. We take the azimuthally-averaged escape probability $\langle p_{\rm esc} \rangle$ and escape fraction $\langle f_{\rm esc} \rangle$ and plot it as a function of $\theta$ in Figure~\ref{fig_escape}. The solid and dashed curves show, respectively, the  Ly$\alpha$ escape probability and escape fraction for \texttt{Quasar-L3e47} at $z \, = \, 6.2$, for different Ly$\alpha$ emission mechanisms.
We see that, irrespective of the emission mechanism, the Ly$\alpha$ flux preferentially escapes along the disc's rotation axis at $90^\circ$.
For all processes, the escape probability (fraction) is $\approx 10 \% \-- 20 \%$ ($\approx 10 \% \-- 35 \%$) along the disc plane. Along the polar direction, this increases to $\approx 35\%$ ($\approx 30 \%$), i.e. by a factor $\approx 3 \-- 4$, for recombination radiation. For BLR photons, the escape probability (fraction) rises to $\gtrsim 100\%$ ($\approx 80 \%$). Ly$\alpha$ photons generated via collisional excitation appear to be the least affected by orientation, varying only by a factor $\lesssim 2$ between disc plane and rotation axis.

In order to understand the link between Ly$\alpha$ escape and elevation angle $\theta$, it is useful to consider the scales where Ly$\alpha$ photons are generated. For BLR scattering, photons are produced in a point source within the cavity located at the centre of the quasar disc (see Figure~\ref{fig:intro}). Selecting only photons generated within $R_{\rm vir}$ at $z \, = \, 6.2$, we further find that $50\%$ of recombination photons are created within $\approx 2.6 \, \rm kpc$ from the quasar, i.e. inside the quasar host galaxy. For collisional excitation we find that $50 \%$ of 
the photons are instead produced within $\approx 25.4 \, \rm kpc$, i.e. at 10 times larger scales and well beyond the quasar host galaxy. These findings suggest that Ly$\alpha$ scattering and dust absorption within the quasar host galaxy drives escape anisotropy, affecting primarily BLR and recombination photons.
We can test this hypothesis: by selecting only photons generated at radial distances of $> 10 \, \rm kpc$ from the quasar (well outside of the galactic disc) we find that the escape fraction of \emph{recombination photons} varies only by a factor $\approx 2$ between disc plane and poles, like in the collisional excitation case.

For recombination radiation and collisional excitation, the net escape fractions are, respectively, $f_{\rm esc} \approx 21\%$ and $ f_{\rm esc} \approx 30\%$. 
For BLR emission, the net escape fraction is $f_{\rm esc} \approx 73\%$. This high escape fraction is a direct consequence of AGN feedback via radiation pressure on dust. In order to ensure efficient momentum transfer and power large-scale outflows, radiation pressure on dust \emph{requires} large dust abundances \citep{Costa:18a}. Above a critical AGN luminosity, this momentum transfer, significantly aided by radiation pressure of trapped infrared photons \citep{Costa:18}, expels the dusty gas layers, allowing optical and UV radiation to escape. This interpretation can be confirmed by computing Ly$\alpha$ escape fractions in \texttt{noQuasar}: $\langle f_{\rm esc} \rangle \, \approx \, 0.1 \%$ for recombination radiation, $\langle f_{\rm esc} \rangle \, \approx \, 17 \%$ for collisional excitation and $\langle f_{\rm esc} \rangle \, = \, 0\%$ for BLR emission, cementing our conclusion that the escape of recombination and BLR photons are sensitively controlled by the properties of the central galaxy.

Comparing the solid and dashed curves in Figure~\ref{fig_escape}, we find that escape probabilities and escape fractions are similar. From this comparison we can, however, see that the redistribution in solid angle caused by Ly$\alpha$ scattering (i) reduces the escape probability along the plane in addition to dust absorption and (ii) enhances escape along the poles. For BLR photons, this effect is particularly dramatic. Values of $\langle p_{\rm esc} \rangle \gtrsim 1$ along the disc poles show that scattering beams the quasar's Ly$\alpha$ flux so efficiently that its luminosity would appear $30 \%$ higher than the true Ly$\alpha$ luminosity along these lines-of-sight. 

Preferred escape directions naturally occur for anisotropic gas density fields. If they initially propagate along the galactic or CGM plane, Ly$\alpha$ photons encounter a higher HI column. These photons thus scatter more frequently. The chance that they are deflected away from the plane is thus also higher.
Escape along the disc becomes unlikely, because escaping photons would have to scatter coherently into the same direction.
If they initially propagate into the polar axis, Ly$\alpha$ photons undergo fewer scatterings and escape more easily. These photons are joined by those deflected away from the disc plane, resulting in up to an order of magnitude enhancement in observed Ly$\alpha$ luminosity.
The vertical grey lines in Figure~\ref{fig_escape} mark the elevation angles associated to the lines-of-sight used in this study\footnote{These are random lines-of-sight. For low $|\theta|$, there are many directions for different azimuthal angles, while along the polar axis, there is only one.}. We see that Ly$\alpha$ luminosities should vary by a factor $\lesssim 10$ due to variations in sight-line.

\subsection{Comparison with observed nebulae}
\label{sec:comparison}
In this Section, we compare the detailed properties of our mock Ly$\alpha$ nebulae with those of observed nebulae at $z > 6$ for the REQUIEM Survey \citep{Farina:19}.

\subsubsection{Surface brightness radial profiles}
\label{sec:SBprofile}

\begin{figure}
    \centering
    \includegraphics[width=0.465\textwidth,trim={0 0.6cm 0 0},clip]{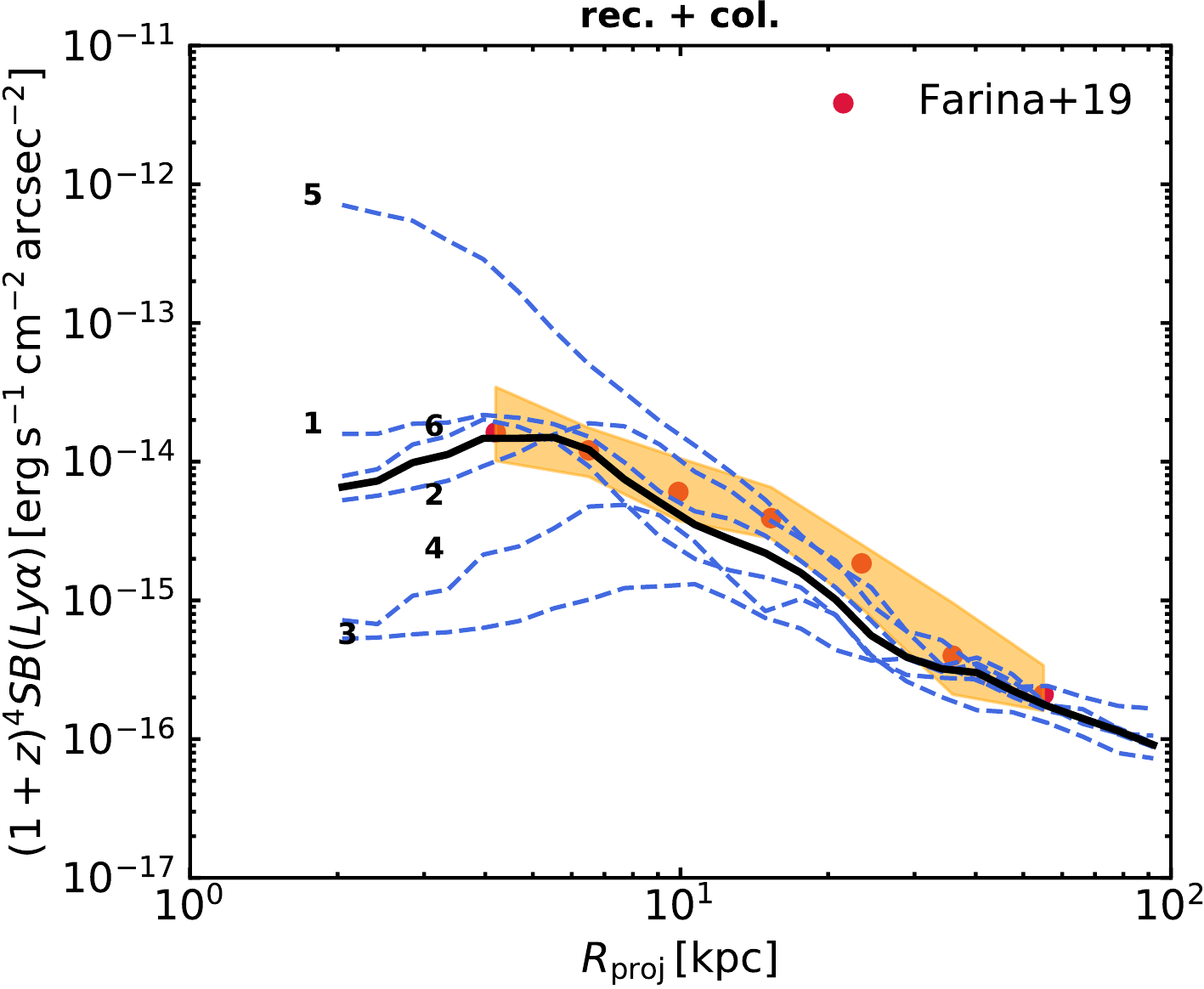}	
    \includegraphics[width=0.465\textwidth,trim={0 0.6cm 0 0},clip]{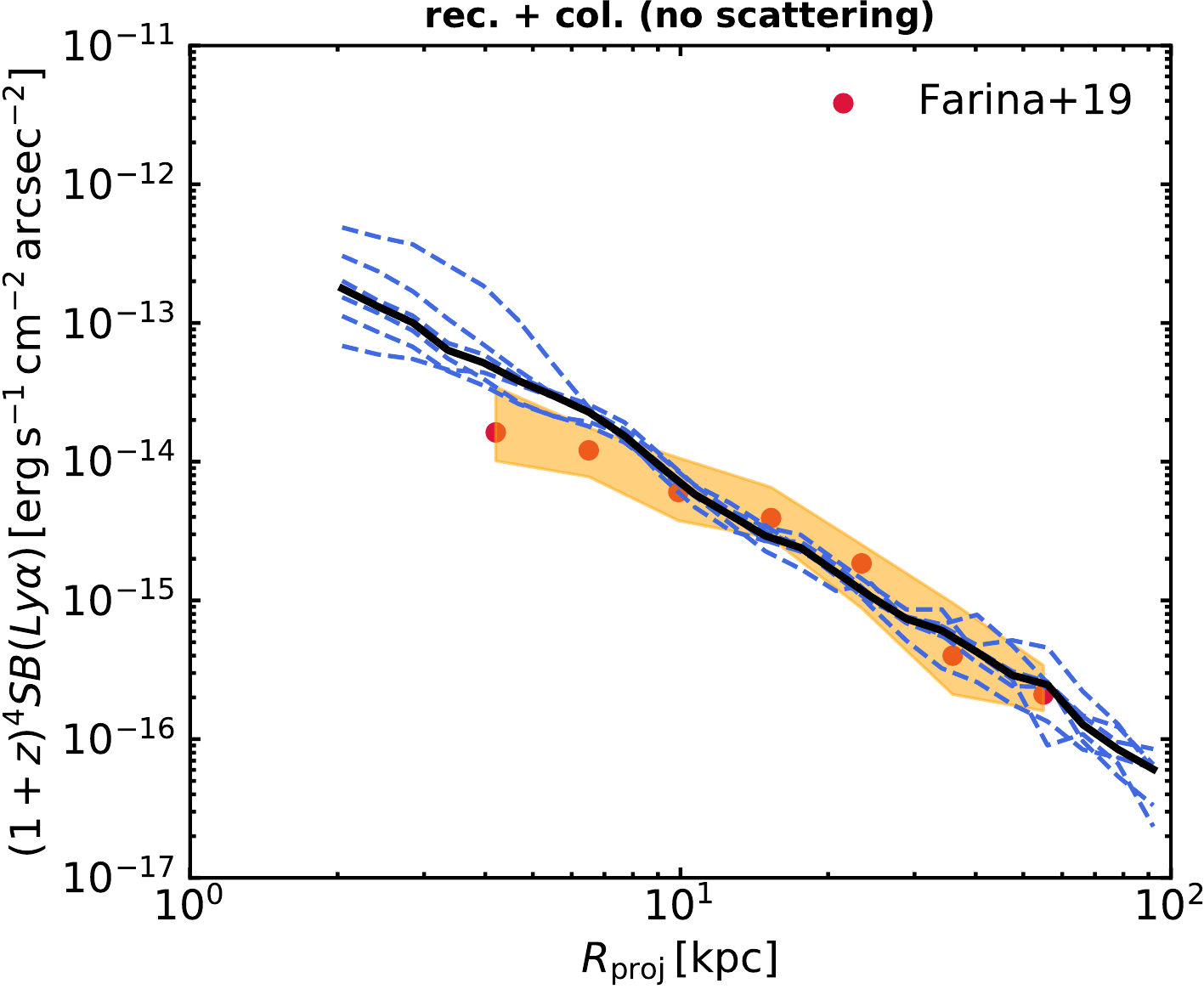}
    \includegraphics[width=0.465\textwidth]{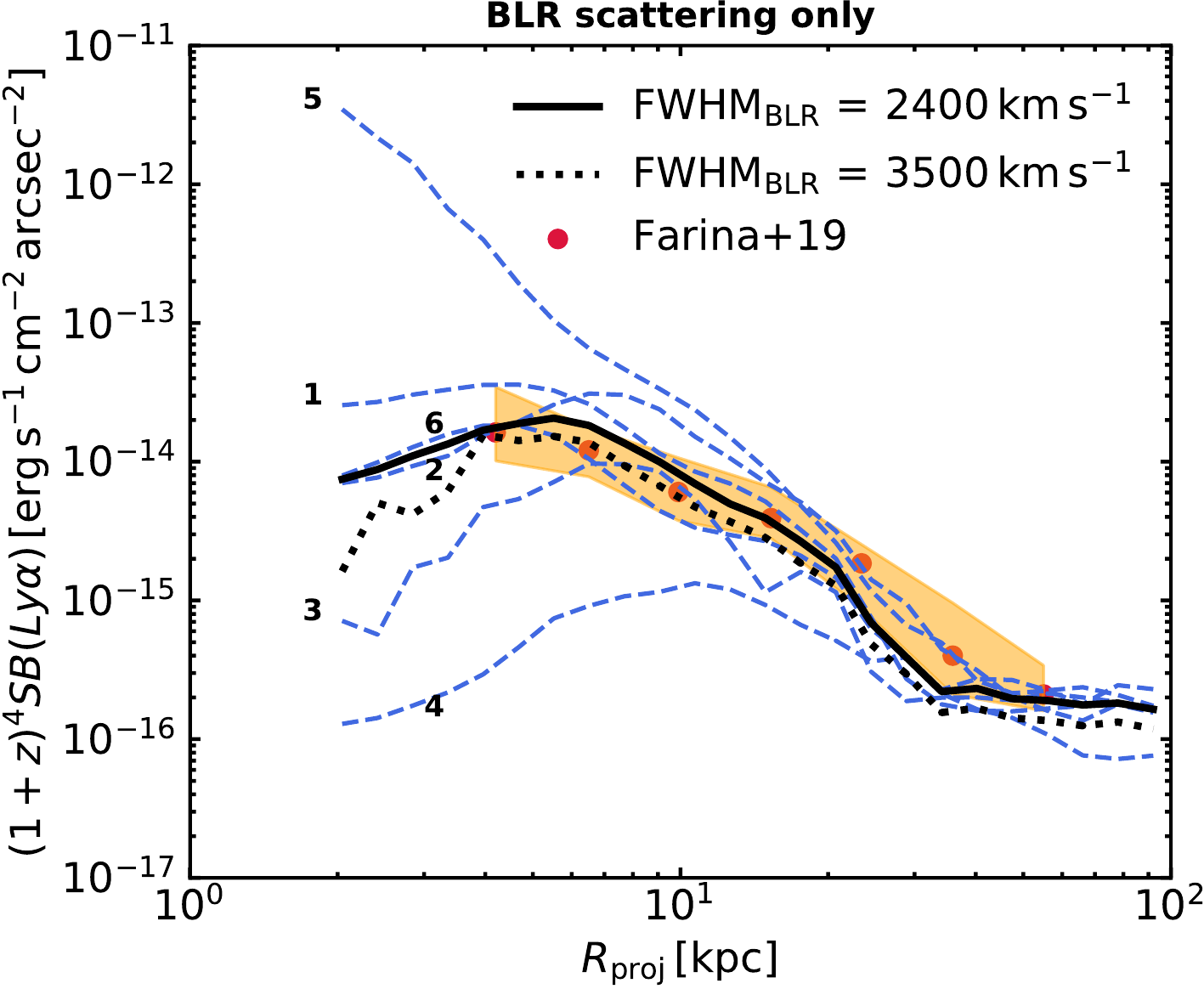}
    \caption{Surface brightness profiles for \texttt{Quasar-L3e47} at $z \, = \, 6.2$ for different lines-of-sight (dashed blue curves) and different Ly$\alpha$ emission processes. Thick black curves show the median obtained from the different lines-of-sight. In the top and middle panels we consider recombination radiation and collisional excitation with and without scattering, respectively. In the bottom panel we show results for BLR scattering only (dashed blue curves and solid black curve for $\rm FWHM_{\rm BLR} \, \approx \, 2400 \, \rm km \, s^{-1}$, and dotted black curve for $\rm FWHM_{\rm BLR} \, \approx \, 3500 \, \rm km \, s^{-1}$). The observed median profile for Ly$\alpha$ nebulae at $z \, > \, 6$ in the REQUIEM Survey is shown with red circles. The orange shade illustrates the range between the $25^{\rm th}$ and $75^{\rm th}$ percentiles of the observed profile distribution. The agreement between data and simulations is very close for all processes considered, though the shape of the profiles agree best if scattering operates.}
    \label{fig_SBprofile}
\end{figure}

Figure~\ref{fig_SBprofile} shows surface brightness radial profiles obtained from the smoothed surface brightness maps for our different lines-of-sight (dashed, blue curves). In order to generate these radial profiles, the origin is placed at the position of the quasar. For each line-of-sight, we then take the spherical average in 32 logarithmically spaced rings in the radial range $[0.5 \, \mathrm{kpc}, 100 \, \mathrm{kpc}]$, taking into account all pixels within each ring. For consistency with \citet{Farina:19}, we collapse our data-cubes only between the velocity channels $-500 \, \rm km \, s^{-1}$ and $500 \, \rm km \, s^{-1}$.

Different panels give surface brightness profiles for different combinations of Ly$\alpha$ sources and radiative transfer properties.
The top panel of Figure~\ref{fig_SBprofile} shows results for the combined effect of recombination radiation and collisional excitation. Individual lines-of-sight produce surface brightness profiles which deviate from the median observed profile by up to 1 dex. Variations in the surface brightness profiles are most prominent at small radii, while different lines-of-sight appear to yield similar surface brightness profiles at radii $\gtrsim 30 \, \rm kpc$.

Observed profiles also display significant object to object variation \citep[see e.g.][]{Drake:19} and individual objects can also deviate significantly from a sample median profile. 
In Figure~\ref{fig_SBprofile}, the median profile obtained in \citet{Farina:19} is shown for comparison with red, filled circles together with the $25^{\rm th}$ to $75^{\rm th}$ percentile range, delimited by the orange shaded region.
We see that, despite individual deviations, the theoretical profiles cluster around the observed median profile.

To perform a fairer comparison with the \citet{Farina:19} median profile, we compute the median profile obtained from our six random lines-of-sight and show it in Figure~\ref{fig_SBprofile} with a thick, black curve.
The agreement of both the shape and normalisation of the median mock profile and the median profile of \citet{Farina:19} is striking, particularly in view of the fact that our simulations capture only one halo, that they are not tuned in any way to yield a realistic CGM and despite our highly-idealised treatment of AGN feedback. 

Inspecting the central panel of Figure~\ref{fig_SBprofile}, where resonant scattering is ignored, we find poorer agreement in the profile shape between the observed median profile and our theoretical surface brightness profile at radii $\lesssim 10 \, \rm kpc$. While the observed median profile flattens out at a radius $10 \-- 20 \, \rm kpc$, the theoretical profile now behaves like a single power law with an exponent $\approx -2$. 

The better agreement with observed radial profile shape seen in our radiative transfer computations that do account for scattering corroborates our previous argument that scattering plays an important role in reconciling theoretical predictions and observations. Scattered Ly$\alpha$ photons could be initially produced via recombination and collisional excitation (as in the first panel of Figure~\ref{fig_SBprofile}), but could also consist entirely of reflected quasar light. The third panel validates even this extreme-case scenario: scattering from a point source positioned at the quasar produces Ly$\alpha$ nebulae with surface brightness profiles which can explain both shape and normalisation of the observed profiles. Figure~\ref{fig_SBprofile} shows median profiles for the two different quasar broad line widths considered: $\rm FWHM_{\rm BLR} \, \approx \, 2400 \, \rm km \, s^{-1}$ (solid curve) and $\rm FWHM_{\rm BLR} \, \approx \, 3500 \, \rm km \, s^{-1}$ (dotted curve). In both cases, we see extended nebulae, with the broader line producing an only somewhat fainter nebula (see Section~\ref{sec:lineprofile} for an explanation).

One difference between simulated and observed radial profiles resides in the scatter around the relation. According to our simulations, scatter decreases with increasing radial distance from the quasar. The observed scatter appears not to change significantly with radius, however. Comparing the first and second panels of Figure~\ref{fig_SBprofile} shows that the scatter seen around our median radial profile is mainly driven by photon scattering. Since this process is most efficient in the central regions of the halo, the scatter is also largest at smaller radii. Extending observational surveys to smaller radii than currently resolved would test our predictions. In order to capture scatter at large radii, simulations likely need to probe an ensemble of massive haloes in order to sample different large scale gas and galaxy satellite configurations.

\subsubsection{The Ly$\alpha$ nebula mechanism}
\label{sec:mechanism}

\begin{figure}
    \centering
    \includegraphics[width=0.475\textwidth]{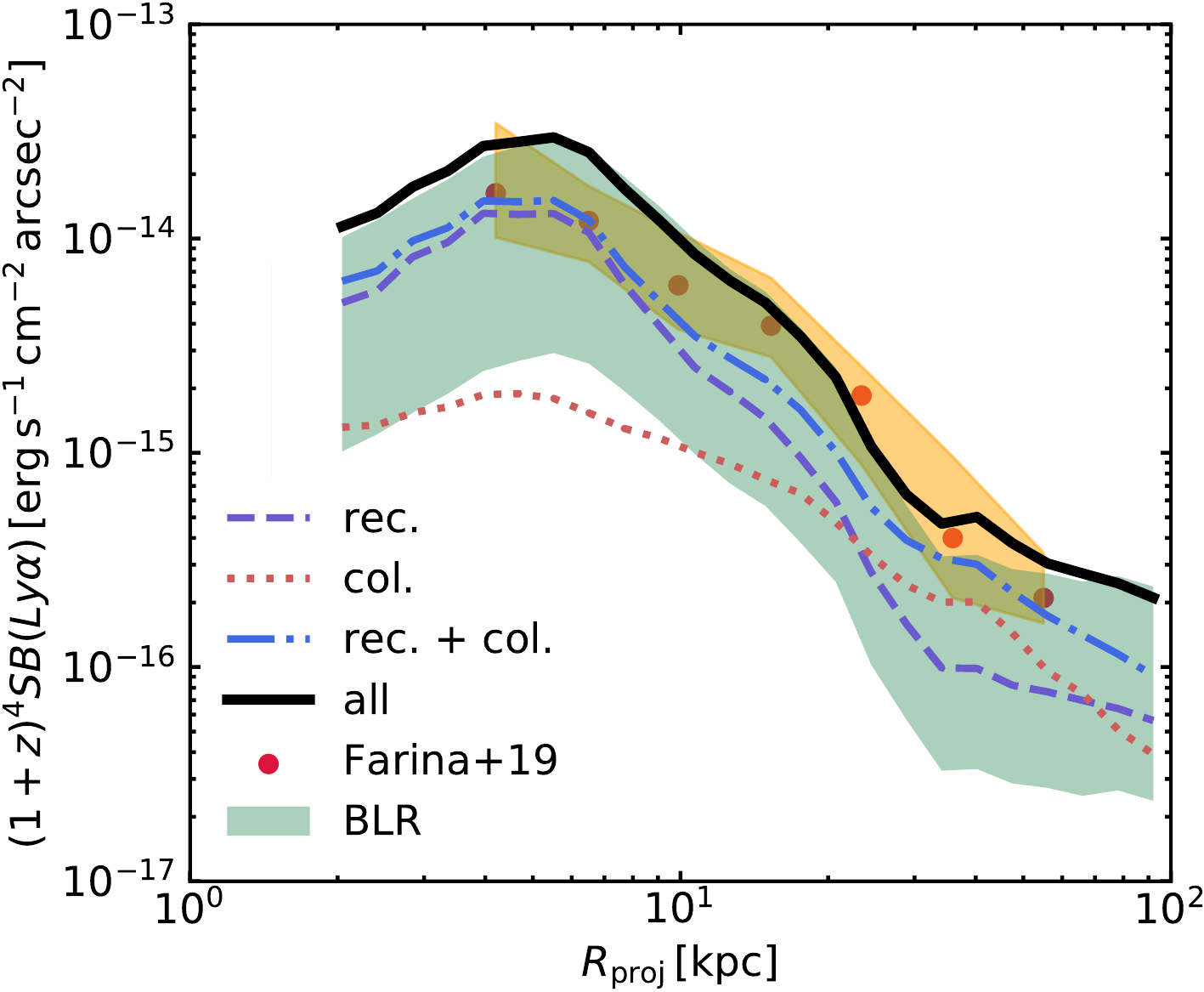}
    \caption{Median radial profiles obtained by averaging the smoothed surface brightness maps over six random lines-of-sight at $z \, = \, 6.2$ in \texttt{Quasar-L3e47}. 
    We show profiles for recombination radiation (dark blue, dashed curve), collisional excitation (red, dotted curve), both processes together (light blue, dot-dashed curve), BLR photon scattering (green, shaded region) and a combination of all processes (thick, black curve). We identify three possible origins for observed Ly$\alpha$ nebulae at $z \, = \, 6$: (i) a combination of recombination cooling and collisional excitation, (ii) scattering from the broad line region and (iii) a combination of all mechanisms.}
    \label{fig_SBprofile_contribution}
\end{figure}

In Figure~\ref{fig_SBprofile_contribution}, we again plot median surface brightness profiles, but now decomposed into different combinations of Ly$\alpha$ sources. The dark blue, dashed curve illustrates the profile obtained considering recombination radiation, the red dotted curve gives results for collisional excitation, while the dot-dashed light blue curve shows the effect of combining both processes. 

On its own, collisional excitation or recombination radiation does not match the observed median profile of \citet{Farina:19}.
On the one hand, recombination radiation closely reproduces the observed profile in the central regions. However, the associated profile is steeper than the observed median profile, underestimating the surface brightness at large radii. Collisional excitation, on the other hand, dominates at larger radii, grazing the observed surface brightness profile at scales $\gtrsim 30 \, \rm kpc$. Due to its flatter profile, collisional excitation becomes less important in the central regions and underestimates the observed surface brightness by about an order of magnitude. 
Interestingly, a combination of both processes results in a closer match to the observed profile (see also top panel in Figure~\ref{fig_SBprofile}), correctly predicting the shape and yielding a normalisation which is close to the observed median surface brightness profile.

The green shaded region shown in Figure~\ref{fig_SBprofile_contribution} illustrates the surface brightness profile that results from considering BLR scattering alone, assuming $\rm FWHM_{\rm BLR} \, \approx \, 3500 \, \rm km \, s^{-1}$ for the input quasar broad line.
In general, we should not expect this mechanism to operate in isolation. However, we consider its individual contribution (i) to test the viability of the scenario in which giant Ly$\alpha$ nebulae are powered via scattering from a single point source, and (ii) to explore how nebulae may form in configurations where the large-scale gas distribution remains neutral despite a bright central quasar, e.g. due to special large-scale gas configurations or AGN light-cone directions.  
The contribution of BLR scattering is very sensitive to the fraction $f_{\mathrm{Ly} \alpha}$ of the quasar bolometric luminosity which is associated to Ly$\alpha$. The associated uncertainty is quantified in Figure~\ref{fig_SBprofile_contribution} with a shaded region, illustrating how the normalisation of the profile changes by varying $f_{\mathrm{Ly} \alpha}$ from $0.001$ to $0.01$. 
BLR scattering can account for (i) the shape of the observed surface brightness profile and (ii) its normalisation, which falls within the plausible range of $f_{\mathrm{Ly} \alpha}$ values. At face value, we see that the propagation of Ly$\alpha$ photons from the BLR via resonant scattering constitutes an equally viable mechanism for generating spatially extended Ly$\alpha$ nebulae, even if operating in isolation. When adding BLR scattering, recombination radiation and collisional excitation, we obtain the black, solid curve. Both its shape and normalisation remain consistent with the observed profile. As explained in Section~\ref{sec:scatteringBLR}, this combination may, however, double-count the Ly$\alpha$ luminosity and should be regarded as an upper limit.

We thus identify three possible origins for observed Ly$\alpha$ nebulae at $z \, = \, 6$: (i) a combination of recombination cooling and collisional excitation, (ii) BLR scattering, (iii) a combination of all processes. We revisit this point in Section~\ref{sec:whyQSO}, where we provide an explanation for why BLR scattering is so efficient in our simulations.  

\subsubsection{Line profile}
\label{sec:lineprofile}
\begin{figure}
    \centering
    \includegraphics[width=0.499\textwidth]{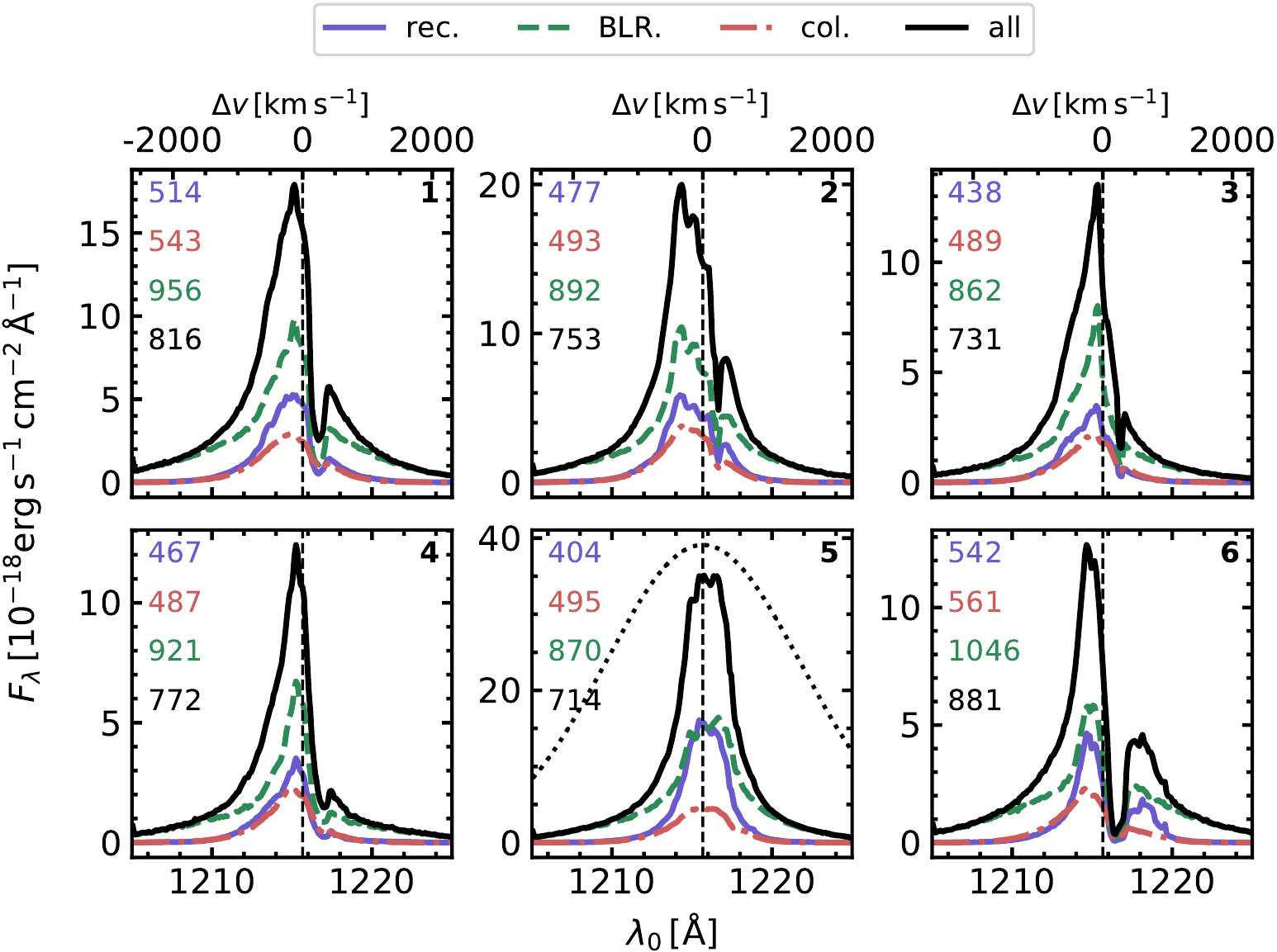}
    \includegraphics[width=0.499\textwidth]{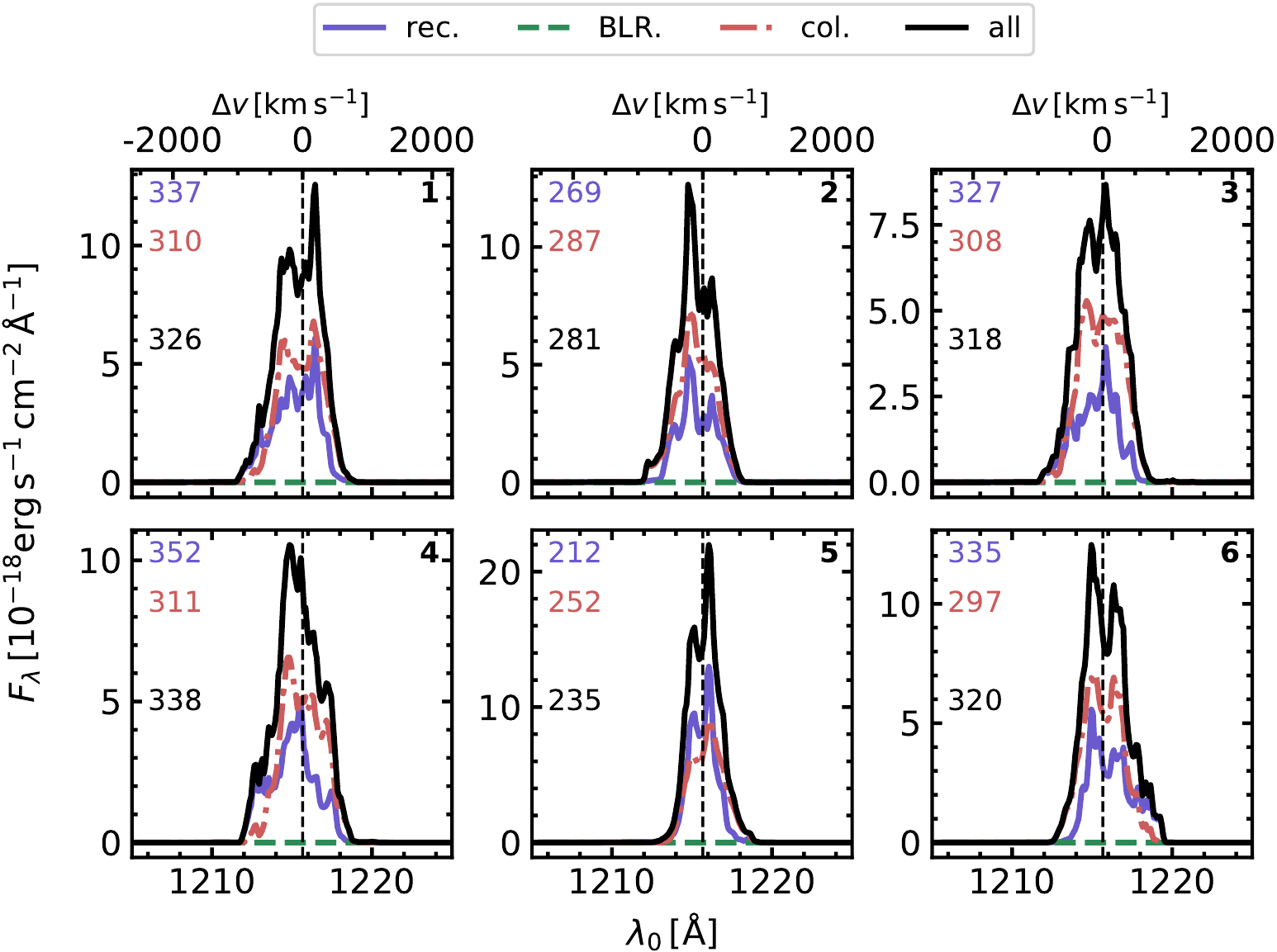}
    \caption{Integrated Ly$\alpha$ spectra obtained for six random lines-of-sight (different panels) at $z \, = \, 6.2$ and for different Ly$\alpha$ source models (different coloured curves), at a spectral resolution of $\Delta \lambda_{\rm obs} \approx 0.7 \, \rm \angstrom$, a factor $\approx 3.5$ better than in MUSE. In the top row, we show results including scattering and illustrate the impact of neglecting resonant scattering in the bottom row. Individual contributions from recombination radiation, collisional excitation, BLR scattering (with $\rm FWHM_{\rm BLR} \, \approx \, 3500 \, \rm km \, s^{-1}$) and a combination of all processes are illustrated with different curves. Scattering broadens the line profiles significantly for all lines-of-sight (see also moment maps in Appendix~\ref{appendix:b}). The numbers indicate the flux-weighted velocity dispersion associated to each spectrum, in $\rm km \, s^{-1}$. Full-widths-at-half-maximum are given in Table~\ref{table_FWHM}. In panel 5 in the top sub-figure, we show the intrinsic BLR spectrum, renormalised arbitrarily for comparison with the emerging spectrum. The spectral line profile associated to extended emission is always narrower than that of the quasar even if the nebula is generated via BLR scattering. Note that the flux from within an aperture with radius $0.5 \, \rm arcsec$ around the quasar is subtracted before producing spectra, such that these correspond to the extended component only. For this reason there is no BLR contribution in the bottom panel, where scattering is neglected.}
    \label{fig:line_profile}
\end{figure}

We present spectral line profiles obtained by integrating our mock surface brightness maps for our six lines-of-sight in Figure~\ref{fig:line_profile}, including and excluding resonant scattering (top and bottom sub-figures, respectively).
Before generating spectra, we subtract all flux from within an aperture with radius $0.5 \, \rm arcsec$ centred on the quasar position to mimic PSF subtraction (see Section~\ref{sec:overview}).
Different curves show how the spectral line profile varies with emission mechanism. Blue curves show the emerging spectra for recombination radiation, the dotted, red curves show results for collisional excitation, and dashed, green curves for BLR scattering (with $\mathrm{FWHM}_{\rm BLR} \, \approx 3500 \, \rm km \, s^{-1}$). Combining all these processes gives the black curves.
We see a variety of line shapes, including single- and double-peaked profiles (e.g. panels 3 and 5 and panel 6, respectively).
For some lines-of-sight, the profile is symmetric around the line-centre (e.g. panel 5), though profiles are often asymmetric and skewed towards short wavelengths.

\begin{table}
    \centering
\begin{tabular}{w{c}{0.2cm} w{c}{1.5cm} w{c}{1.5cm} w{c}{1.5cm} w{c}{1.5cm}}
 \hline
 l.o.s. & $\rm FWHM_{\rm all}$ & $\rm FWHM_{\rm all}^{\rm HR}$ & $\rm FWHM_{\rm QSO}^{\rm HR}$ & $\rm FWHM_{\rm REC+COL}^{\rm HR}$ \\
 & \multicolumn{4}{c}{$\rm [km \, s^{-1}]$}\\
 \hline
1 & 765 (765) & 691 (691) & 666 (666) & 716 \\
2 & 814 (814) & 716 (716) & 691 (691) & 765 \\  
3 & 691 (691) & 617 (617) & 444 (469) & 716 \\
4 & 617 (617) & 543 (543) & 444 (444) & 562 \\
5 & 888 (913) & 790 (790) & 937 (913) & 716 \\
6 & 543 (543) & 494 (494) & 494 (494) & 469 \\
 \hline

\end{tabular}
\caption{Full-widths-at-half-maximum for different lines-of-sight (first column) from the integrated spectra for all Ly$\alpha$ sources at MUSE resolution (second column) and at a higher spectral resolution of $\Delta \lambda_{\rm obs} \, = \, 0.7 \angstrom$ for all Ly$\alpha$ sources (third column), for only BLR scattering (fourth column) and for a combination of collisional excitation and recombination cooling (fifth column). Values in brackets are obtained using $\rm FWHM_{\rm BLR} \, \approx \, 2400 \, \rm km \, s^{-1}$, while non-bracketed values are computed using $\rm FWHM_{\rm BLR} \, \approx \, 3500 \, \rm km \, s^{-1}$ for the intrinsic spectrum of the quasar Ly$\alpha$ line. The values given in this table are computed after subtracting the flux from within an aperture with radius $0.5 \, \rm arcsec$ from the quasar, as in the observations. They therefore correspond to extended emission only.}
\label{table_FWHM}
\end{table}

Asymmetries in the integrated line profiles exist even in the absence of resonant scattering (bottom sub-figure in Figure~\ref{fig:line_profile}), but they are greatly amplified if scattering is accounted for. In most cases (except in panel 5), we see much that blue peaks are far more pronounced, indicating that Ly$\alpha$ photons are mostly processed by infalling material \citep[as also found in][]{Mitchell:21}, as shown qualitatively in Section~\ref{sec:overview}. The absence of a pronounced red wing indicates that outflowing gas, even if present (see Figure~\ref{fig:intro}), either is too fast or does not provide a high enough HI covering fraction (see Section~\ref{sec:openquestions} for potential explanations) at $z \, = \, 6.2$, and, as we verified, also at $z \, = \, 6.3$ and $z \, = \, 6$. Observed spectral line profiles in REQUIEM do not display symmetric, double-peaked profiles and are broadly consisted with our mock spectra, though distinct peaks may not be detected due to high levels of noise, which we have not attempted to model here.

The numbers given in every panel of Figure~\ref{fig:line_profile} give the flux-weighted velocity dispersion (second moment of flux distribution) for each Ly$\alpha$ source model (see also moment maps in Appendix~\ref{appendix:b}). These numbers are coloured according to the emission mechanism, following the same convention as the coloured curves.
For recombination radiation and collisional excitation, second moments range from $\approx 400 \, \rm km \,s^{-1}$ to $\approx 560 \, \rm km \,s^{-1}$. For BLR scattering, spectral lines are generally broader, with second moments ranging from $\approx 860 \, \rm km \, s^{-1}$ to $\approx 1050 \, \rm km \,s^{-1}$. For closer comparison with \citet{Farina:19}, we also quantify line-widths through a full-width-at-half-maximum (FWHM). We compute FWHMs for each line-of-sight and for various combinations of Ly$\alpha$ emission mechanisms, listing the results in Table~\ref{table_FWHM}. When combining all emission processes, FWHMs range from $540 \, \rm km \, s^{-1}$ to $910 \, \rm km \, s^{-1}$, consistent with \citet{Farina:19}, where FWHMs follow an approximately flat distribution ranging from $\approx 300 \, \rm km \, s^{-1}$ to $\approx 1800 \, \rm km \, s^{-1}$. Table~\ref{table_FWHM} also indicates that FWHM are likely overestimated even at MUSE resolution. Comparing the first and second columns, we see that decreasing the spectral resolution from $2.6 \, \rm \angstrom$ to $0.7\, \rm \angstrom$ results in FWHMs which are narrower by $50 \-- 100 \, \rm km \, s^{-1}$.

Panel 5 in Figure~\ref{fig:line_profile} shows the shape of the input quasar Ly$\alpha$ line (dotted, black curve), which we have re-normalised in order to more closely compare with the emerging spectrum. This input spectrum is much broader than the spectral line associated to extended emission. 
For pure BLR scattering alone, we find FWHMs $\lesssim 940 \, \rm km \, s^{-1}$ for extended emission (see Table~\ref{table_FWHM}). Interestingly, the FWHM associated to extended emission does not appear to change significantly by increasing the $\rm FWHM_{\rm BLR}$ of the quasar Ly$\alpha$ line from of $2400 \, \rm km \, s^{-1}$ (bracketed values) to $3500 \, \rm km \, s^{-1}$. 

 \emph{Extended Ly$\alpha$ nebulae characterised by much narrower line-widths than the quasar's Ly$\alpha$ line \citep[e.g.][]{Ginolfi:18} are thus not inconsistent with a BLR scattering origin.} Photons belonging to the broad wings of the quasar emission line are not absorbed efficiently and stream freely without scattering. These photons are seen as a point source, but do not contribute to \emph{extended} emission. Those photons that do scatter and create a Ly$\alpha$ nebula are those that have $|\Delta v| \lesssim 1000 \, \rm km \, s^{-1}$. Assuming a constant luminosity, an intrinsically broader quasar Ly$\alpha$ line can still power a large nebula (see Fig.~\ref{fig_nebulae_los}) with a narrow spectral line, with the main difference being that the nebula becomes somewhat fainter due to the fact that the quasar flux is more widely distributed in frequency space (see Fig.~\ref{fig_SBprofile}). 

One may try to compare the line-widths of the integrated spectra to the velocity dispersion of the dark matter halo hosting the bright quasar in our simulations. A direct connection between gas dynamics and the line-width can exist if optical depths are relatively low. At radii of $\approx 10 \-- 100 \, \rm kpc$, the circular velocity associated to the dark matter component is $\approx 400 \, \rm km \, s^{-1}$, which is close to the mean dispersion values $\approx 335 \, \rm km \, s^{-1}$ obtained in the absence of scattering (bottom sub-figure in Figure~\ref{fig:line_profile}). However, we can see that scattering broadens the spectral lines significantly, yielding FWHMs that can exceed the halo's circular velocities by factors $\approx 1.5$.

\subsubsection{Nebula luminosities and sizes}
\label{sec:luminosities}

\begin{figure*}
    \centering
    \includegraphics[width=0.475\textwidth]{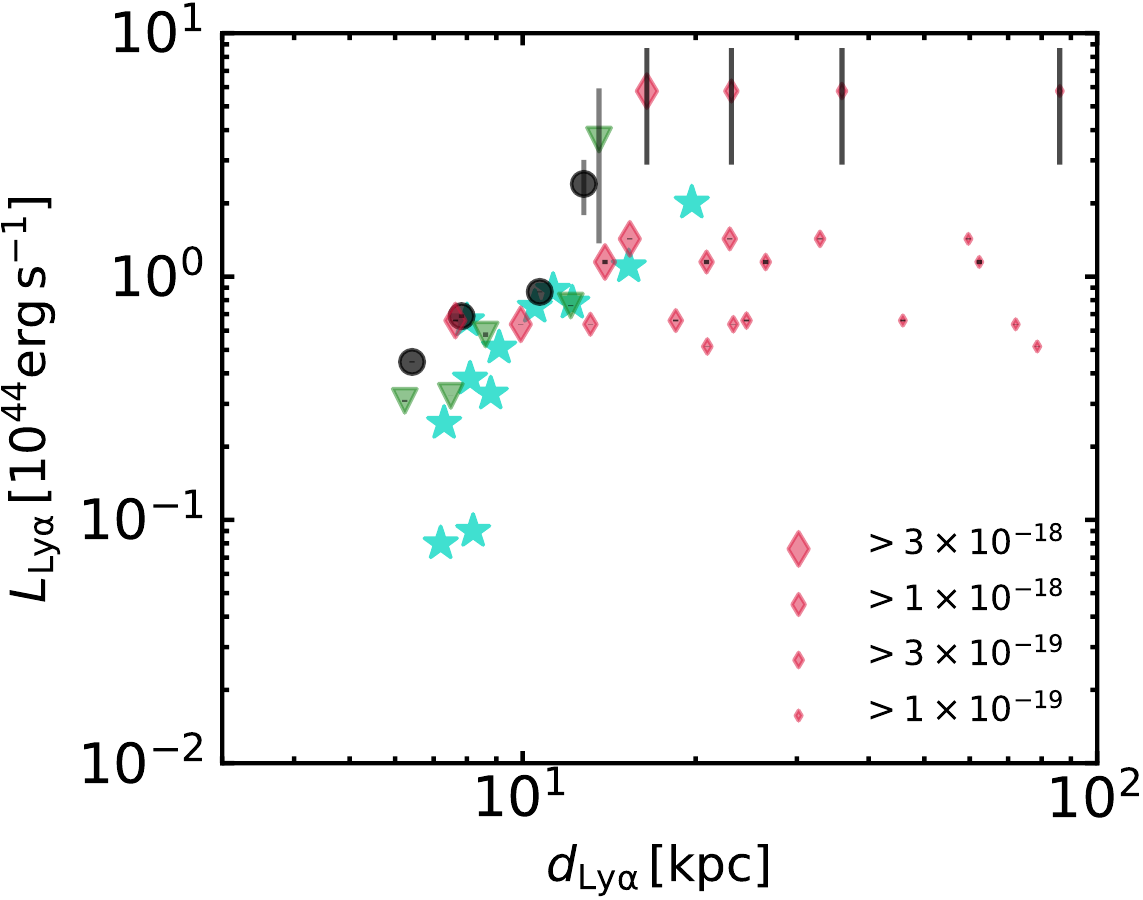}
    \includegraphics[width=0.475\textwidth]{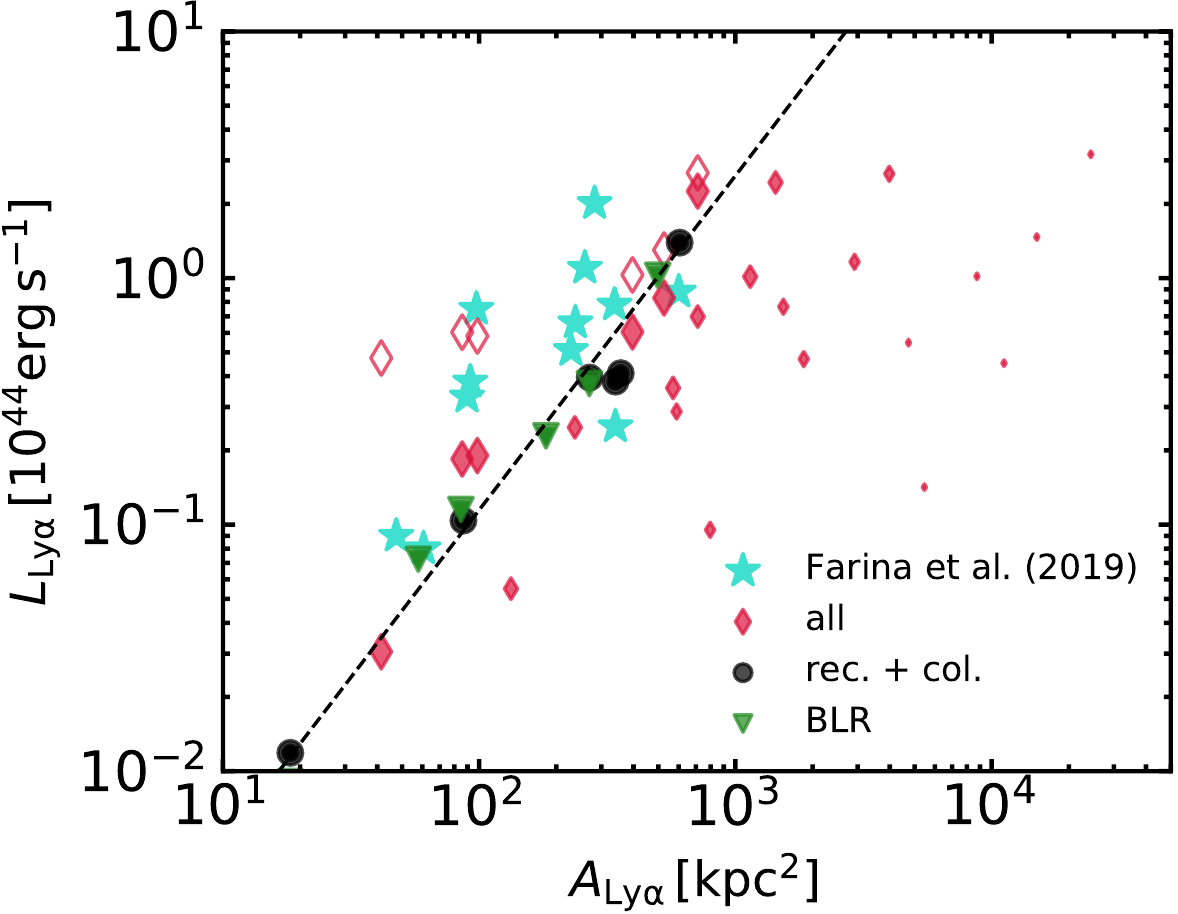}
    \caption{{\bf Left:} Relation between the nebula luminosities for different lines-of-sight and the size of the nebulae evaluated by measuring the radius at which the spherically-averaged SB profile falls below different surface brightness limits (different plot marker sizes). Comparison between data (cyan stars) and simulations (other symbols) should be performed for the large symbols only, which correspond to the depth of the REQUIEM survey observations. Simulation results are shown with black circles (for the combined recombination radiation and collisional excitation scenarios), with green triangles for the BLR scattering scenario and with red diamonds for a combination of all Ly$\alpha$ sources. Error bars quantify the luminosity change that occurs when subtracting the flux from within a circular aperture with radius $0.5 \, \rm arcsec$ to mimic PSF subtraction. {\bf Right:} Relation between the nebula luminosities and projected area for different  lines-of-sight and surface brightness cut-offs. There is a monotonic relation between nebula size and luminosity. The normalisation of this relation depends on how the luminosity is quantified. The open symbols give luminosities as obtained by integrating the surface brightness map within an aperture of $15 \, \rm arcsec$, while the filled symbols show the values obtained by integrating only over pixels above a given surface brightness threshold. Results for surface brightness thresholds lower than $3 \times 10^{-18} \, \rm erg \, s^{-1} \, cm^{-2} \, arcsec^{-2}$ are only shown for the scenario where BLR scattering, collisional excitation and recombination radiation all contribute for clarity. The dotted line gives a linear fit for the relation found in the simulations (here shown for a combination of all Ly$\alpha$ emission processes) for the highest surface brightness cut-off considered. Table~\ref{tab:linear_regression} lists best-fit parameters for this and other surface brightness thresholds.}
    \label{fig_size_luminosity}
\end{figure*}

In the REQUIEM Survey, detected nebulae have Ly$\alpha$ luminosities that range from $10^{43} \, \rm erg \, s^{-1}$ to $\approx 2 \times 10^{44} \, \rm erg \, s^{-1}$. 
Nebula sizes vary depending on how they are defined.
Nebulae are typically identified by finding connected regions above a given signal-to-noise ratio. The nebula's size can then, for instance, be estimated by measuring the maximum diameter distance. In \citet{Farina:19}, this definition yields sizes ranging from $\approx 15 \, \rm kpc$ to $\approx 45 \, \rm kpc$. Sizes obtained using this definition depend on the depth of the data.

In order to more closely compare with predictions from our simulations, we adopt a different definition for nebula size, which is less sensitive to variable signal-to-noise ratios.
A uniform measurement across all observed nebulae involves measuring the radius at which the spherically averaged surface brightness profile falls below a certain threshold. In \citet{Farina:19}, this definition yields smaller nebula sizes, ranging from $\approx 7 \, \rm kpc$ to $\approx 20 \, \rm kpc$ for a surface brightness threshold of $3 \times 10^{-18} \, \rm erg \, s^{-1} \, cm^{-2} \, arcsec^{-2}$. 

In Figure~\ref{fig_size_luminosity}, we plot our simulation predictions for nebula sizes and luminosities for our six lines-of-sight at $z \, = \, 6.2$ on the left-hand panel. The Ly$\alpha$ luminosity is here obtained by measuring the total flux within a circular aperture of radius $15 \, \rm arcsec$ and the nebula size is estimated using the same surface brightness-based size definition as in the REQUIEM survey. As in \citet{Farina:19}, the surface brightness is obtained by integrating the mock data-cube between the velocity channels  $-500 \, \rm km \, s^{-1}$ and $500 \, \rm km \, s^{-1}$.
Data from {\sc REQUIEM} is shown with cyan stars, while data from the simulations is shown with black circles for the combined recombination radiation and collisional excitation scenarios, with green triangles for the BLR scattering scenario (with $\rm FWHM_{\rm BLR} \, \approx \, 3500 \, \rm km \, s^{-1}$) and with red diamonds for the combination of all three processes. Different plot symbol sizes show how nebula sizes vary when the surface brightness limit is modified, with the largest symbols corresponding to the same surface brightness limit as in REQUIEM. Comparison between data and simulations should be performed for these large symbols only. In order to assess how nebula sizes might change with future, deeper observations, smaller symbols show results for lower surface brightness limits for the combined BLR scattering, collisional excitation and recombination radiation scenario. We verify that all source models follow the same qualitative trends. 

At the $3 \times 10^{-18} \, \rm erg \, s^{-1} \, cm^{-2} \, arcsec^{-2}$ sensitivity of \citet{Farina:19}, we find a close match between simulations and observations for all Ly$\alpha$ processes investigated. Nebula sizes range from $\approx 4 \, \rm kpc$ to $\lesssim 20 \, \rm kpc$.
While the nebula luminosity varies significantly with the line-of-sight, as shown in Section~\ref{sec:Ly-escape}, it changes only by a factor $\lesssim 2$ when including or excluding the central $0.5 \, \rm arcsec$ (error bars). 
The nebula size instead depends strongly on the sensitivity of the observations. For a sensitivity of $SB > 3 \times 10^{-19} \, \rm erg \, s^{-1} \, cm^{-2} \, arcsec^{-2}$, we find a nebular size range of $20 \-- 40 \, \rm kpc$ and, for $SB > 10^{-19} \, \rm erg \, s^{-1} \, cm^{-2} \, arcsec^{-2}$, a range of $40 \-- 100 \, \rm kpc$, up to more than a factor 2 larger than the virial radius of the quasar host halo.  
If nebulae properties did not experience significant redshift evolution down to $z \approx 3$, the strong $(1+z)^{-4}$ scaling associated to SB-dimming means that a Ly$\alpha$ nebula at $z \approx 6$ would been seen with a size of up to $\approx 100 \, \rm kpc$ around a $z \approx 3$ quasar if probed down to the same surface brightness level as in \citet{Farina:19}. Current observations of Ly$\alpha$ nebulae at $z > 6$ thus only likely probe a small fraction of their true extent.

\begin{table}
    \centering
\begin{tabular}{w{c}{1cm} w{c}{2cm} w{c}{1cm} w{c}{1cm} w{c}{1cm}}
 \hline
 \multicolumn{5}{c}{Best linear fit parameters} \\
 \hline
   & SB limit & slope & intercept & p-value\\
   & $\rm [erg \, s^{-1} \, cm^{-2} \, arcsec^{-2}]$ & & & \\

 \hline
    &$3 \times 10^{-18}$   & 1.27 & -3.51 & 0.00014 \\
    \multirow{2}{*}{rec. + col.}   &$1 \times 10^{-18}$   & 1.44 & -4.41 & 0.00523 \\
    &$3 \times 10^{-19}$   & 1.53 & -5.31 & 0.01006 \\
    &$1 \times 10^{-19}$   & 1.70 & -6.69 & 0.00106 \\
    
 \hline
 
    &$3 \times 10^{-18}$  & 1.36 & -3.65 & 0.00011 \\
    \multirow{2}{*}{BLR}  &$1 \times 10^{-18}$   & 1.36 & -4.08 & 0.00029 \\
    &$3 \times 10^{-19}$  & 1.32 & -4.30 & 0.00665 \\
    &$1 \times 10^{-19}$  & 1.65 & -6.17 & 0.10504 \\
    
 \hline
  
    &$3 \times 10^{-18}$  & 1.25 & -3.34 & 0.00176 \\
    \multirow{2}{*}{all}  &$1 \times 10^{-18}$   & 1.37 & -4.07 & 0.00176 \\
    &$3 \times 10^{-19}$  & 1.38 & -4.66 & 0.02158 \\
    &$1 \times 10^{-19}$  & 1.41 & -5.78 & 0.04810 \\
    
 \hline
 
     &$3 \times 10^{-18}$  & 1.15 & -3.18 & 0.08777 \\
    \multirow{2}{*}{all (no scat.)}  &$1 \times 10^{-18}$ & 0.36 & -1.37 & 0.78678 \\
    &$3 \times 10^{-19}$  & 0.88 & -3.18 & 0.14315 \\
    &$1 \times 10^{-19}$  & 0.80 & -3.10 & 0.07018 \\
    
 \hline
\end{tabular}
    \caption{Best-fit parameters for linear regression between logarithmic luminosity and logarithmic nebula area and for different surface brightness limits and different nebula mechanisms.}
    \label{tab:linear_regression}
\end{table}

\begin{figure*}
    \centering
    \includegraphics[width=0.475\textwidth,trim={0 0.75cm 0 0},clip]{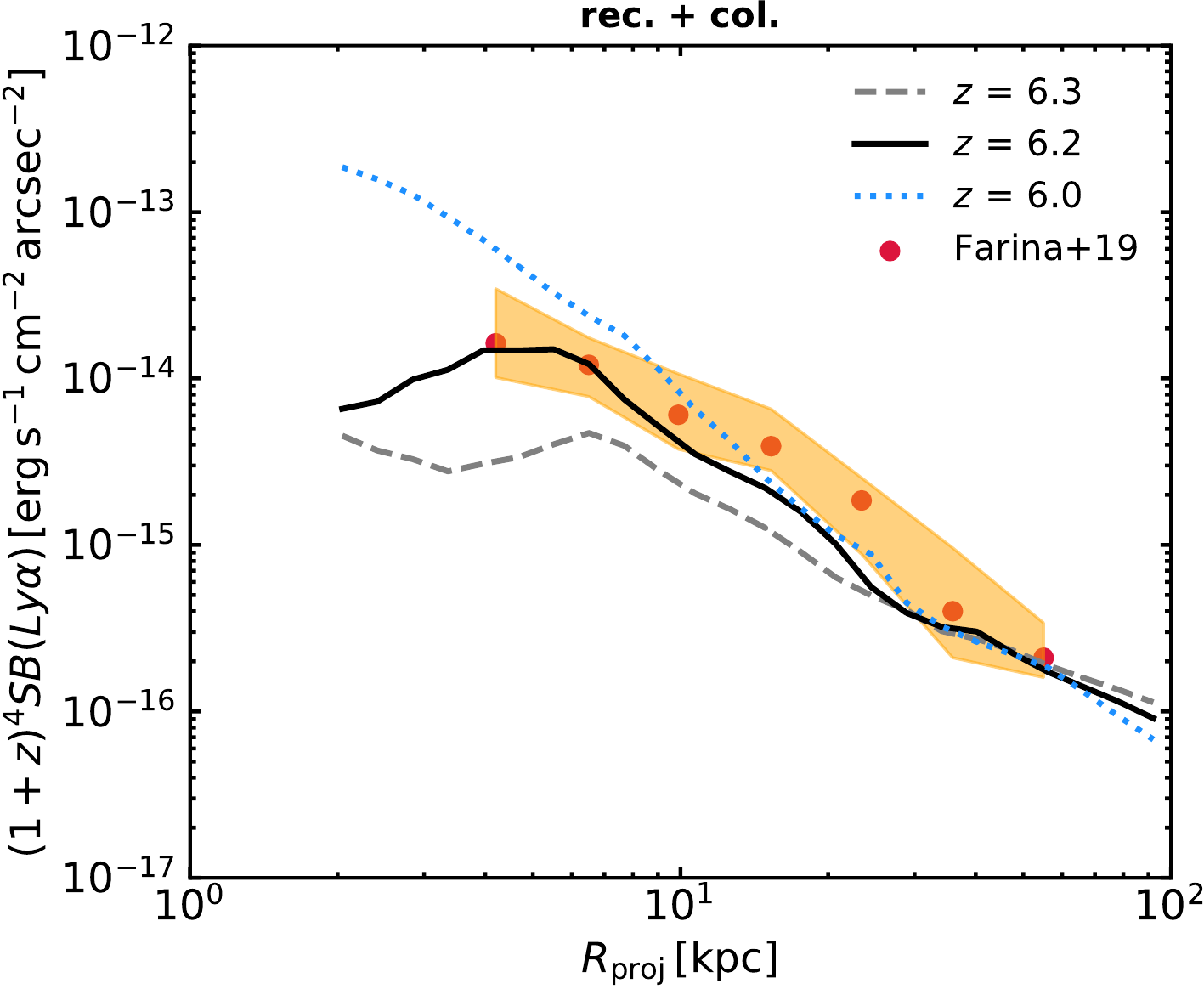}
    \includegraphics[width=0.475\textwidth,trim={0 0.75cm 0 0},clip]{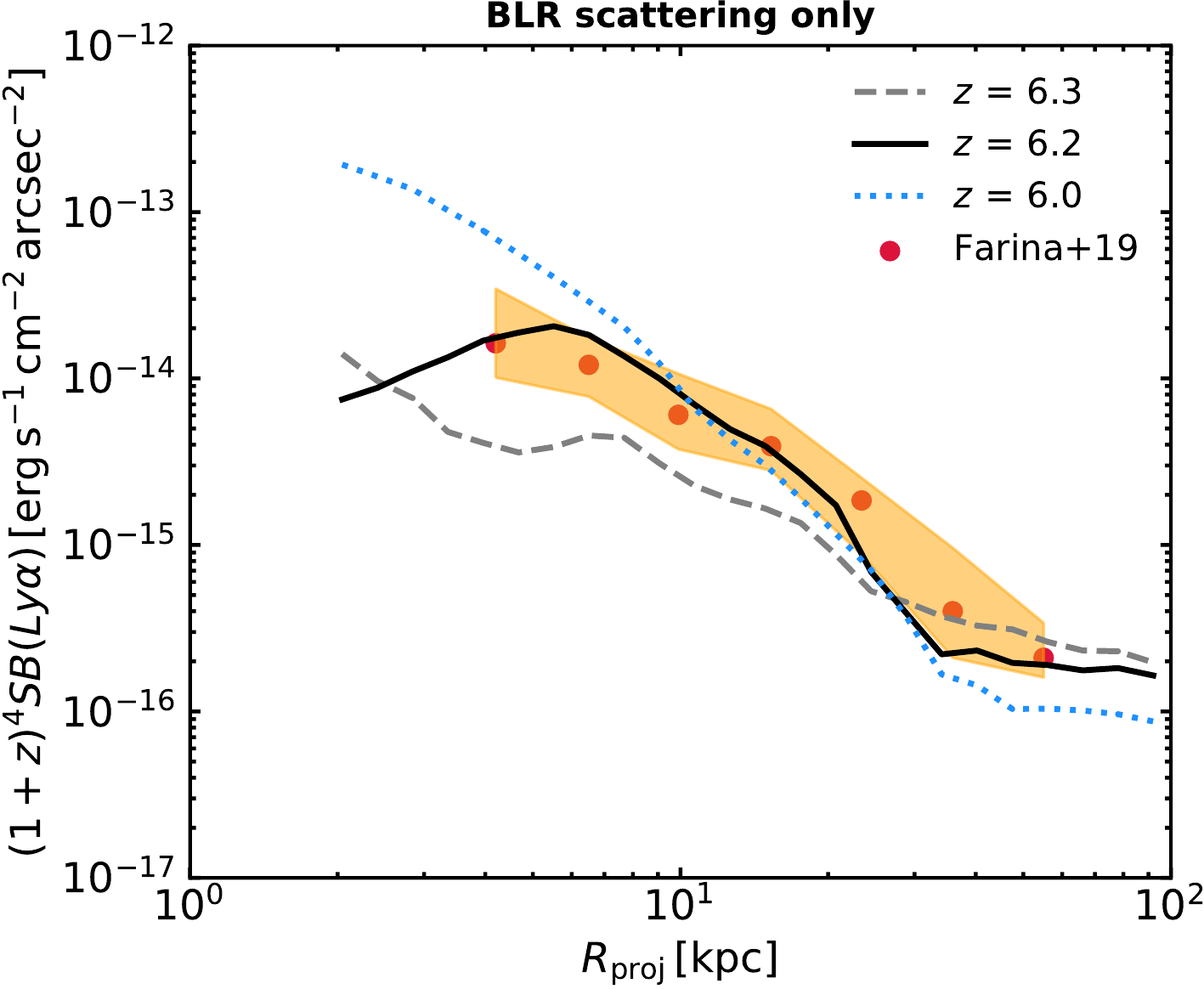}
    \includegraphics[width=0.475\textwidth]{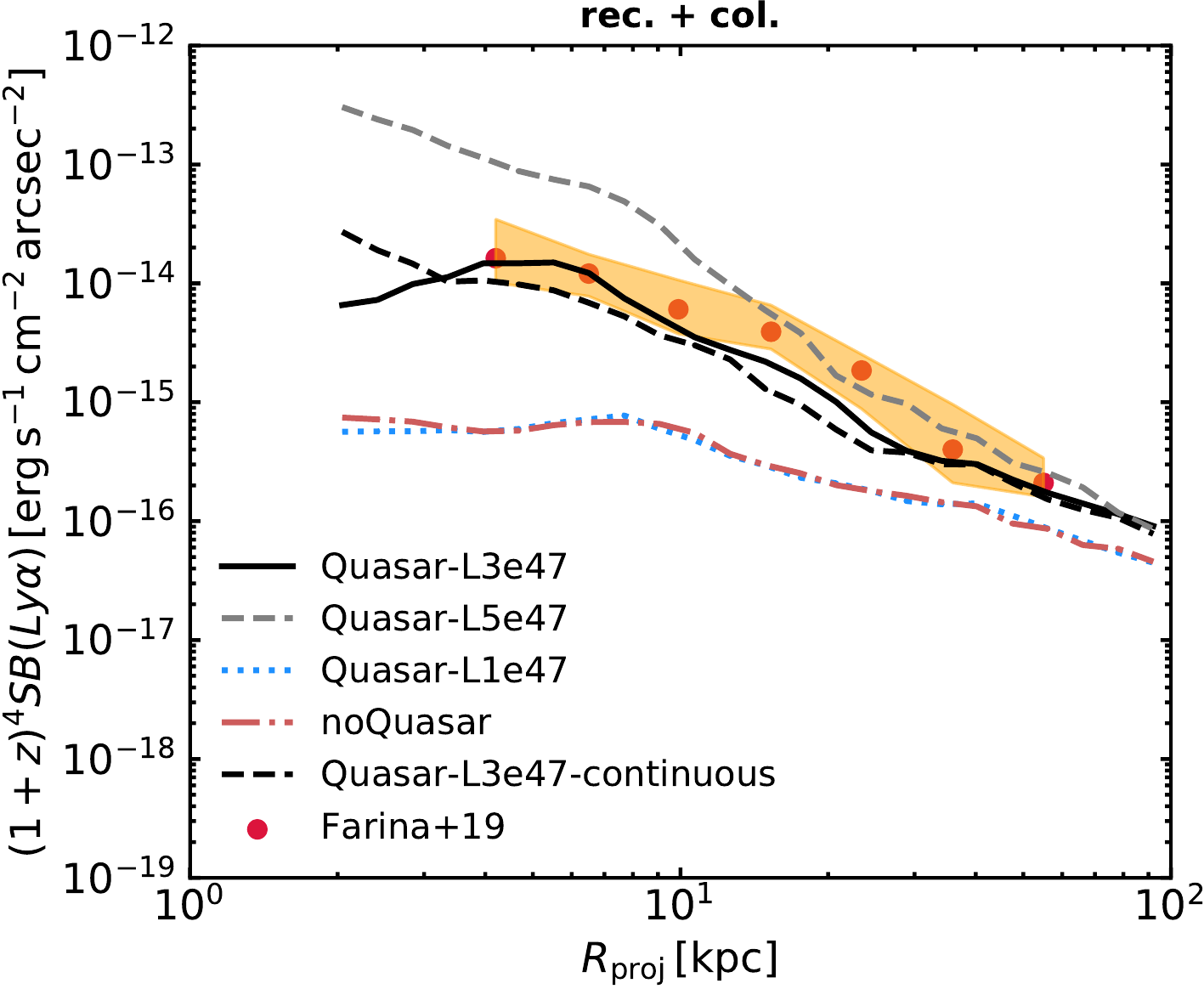}
    \includegraphics[width=0.475\textwidth]{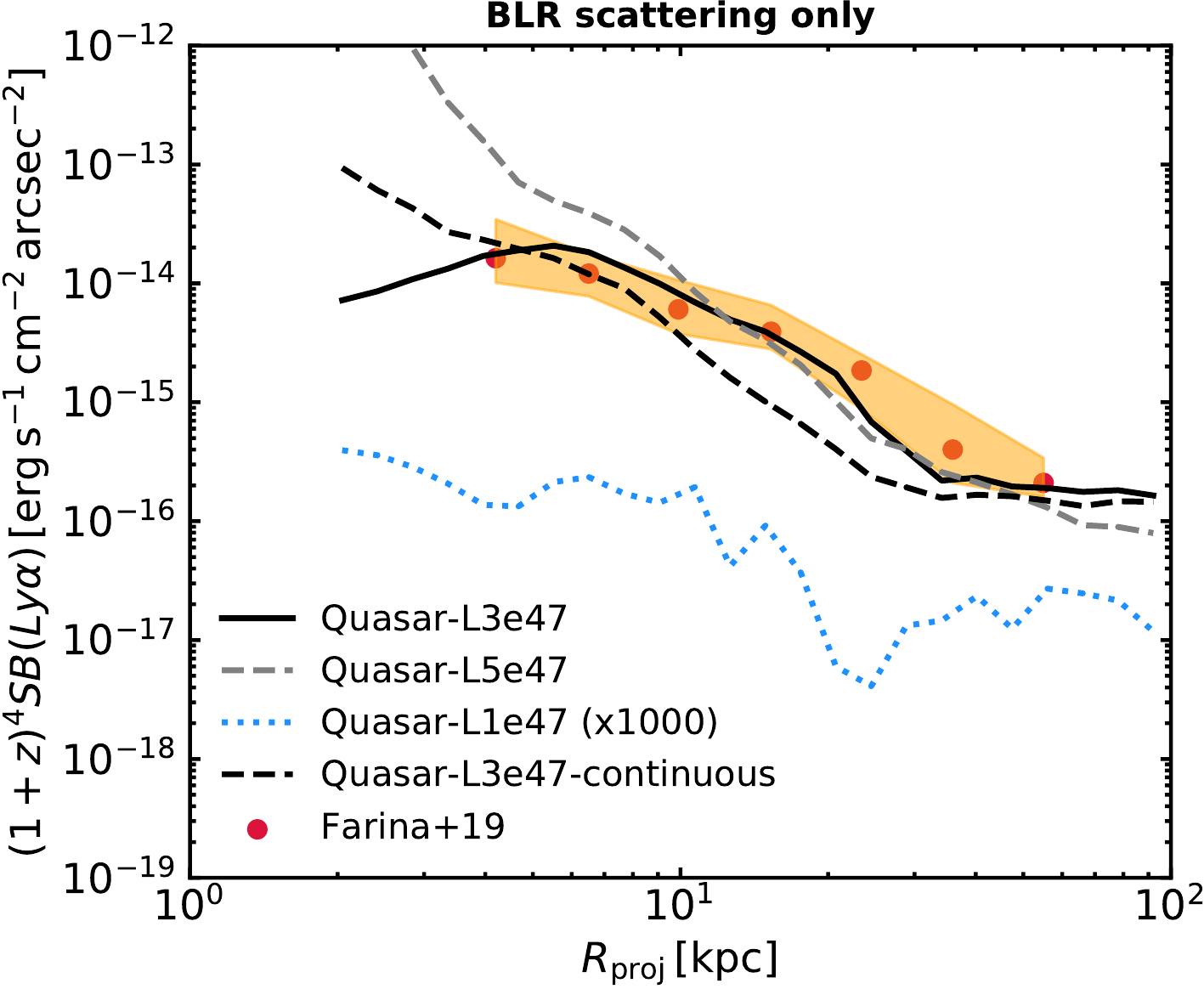}
    \caption{Median surface brightness profiles for a combination of collisional excitation and recombination radiation (left-hand panels) and for BLR scattering with $\rm FWHM_{\rm BLR} \, = \, 2400 \, \rm km \, s^{-1}$ (right-hand panels). Different curves illustrate the profiles obtained by varying the redshift (top row) and the AGN bolometric luminosity (bottom row). Brighter AGN and later times are associated to higher net AGN injected energies, which result in more destruction of dust and HI gas in the central regions of the halo. Since resonant scattering becomes less efficient, the surface brightness profile steepens if AGN feedback is stronger. On the bottom right panel, the surface brightness for \texttt{Quasar-L1e47} is multiplied by a factor of $1000$, since the escape fraction for BLR photons is only $\approx 0.03 \%$.}
    \label{fig_SBprofile_time}
\end{figure*}

On the right-hand panel of Figure~\ref{fig_size_luminosity}, we plot nebula area as a function of luminosity at $z \, = \, 6.2$. We show results for all Ly$\alpha$ source models at surface brightness $> 3 \times  10^{-18} \, \rm erg \, s^{-1} \, \rm cm^{-2} \, arcsec^{-2}$ and for the model accounting for combined BLR scattering, collisional excitation and recombination radiation for lower surface brightness limits.
In order to measure nebula areas, we first identify all connected pixels above a given surface brightness threshold, and then add up their individual areas to obtain the nebula's total area $A_{\mathrm{Ly \alpha}}$. Instead of measuring the total flux within some aperture, as performed for the left-hand panel, we here experiment defining the nebula's luminosity by adding only the contributions of pixels above a given surface brightness limit.
At any fixed surface brightness limit, we find a monotonic relation between $A_{\mathrm{Ly \alpha}}$ and luminosity. We fit a power law through each of the four sets of points assuming $\log_{\rm 10}{\left( L_{\mathrm{Ly} \alpha}/ \mathrm{[10^{44} \, erg \, s^{-1}]} \right)} \, = \, m \log_{\rm 10}{\left( A_{\mathrm{Ly} \alpha} / \mathrm{[kpc^{2}]} \right)}+ C$ and provide the best-fit parameters in Table~\ref{tab:linear_regression}. It is interesting to observe that at surface brightness limits $\gtrsim 10^{-18} \, \rm erg \, s^{-1} \, \rm cm^{-2} \, arcsec^{-2}$, roughly the same area -- luminosity relation is shared between BLR scattering, combined collisional excitation and recombination radiation, or all these processes together.

We also see on the right-hand panel of Figure~\ref{fig_size_luminosity} that simulated nebula areas are largely consistent with observational estimates for surface brightness limits $\gtrsim 3 \times 10^{-18} \, \rm erg \, s^{-1} \, \rm cm^{-2} \, arcsec^{-2}$ (closed, diamon symbols). At fixed area, however, observed nebulae sometimes appear brighter than our simulations. This discrepancy is caused by different nebula luminosity definitions. While \citet{Farina:19} gives the total flux within an aperture of radius ranging from $9 \, \rm kpc$ to $30 \, \rm kpc$ depending on the quality of the data, we quote the luminosity of pixels with a surface brightness above a certain threshold. Were we to adopt a definition closer to that used in \citet{Farina:19}, as on the left-hand panel of Figure~\ref{fig_size_luminosity}, we would obtain the open diamonds on the right-hand panel of Figure~\ref{fig_size_luminosity}. The corresponding luminosities can be considerably higher, bracketing the observed values. This discrepancy caused by different luminosity conventions disappears as we decrease the surface brightness threshold and fainter pixels are accounted for when evaluating the nebula's luminosity.

\subsection{The effect of the quasar luminosity}
\label{sec:qsoluminosity}

In order to investigate the time and luminosity dependence of the surface brightness profiles presented in Section~\ref{sec:SBprofile}, we plot in Figure~\ref{fig_SBprofile_time} median profiles obtained at different redshifts (top panels) and for different AGN luminosities at $z \, = \, 6.2$ (bottom panels). On the left-hand panels, we show surface brightness profiles for a combination of collisional excitation and recombination photons. On the right-hand panels, we show results for BLR scattering alone, using $\rm FWHM_{\rm BLR} \, \approx \, 2400 \, \rm km \, s^{-1}$ for the quasar broad line.
The agreement between simulated and observed median profiles remains close within the redshift range of $6 < z < 6.3$, in particular at large radial distances from the quasar, for both emission processes. Some systematic time evolution can, however, be seen. Profiles tend to steepen with time, increasing in the central $\sim 10 \, \rm kpc$ and dropping above radii of $\gtrsim 30 \, \rm kpc$.

On the bottom panels, we see that increasing the AGN luminosity produces an analogous effect as looking at later times: the surface brightness profile becomes steeper. In \texttt{Quasar-L5e47}, simulated and observed profiles share similar shapes, but the normalisation of the theoretical profiles is higher in the central $10 \, \rm kpc$ than observed. Allowing the quasar to shine for longer, as in \texttt{Quasar-L3e47-continuous}, also boosts the central surface brightness.
The higher quasar luminosity of \texttt{Quasar-L5e47} results in stronger feedback, expelling more material from the galactic nucleus. The escape fractions are $f_{\rm esc} \approx 39 \%$ for recombination radiation, $f_{\rm esc} \approx 30\%$ for collisional excitation and $f_{\rm esc} \approx 87 \%$ for BLR photons. Recall that $f_{\rm esc} \approx 20 \%$ for recombination radiation, $f_{\rm esc} \approx 30\%$ for collisional excitation and $f_{\rm esc} \approx 73 \%$ for BLR photons in \texttt{Quasar-L3e47} (Section~\ref{sec:Ly-escape}). Brighter AGN also produce more ionising photons, increasing the intrinsic recombination and BLR emissivities.
Reducing the AGN luminosity, in turn, suppresses the surface brightness. In \texttt{Quasar-L1e47}, the momentum flux associated to radiation pressure is barely sufficient to overcome the gravitational force binding gas to the galactic nucleus \citep{Costa:18}. Consequently a large dusty gas reservoir persists in the galactic nucleus, preventing recombination and BLR photons from escaping efficiently. Escape fractions for both processes drop to, respectively, $f_{\rm esc} \approx 0.5\%$ and $f_{\rm esc} \approx 0.03\%$. Even if they do escape, high HI optical depths cause these photons to scatter beyond the spectral window of $-500 \, \mathrm{km \, s^{-1}} < \Delta v < 500 \, \mathrm{km \, s^{-1}}$ used to construct the surface brightness profiles (Section~\ref{sec:SBprofile}), further diminishing their contribution. The resulting profiles thus become similar to those obtained in \texttt{noQuasar} (see bottom left panel) and most escaping flux is generated via collisional excitation, for which $f_{\rm esc} \approx 24\%$, outside of the host galaxy.

\begin{figure}
    \centering
    \includegraphics[width=0.45\textwidth]{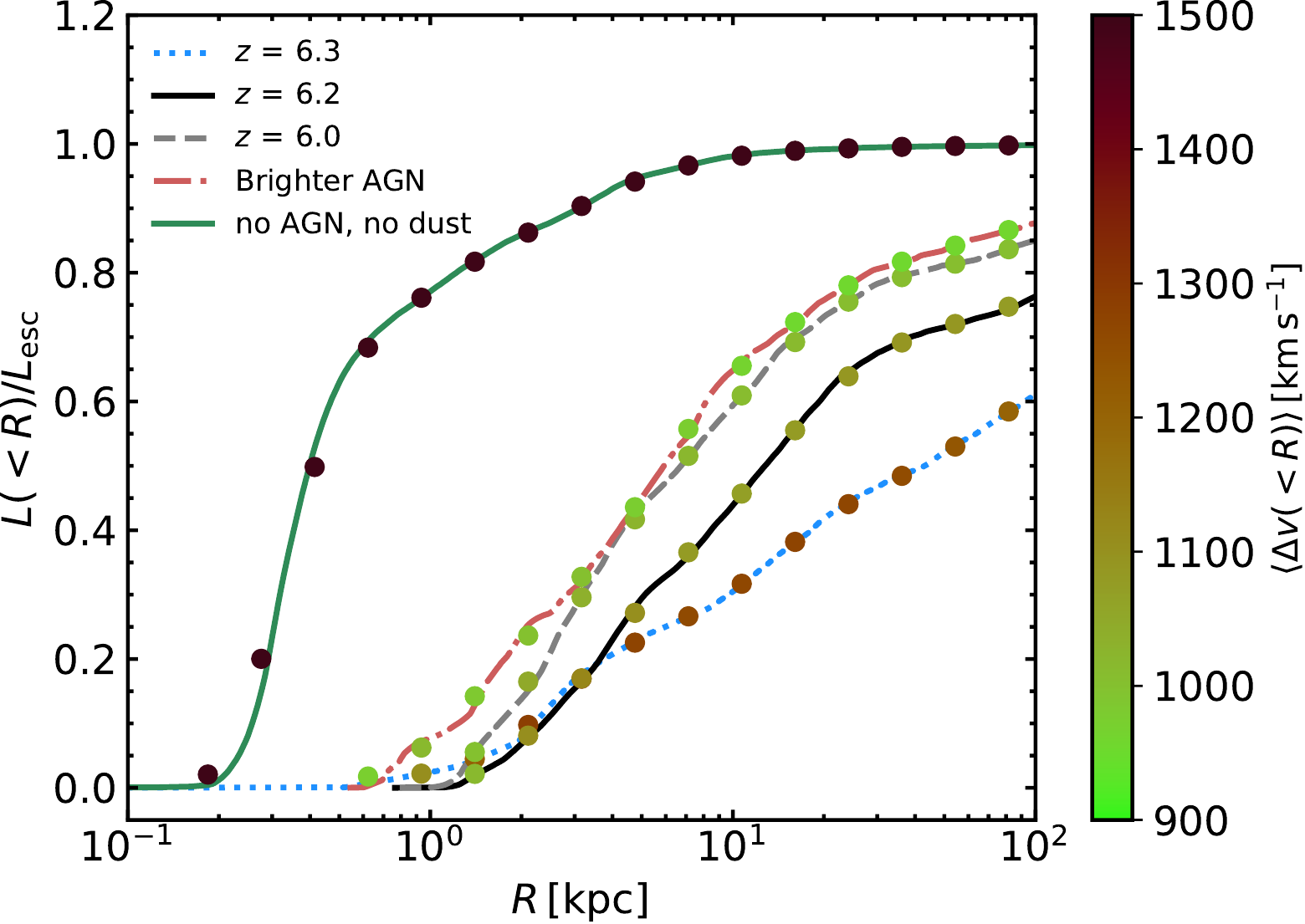}
    \caption{Cumulative luminosity of escaping Ly$\alpha$ photons that have undergone their last scattering prior to reaching radius $R$ for BLR scattering. The cumulative luminosity is plotted as a function of radius in our various simulations, and is normalised to the total escaping luminosity. The coloured circles indicate the $75^{\rm th}$ percentile of the velocity shift associated to escaping photons below radius $R$. In order to produce an extended nebula, BLR photons have to scatter out to large radii. This condition is best satisfied for \texttt{Quasar-L3e47} at $z \, = \, 6.2$ (black curve) and at $z \, = \, 6.3$ (blue, dotted curve). Brighter quasar (dot-dashed, red curve) or an AGN operating for longer (dashed, grey curve) result in lower HI optical depths and reduce the efficiency of scattering, causing more compact nebulae. The absence of AGN feedback (green curve), shown here for a calculation neglecting dust absorption results in too much scattering. In this case, photons stream way from the system on a single fly-out directly from the quasar host galaxy without producing an extended nebula.}
    \label{fig:nscats}
\end{figure}

In the following, we explain why brighter AGN and later simulation times appear to be correlated with lower surface brightness \emph{at very large radii}, focussing on BLR scattering, where this effect is particularly clear.
In Figure~\ref{fig:nscats}, we plot the cumulative luminosity of photons that have undergone their last scattering event prior reaching out to a radius $R$. These photons no longer scatter at radii $> R$ and therefore do not contribute to extended emission beyond that point. Recall that in order to generate an extended Ly$\alpha$ nebula, BLR photons need to scatter.
The cumulative luminosity is plotted as a function of radius in our various simulations, and is normalised to the total escaping luminosity.
The coloured circles further indicate the $75^{\rm th}$ percentile of the velocity shift associated to escaping photons below radius $R$. If this velocity shift is high, then photons streaming away from the system without further interaction have experienced a high number of scatterings and thus encountered a high HI column. A low velocity shift conversely indicates a small number of scatterings and lower HI optical depths.
In Figure~\ref{fig:nscats}, we find that the lowest velocity shifts occur for \texttt{Quasar-L5e47} at $z \, = \, 6.2$ (red, dot-dashed curve) and for \texttt{Quasar-L3e47} at $z \, = \, 6$ (grey, dashed curve). In both cases, $80\%$ of ``last scattering events'' occur below $R \, \approx \, 30 \, \rm kpc$. Correspondingly, the associated Ly$\alpha$ nebulae are the least extended, in agreement with Figure~\ref{fig_SBprofile_time}.
Higher velocity shifts occur for \texttt{Quasar-L3e47} at $z \, = \, 6.2$. For this simulation, most last scatterings occur at larger radii than in e.g. \texttt{Quasar-L5e47}, and the associated nebula is, correspondingly, more extended. Yet larger velocity shifts occur for \texttt{Quasar-L3e47} at $z \, = \, 6.3$. Here, scattering is particularly efficient and therefore able to transport photons from the BLR to scales $> 100 \, \rm kpc$, likely because there has been less time for AGN feedback to destroy HI gas in the CGM.

Beyond a critical point, however, scattering becomes so efficient that the associated frequency shifts prevent photons from interacting further. The green curve in Figure~\ref{fig:nscats} shows results for \texttt{noQuasar}. Since $f_{\rm esc} \, = \, 0 \%$ for BLR photons in \texttt{noQuasar} (if dust absorption is accounted for), we show results from a Ly$\alpha$ radiative transfer calculation in which we neglect dust absorption. Scattering is here so efficient that most photons stream away from the host galaxy on a single fly-out already at scales $\lesssim 500 \, \rm pc$. 

Figure~\ref{fig:nscats} underlines the central role of AGN feedback in shaping the properties of Ly$\alpha$ nebulae.
If more efficient, either because the AGN is brighter or if it has been active for a longer time, AGN feedback reduces the central optical depths. More Ly$\alpha$ radiation leaves the system without scattering in the central regions and thus fewer photons scatter our to large radii: the surface brightness profile steepens and the Ly$\alpha$ nebula shrinks. Less efficient AGN feedback (i) allows photons to escape without being absorbed by dust and (ii) makes it possible for HI scatterers to survive and efficiently transport photons to large radii, producing the most extended nebulae.   
At face value comparison between the theoretical surface brightness profiles and the median observed profile seem to disfavour strong AGN feedback, as the resulting surface brightness profiles become steeper than observed. But some AGN feedback is clearly required.

\subsection{Ly$\alpha$ nebulae in $z > 7$ quasars?}
\label{sec:lyalpha7}
In this section, we consider whether the most distant quasars at $z > 7$ should also exhibit observable Ly$\alpha$ nebulae. We perform a new cosmological simulation targeting the same massive halo, but injecting quasar radiation starting at $z \, = \, 7.7$, at a constant bolometric luminosity of $L_{\rm bol} \, = \rm 10^{47} \, \rm erg \, s^{-1}$. These values are chosen in order to mimic the properties of the most distant quasar ($z \, \approx \, 7.6$) known \citep{Wang:21}.
These simulations are then post-processed with {\sc RASCAS}, again accounting for recombination cooling, collisional excitation and the quasar BLR as Ly$\alpha$ sources.

\begin{figure}
    \centering
    \includegraphics[width=0.45\textwidth,trim={0 1cm 0 0},clip]{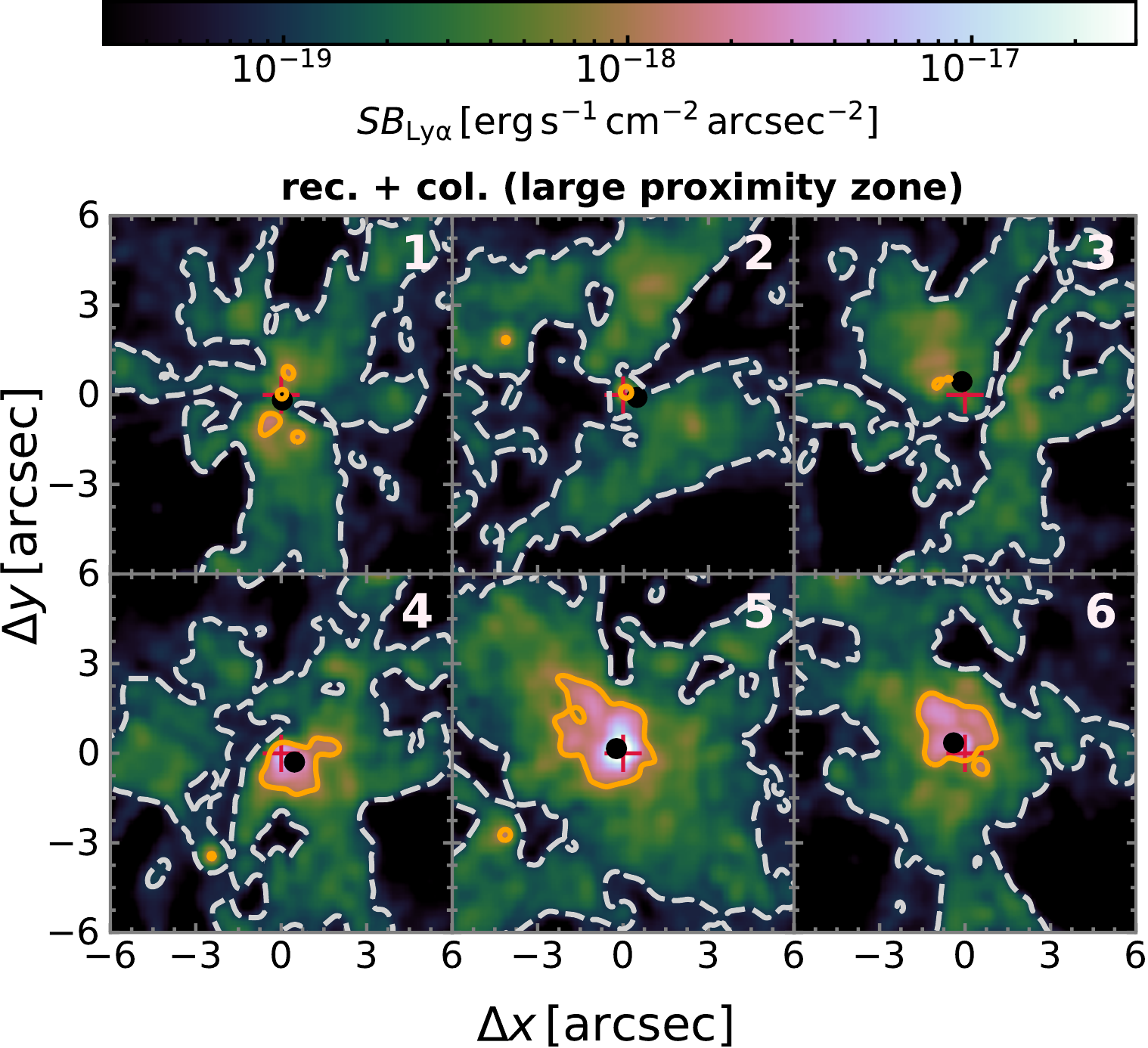}
    \includegraphics[width=0.45\textwidth,trim={0 1cm 0 2cm},clip]{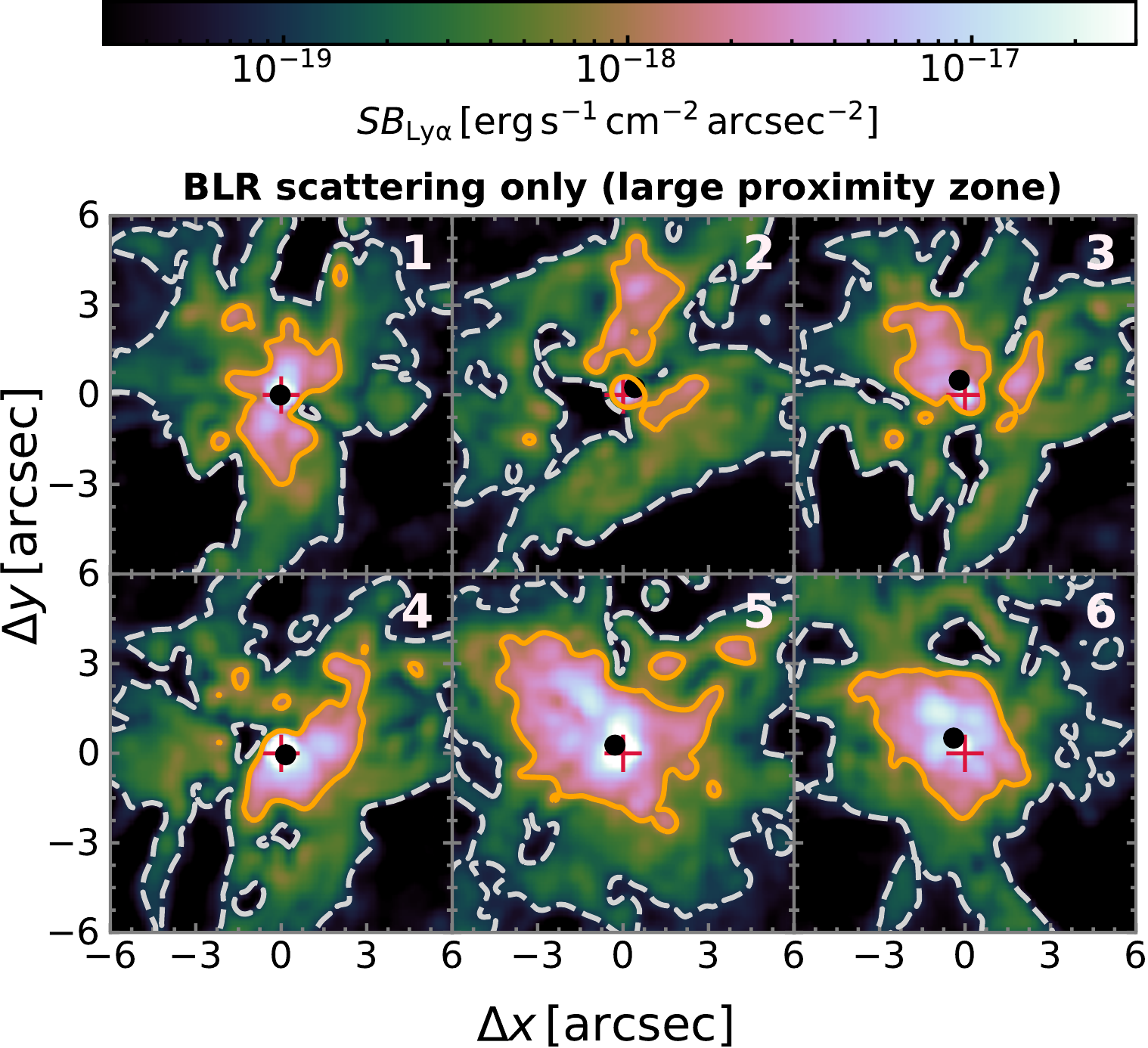}
    \includegraphics[width=0.45\textwidth,trim={0 0 0 2cm},clip]{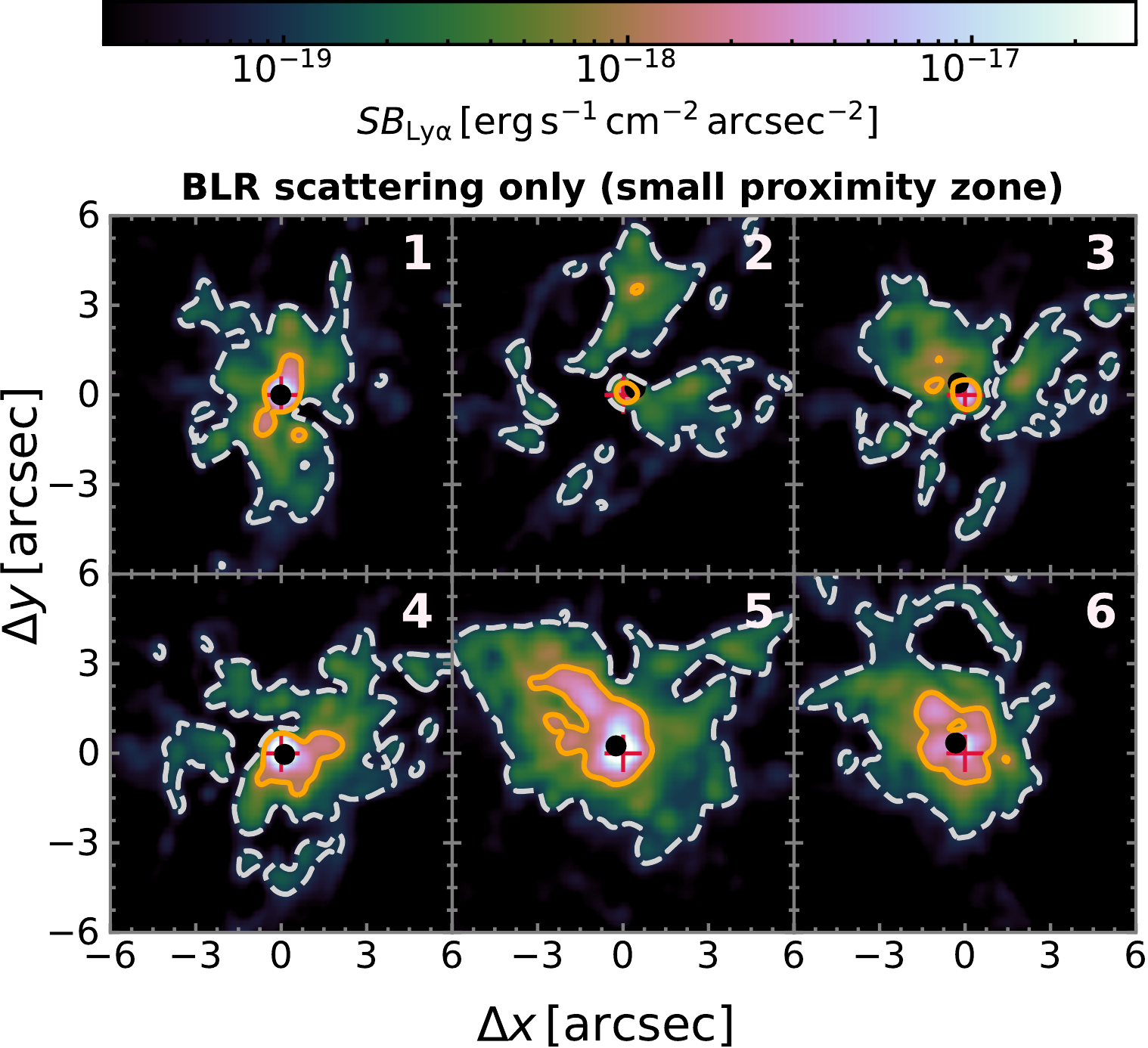}
    \caption{Ly$\alpha$ surface brightness maps centred on a quasar at $z \, = \, 7.6$ plotted for different lines-of-sight and different combinations of Ly$\alpha$ sources. In the top two sub-figures, we assume a proximity zone with radius $R_{\rm p} \, = \, 3 \, \rm Mpc$, while we adopt $R_{\rm p} \, = \, 0.5 \, \rm Mpc$ in the bottom sub-figure. Solid orange and black dashed contours respectively show $10^{-18} \, \rm erg \, s^{-1} \, \rm kpc^{-2} \, \rm arcsec^{-2}$ and  $10^{-19} \, \rm erg \, s^{-1} \, \rm kpc^{-2} \, \rm arcsec^{-2}$ isophotes. Spatially extended nebulae at scales of several arcsec ($1 \, \mathrm{arcsec} \, \approx \, 5 \, \rm kpc$ at $z \, = \, 7.6$) should be observable if observations reach surface brightness limits of $10^{-18} \, \rm erg \, s^{-1} \, \rm kpc^{-2} \, \rm arcsec^{-2}$. As at $z \, = \, 6$, the nebula's morphology varies strongly with the line-of-sight. Quasars embedded in smaller proximity zones should exhibit fainter and smaller nebulae, but some residual extended emission should remain detectable.}
    \label{fig:SBmaps_z}
\end{figure}

Figure~\ref{fig:SBmaps_z} shows Ly$\alpha$ surface brightness maps at $z \, = \, 7.6$ for six random lines-of-sight for (i) recombination radiation and collisional excitation (first set of panels) and (ii) BLR scattering only (second and third sets of panels).
In the first two panel sets, we assume a proximity zone (the volume assumed to be fully ionised by the quasar in our analytic model, see Section~\ref{sec:igmabsorption}) with radius $R_{\rm p} \, = \, 3 \, \rm Mpc$. In the bottom set of panels, we explore using $R_{\rm p} \, = \, 0.5 \, \rm Mpc$, to illustrate a worst-case scenario.
Similarly to $z \, \approx \, 6$, extended Ly$\alpha$ nebulae  surround the central quasar in every case. Nebulae are fainter than at $z \, = \, 6$ due to stronger cosmological dimming, but also, due to lower intrinsic luminosities $\approx (1 \-- 3) \times 10^{43} \, \rm erg \, s^{-1}$.

Extended emission can be seen on scales of several arcsec, corresponding to physical scales of $\sim 20 \, \rm kpc$, down to surface brightness levels of $10^{-18} \, \rm erg \, s^{-1} \, cm^{-2} \, arcsec^{-2}$, though not for every line-of-sight (e.g. panels 1 - 3 for recombination radiation and collisional excitation). If the quasar proximity zone is small ($\lesssim 1 \, \rm Mpc$), the case shown in the third set of panels for BLR scattering, IGM absorption dims nebular emission considerably. In such a case, detecting extended emission might still be possible at surface brightness levels of $10^{-18} \, \rm erg \, s^{-1} \, cm^{-2} \, arcsec^{-2}$, but would likely require reaching surface brightness levels of $10^{-19} \, \rm erg \, s^{-1} \, cm^{-2} \, arcsec^{-2}$ (white, dashed contours). 

\section{Discussion}
\label{sec:discussion}
In this section, we explore the consequences of the findings presented in this paper. We discuss the role of resonant photon scattering in shaping the morphologies, surface brightness profiles and spectra of observed Ly$\alpha$ nebulae (Section~\ref{sec:reflectedlight}). We comment on the extent to which nebulae encode information about quasar feedback (Sections~\ref{sec:whyQSO} and~\ref{sec:insights}), and why the Ly$\alpha$ nebula mechanisms we identify for $z > 6$ quasars may be generalised to giant Ly$\alpha$ nebulae at lower redshift (Section~\ref{sec:openquestions}). In Section~\ref{sec:openquestions}, we also discuss limitations in our modelling as well as important future avenues of research.

\subsection{Giant Ly$\alpha$ nebulae as reflected light}
\label{sec:reflectedlight}
Resonant scattering appears to play a vital role in reconciling simulated-based Ly$\alpha$ nebulae with observations at $z > 6$. 
In Section~\ref{sec:Ly-escape}, we showed that large spatial offsets between the position of the centroid of the nebula and the position of the quasar shrink significantly if scattering is neglected and nebulae become more symmetric. Such offsets and asymmetries are, however, often seen in observed Ly$\alpha$ nebulae at $z > 6$ \citep[e.g.][]{Drake:19}.
Scattering is probably not a unique explanation for observations. For instance, an inhomogeneous ionised gas or dust distribution may also result in an asymmetric nebula, though this appears not to be the case in the halo we target in our simulations.
Performing Ly$\alpha$ radiative transfer on a larger, statistical halo sample will be important to quantify the incidence of nebula asymmetries and verify the robustness of our proposition that large spatial offsets between quasar and surface brightness peak provide evidence for scattering.

Other clues gathered in our paper, however, point to the important role of resonant scattering. 
For instance, we have also seen in Section~\ref{sec:SBprofile} that the observed median surface profile for $z > 6$ quasars flattens out a scales $\lesssim 10 \rm kpc$ due to scattering. In some observed nebulae, the surface brightness profile even drops at small radii \citep[e.g.][]{Ginolfi:18}, mimicking the behaviour of some of our mock profiles.
As shown in Section~\ref{sec:SBprofile}, our numerical experiments can only reproduce this feature if resonant scattering is efficient. Neutral gas densities increase towards the central regions of the halo (see e.g. Figure~\ref{fig:intro}) and Ly$\alpha$ photons generated via recombination or produced in the broad line region have a low escape probability. Instead of streaming towards the observer directly, as would occur in the absence of scattering, these photons diffuse outwards, enhancing the surface brightness at large radii. 

The importance of scattering highlighted by our models is associated with various testable predictions:
\begin{enumerate}
    \item Due to more efficient scattering, AGN residing in galaxies with an edge-on orientation should in general (1) produce fainter Ly$\alpha$ nebulae, (2) be more likely to exhibit detectable spatial offsets between quasar position and surface brightness peak or flux-weighted centroid, (3) display more asymmetric and irregular nebulae and (4) result in flatter central surface brightness profiles,
    \item deeper observations probing a surface brightness limit of $10^{-19} \, \rm erg \, s^{-1} \, cm^{-2} \, arcsec^{-2}$ in the environments of $z > 6$ quasars should detect extended emission at scales $\sim 100 \, \rm kpc$. Our calculations without scattering would predict no extended emission beyond $\approx 60 \, \rm kpc$,
    \item Surface brightness profiles for non-resonant lines such as H$\alpha$ or He\ensuremath{\,\textsc{ii}} should be more compact than for Ly$\alpha$.
\end{enumerate}

An additional clue that scattering shapes Ly$\alpha$ nebulae around $z \approx 6$ quasars is provided by the spectra presented in Section~\ref{sec:lineprofile}. In the absence of scattering, line-widths are $\lesssim 300 \, \rm km \, s^{-1}$, on the lower end of the values found in \citet{Farina:19}. Scattering broadens these profiles significantly, producing line-widths of $500 \-- 1000 \, \rm km \, s^{-1}$ and moment maps (see Appendix~\ref{appendix:b}) in better agreement with observations.

\subsection{How quasar feedback allows resonant scattering to light-up giant nebulae}
\label{sec:whyQSO}

In order to interpret their observation of a giant $\sim 400 \, \rm kpc$-sized Ly$\alpha$ nebula, \citet{Cantalupo:14} perform a numerical simulation with the hydrodynamic code {\sc Ramses}. This simulation follows a cosmological box of $\left( 40 \, \rm cMpc \right)^3$, where $\rm cMpc$ denotes comoving Mpc, focusing on a $\left( 10 \, \rm cMpc \right)^3$ higher-resolution sub-volume centred on a $M_{\rm vir} \approx 3 \times 10^{12} \, \rm M_\odot$ halo at $z \, = \, 2.3$. \citet{Cantalupo:14} include radiative cooling, star formation, supernova feedback and a spatially uniform UV background.
These processes are also captured in our simulations, though they are modelled differently, but, in addition, we also solve the fully coupled radiation-hydrodynamic equations. 
A critical process in our simulations is radiative feedback from AGN \citep{Costa:18}. Despite strong stellar feedback, the escape of the AGN ionising flux depends on AGN feedback, as also found in \citet{Costa:18} with a different supernova feedback model (``delayed cooling'').
Above a critical luminosity established by the balance of gravitational and radiation pressure forces, AGN radiation pressure launches large-scale outflows \citep{Costa:18}. Besides modifying the temperature and density structure of halo gas, these outflows result in a dramatic drop in the HI and dust optical depths in the host galaxy, facilitating the escape of Ly$\alpha$ from the galactic nucleus. 

\begin{figure*}
    \centering
    \includegraphics[width=0.475\textwidth]{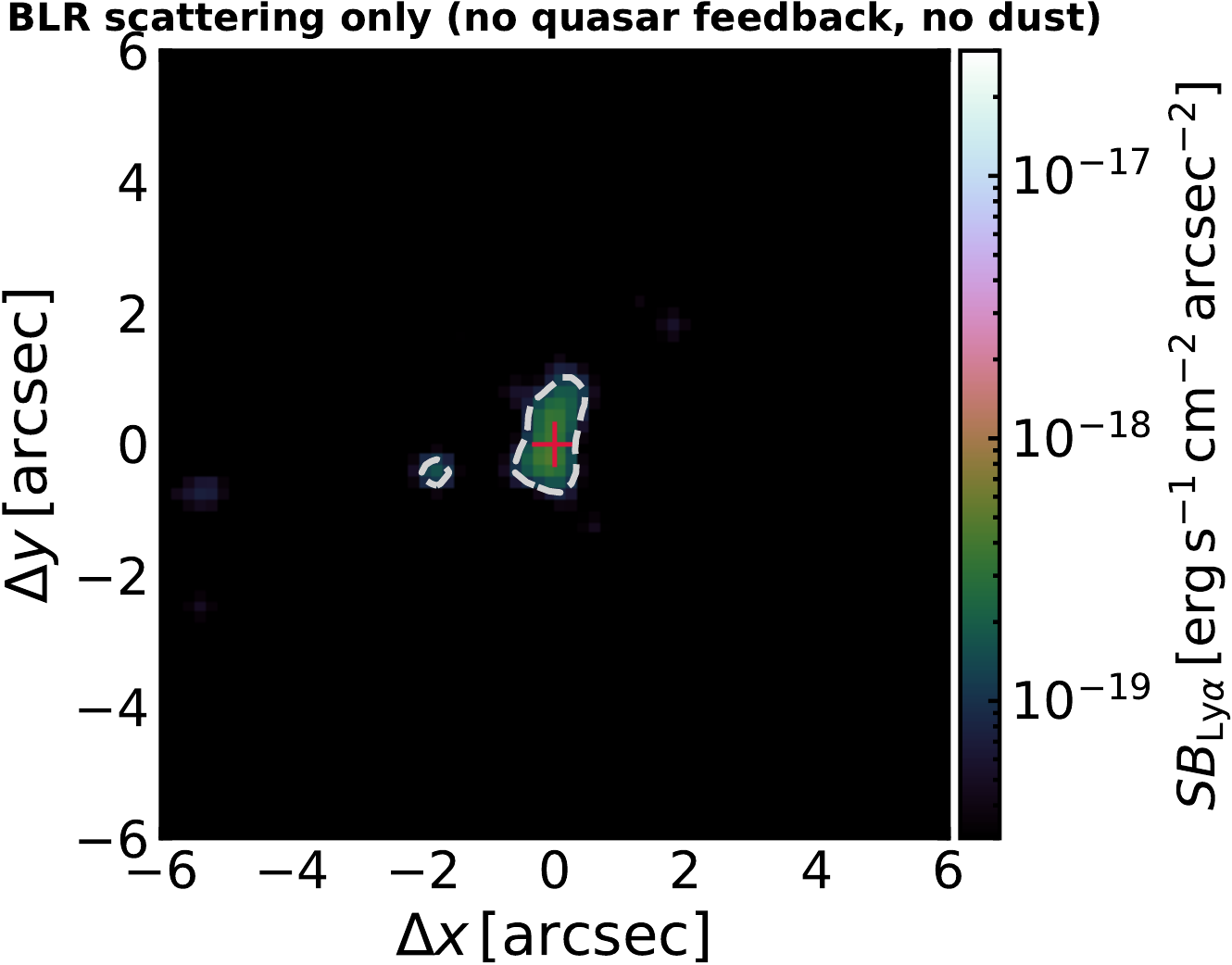}
    \includegraphics[width=0.475\textwidth]{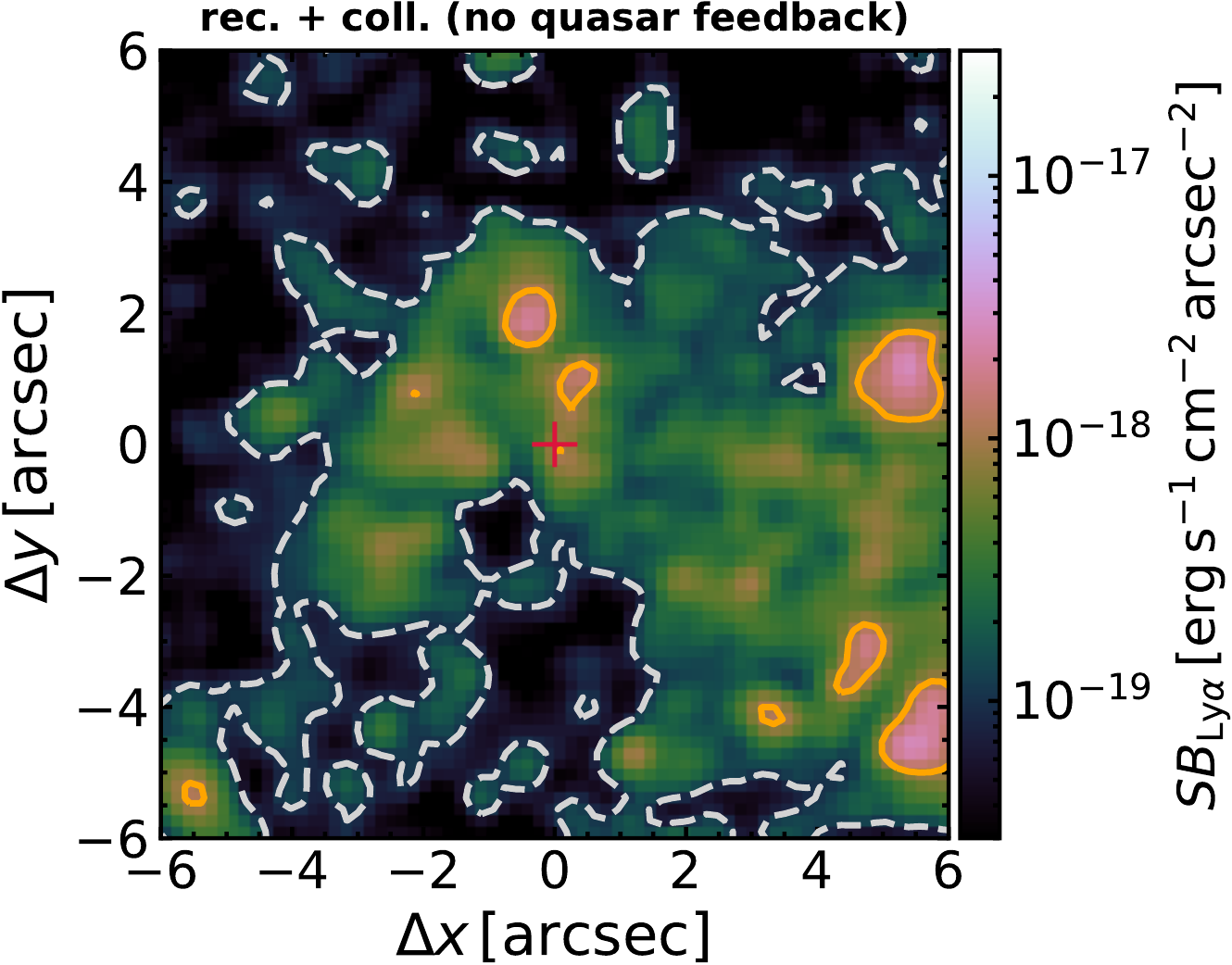}
    \caption{{\bf Left:} Ly$\alpha$ surface brightness image for \texttt{noQuasar}, accounting only for scattering from the BLR. Even if destruction by dust absorption is neglected, no extended nebula forms in our simulations via scattering from the BLR if AGN feedback is not included in the simulation. {\bf Right:} Ly$\alpha$ surface brightness image for \texttt{noQuasar}, accounting for both recombination radiation and collisional excitation (with dust absorption). An extended nebula forms, showing that a residual level of Ly$\alpha$ emission exists in the haloes hosting supermassive black holes at $z > 6$. However, the associated emission produces nebulae which are fainter and surface brightness profiles which are inconsistent with those observed around bright quasars at $z > 6$.} 
    \label{fig_nebulae_noAGN}
\end{figure*}

AGN feedback affects all Ly$\alpha$ emission mechanisms. 
Figure~\ref{fig:nscats} compares the distribution of radii of last scattering in \texttt{Quasar-L3e47} with that of \texttt{noQuasar} (ignoring dust absorption, to test a best-case scenario). In the absence of quasar radiative feedback, last scatterings typically occur at $R \approx 500 \, \rm pc$, within the quasar host galaxy, with only a thinly-populated tail scattering out to $R \sim 10 \, \rm kpc$. With radiative quasar feedback, last scatterings occur at $R > 1 \, \rm kpc$ and typically at $R \, \approx \, 10 \-- 100 \, \rm kpc$, even beyond the virial radius.
The ability of photons to scatter out to $\sim 100 \, \rm kpc$ in \texttt{Quasar-L3e47} is not connected with an increase in HI column densities in the halo in this simulation, as could be expected, for instance, if the quasar-driven outflow transports cold neutral gas. The ability of these photons to scatter out beyond the virial radius is driven by the fact that they scatter less efficiently in the central regions and are thus still in resonance with halo gas. In \texttt{noQuasar}, the quasar is buried within massive, optically thick HI layers. BLR photons are resonantly trapped and escape only becomes possible when their rest-frame wavelength has shifted by several $\sim 10 \rm \angstrom$ away from line centre, corresponding to velocity shifts of several thousand $\rm km \, s^{-1}$ These photons no longer couple to HI in the halo, as proposed in \citet{Cantalupo:14}, who do not model AGN feedback in their simulations.
The left-hand panel of Figure~\ref{fig_nebulae_noAGN} shows the resulting surface brightness map: \emph{no spatially extended nebula forms with BLR scattering in the absence of AGN feedback even if dust absorption is ignored}. The resulting surface brightness map should be compared directly to the bottom set of panels in Figure~\ref{fig_nebulae_los}, where BLR scattering is shown to produce giant, bright nebulae in \texttt{Quasar-L3e47}. 
By destroying HI gas, quasar radiation, however, AGN feedback damps resonant scattering in the central regions. With velocity shifts of $\lesssim 1000 \, \rm km \,s^{-1}$, photons still scatter off HI in the halo. 

As shown in Figure~\ref{fig_SBprofile_time}, recombination radiation, but, due to is higher escape fraction, mainly \emph{collisional excitation is able to power an extended Ly$\alpha$ even in our simulations performed without quasar feedback and radiation.}
The associated surface brightness profile, however, disagrees with the observed median profile of \citet{Farina:19}.
The right-hand panel of Figure~\ref{fig_nebulae_noAGN} shows a surface brightness map for Ly$\alpha$ radiative transfer calculation performed on \texttt{noQuasar}, accounting for recombination radiation and collisional excitation (including dust absorption). The existence of a giant nebula is clear, though, upon comparison with e.g. Figure~\ref{fig_nebulae_los}, strong differences stand out: (i) the resulting nebula is not centrally concentrated around the quasar, as its formation has a different origin and is connected to halo gas cooling, (ii) it is fainter with a surface brightness of $< 10^{-18} \, \rm erg \, s^{-1} \, cm^{-2} \, arcsec^{-2}$, and (iii) it has a very different morphology, appearing patchier and more irregular.

\subsection{Quasar feedback and its limits}
\label{sec:insights}

Our results reveal a connection between the formation and properties of extended Ly$\alpha$ nebulae around bright quasars and AGN feedback.
In our simulations, feedback occurs via radiation pressure on dusty gas \citep[see][]{Costa:18}. This form of feedback operates particularly efficiently on cold, dense material, mediating a transition between obscured- and unobscured quasar states.
Expelling the optically thick layers is unlikely to succeed with purely stellar feedback. \citet{Costa:18} we show that introducing supernova feedback, modeled there through delayed cooling in gas heated by supernova explosions, does not prevent the central galaxy from being completely obscured. In \citet{Costa:19}, stellar radiation is found to reduce the efficiency of supernova feedback in massive haloes, indirectly enhancing the dense gas abundance in the central galaxy. In this study, supernova feedback, now modelled following the mechanical model of \citet{Kimm:15}, also clearly fails to enable Ly$\alpha$ escape. 

The link between AGN feedback and Ly$\alpha$ is indirect, as outflowing material takes paths of least resistance and largely decouples from the dense, cool CGM at scales $\gtrsim 10 \, \rm kpc$ (see Figure~\ref{fig:intro}). The existence of a largely undisturbed CGM is thus not inconsistent with the presence of large-scale AGN outflows \citep[see also][]{Lau:22}. Recent observations reported by \citet{Li:21} provide an example of a large-scale outflow powered by a radio-loud quasar at $z \approx 2$, propagating along a direction perpendicular to a giant $\sim 100 \, \rm kpc$ Ly$\alpha$ nebula. \citet{Li:21} suggest a separate origin for the outflow, which is driven by the quasar, and the Ly$\alpha$ nebula, which could instead trace gas infall, exactly the scenario favoured by our simulations (Section~\ref{sec:lineprofile}). 

At face value, the detection of extended Ly$\alpha$ nebulae at $z \, = \, 6$ \citep{Drake:19,Farina:19} indicates that AGN feedback already operates in $z > 6$ quasars, shaping galaxy evolution since the first Gyr of the Universe's evolution. This conclusion may appear to be in tension with scarce evidence of large-scale outflows in $z > 6$ quasar hosts \citep{Novak:20, Meyer:22}. The detection of large-scale outflows in this regime has been attempted mainly with interferometric observations tracing molecular gas (e.g. via CO emission), and cold, atomic gas (via \lbrack C\ensuremath{\,\textsc{ii}}\rbrack\,158\,$\mu$m). With a typical resolution of $\gtrsim 100 \, \rm pc$, most cosmological simulations, however, cannot resolve cold molecular gas. In addition, many such simulations exclude on-the-fly models for molecular chemistry and radiative transfer, which are required for a robust prediction of molecular gas properties. Such cosmological simulations predict mostly hot $T \gtrsim 10^6 \, \rm K$, low-density outflows on scales $\gtrsim 10 \, \rm kpc$ \citep{Costa:15}. In simulations with exceptionally high-resolution \citep[see][for 5pc resolution ``zoom-in'' simulations targeting quasars at $z \, = \, 7$]{Lupi:21} no sustained molecular outflows are predicted on kpc scales.

The cosmological, radiation-hydrodynamic simulations of \citet{Costa:18} however, predict that warm ($T \sim 10^4 \, \rm K$), dense ($n_{\rm H} > 1 \, \rm cm^{-3}$) gas dominates the outflow mass budget at kpc scales. If ionised by the central quasar, such outflow phase should be bright in e.g. \lbrack O\ensuremath{\,\textsc{iii}}\rbrack, a line which has been used to detect powerful quasar outflows at $z \, = \, 1 \--4$ \citep[e.g.][]{Harrison:16, Zakasmka:16, Bischetti:17}. The James Webb Space Telescope (JWST) will directly probe the warm ionised gas phase in $z > 6$ quasars and quantify the incidence of large-scale outflows in the first quasars. Our results suggest that those quasars presenting evidence of extended Ly$\alpha$ nebulae would constitute good candidates for the detection of large-scale outflows, which, according to our models, must have cleared out the galactic nucleus at some point in the past.

In Section~\ref{sec:qsoluminosity} we have also seen that if too effective, AGN feedback promotes too much Ly$\alpha$ escape in the central $\sim 10 \, \rm kpc$ and reduces the importance of resonant scattering, causing the Ly$\alpha$ surface brightness profile to become steeper than observed. Based on this finding, we suggest that quasars showing evidence of energetic outflows on kpc scales should be associated to Ly$\alpha$ nebulae with steeper surface brightness profiles than older quasars that have been unobscured for a longer time. A stronger outflow impact, potentially revealed through cavities in the galactic nucleus \citep[e.g.][]{Cano-Diaz:12}, might thus be associated with steeper Ly$\alpha$ surface brightness profiles.

\subsection{Open questions}
\label{sec:openquestions}

This study is subject to various uncertainties. 
Insufficient resolution limits our ability to resolve structure in the CGM and, in particular, likely leads to an underestimate in the HI column density in the halo. As shown in Section~\ref{sec:SBprofile}, our simulations appear to recover the observed profiles nevertheless. Likely, this match occurs because current observations of $z \approx 6$ Ly$\alpha$ nebulae only probe the innermost regions of the galactic haloes hosting bright quasars, where our simulations provide the highest resolution. We can anticipate that our simulations might underestimate the surface brightness profile at large scales $\sim 100 \, \rm kpc$ as probed by future, deeper observations.
It will thus be crucial to test our findings with new suites of radiation-hydrodynamic cosmological simulations including refinement techniques tailored to resolve the CGM in detail \citep{vandeVoort:19,Hummels:19,Bennett:20}. These studies unanimously point to an increasing trend in HI column densities as CGM resolution improves, suggesting that scattering may be even more efficient than our study suggests.

If performed at much higher resolution, such simulations may predict that the CGM is composed by a fog-like distribution of very dense $n_{\rm H} \sim 100 \, \rm cm^{-3}$ cloudlets \citep{McCourt:18}. If photo-ionised, these cloudlets could more efficiently generate recombination radiation \citep{Cantalupo:14} than predicted by our current simulations. What our simulations clearly show is that \emph{a fog-like structure is not necessarily required to produce giant Ly$\alpha$ nebulae}, a result that extends earlier findings by \citet{Gronke:17} to the giant Ly$\alpha$ nebulae surrounding the first quasars.

While it is precisely the idealised nature of our simulations and, in particular, their treatment of AGN luminosities and light-curves, that has allowed us to clearly reveal a relation between Ly$\alpha$ nebulae and AGN feedback strength, future simulations should test our results with self-consistent black hole growth models such as employed in \citet{Dubois:13} or \citet{Costa:14}. Existing models are notoriously uncertain for a number of reasons, including: 
\begin{enumerate}
    \item difficulties in resolving characteristic scales, such as the Bondi radius, which can result in order of magnitude uncertainties in the black hole's self-regulation mass and growth history \citep[e.g.][]{Curtis:15}, 
    \item uncertainties associated with computing the black hole accretion rate based on gas properties at scales $\gtrsim 10 \, \rm pc$, well beyond the black hole's sphere of influence. Self-regulation may occur on smaller scales than envisaged by such models, with the consequence that the injection rate of AGN energy and momentum at scales $\gtrsim 10 \, \rm pc$ might be decoupled, and
    \item the absence of a treatment of angular momentum in accretion flows, which may result in significant time lags between accretion events onto the black hole's accretion disc and actual accretion onto the black hole.
\end{enumerate}
Promising solutions to these long-standing challenges include refinement techniques targeting black holes in cosmological simulations \citep{Curtis:15, Angles-Alcazar:21}.

Another open question is whether we can extend our findings to Ly$\alpha$ nebulae at lower redshift.
When accounting for cosmic expansion, the median surface brightness profile of \citet{Farina:19} matches that observed for $z \approx 3$ quasars \citep{Arrigoni-Battaia:19}. This lack of evolution may suggest that the same physical mechanisms producing Ly$\alpha$ nebulae operate both at $z \, = \, 6$ and at $z \, = \, 3$. 
Until $z \approx 2$, massive dark matter haloes with $M_{\rm vir} \sim 10^{12} \, \rm M_\odot$ grow predominantly via cold flow accretion (see Figure~\ref{fig:intro}), and the gas environments characterising bright quasars retain the same properties. In this regime, we may expect our findings to hold.
As soon as haloes acquire stable hot atmospheres, however, cold and dense gas in the halo disperses \citep[see e.g.][]{vanderVlugt:19} and we might expect Ly$\alpha$ nebulae to shrink and become fainter, as reported in \citet{Cai:19, Arrigoni-Battaia:19, O'Sullivan:20}.

While our simulations have emphasised the role of radiation pressure on dust as an AGN feedback mechanism, other AGN feedback mechanisms may play an analogous role in clearing out gas from the vicinity of AGN. Winds generated at the scale of AGN accretion discs \citep{Costa:20} or relativistic jets, as is the case in the observations of \citet{Li:21}, are predicted to inflate hot bubbles that may clear out gas from the galactic nucleus \citep{Costa:14, Talbot:21}. The interesting question is whether such models disagree on the impact of AGN feedback at halo scales, as suggested in \citet{Costa:18}. Energy-driven outflows, for instance, effectively eject halo gas, and, in some studies \citep{Dubois:13}, have been suggested to destroy dense, cool gas as well. A question that future studies should thus explore is the extent to which observations of extended Ly$\alpha$ nebulae allow for destructive AGN feedback models, such as energy-driven outflows and jets, or whether they argue for ``gentler'' AGN feedback channels.

\section{Conclusions}
\label{sec:Conclusions}

We present Ly$\alpha$ radiative transfer calculations performed in post-processing on a suite of cosmological, radiation-hydrodynamic simulations targeting a rare $M_{\rm vir} \, = \, 2.6 \times 10^{12} \, \rm M_\odot$ halo capable of hosting a bright quasar at $z \, = \, 6$. 

Resonant scattering (i) broadens the Ly$\alpha$ surface brightness profile irrespective of emission mechanism and (ii) flattens the profiles in the central regions of the halo (Figure~\ref{fig_nebulae_los}). Despite no attempt to fine-tune our cosmological simulations, or to increase spatial resolution in the CGM, the shape and normalisation of predicted surface brightness profiles are in strikingly close agreement with observational constraints at $z > 6$, particularly when resonant scattering is taken into account (Figure~\ref{fig_SBprofile_contribution}).

We unveil three physical mechanisms that stand out in their ability to produce the close agreement between theory and data: (i) a combination of collisional excitation and recombination radiation, (ii) resonant scattering of Ly$\alpha$ photons from the broad line region, even if operating on its own, or (iii) a combination of these three processes (Figure~\ref{fig_SBprofile_contribution}).
All light-up Ly$\alpha$ nebulae with sizes of up to $\approx 30 \, \rm kpc$ at surface brightness levels of $10^{-18} \, \rm erg \, arcsec^{-2} \, \rm cm^{-2}\, s^{-1}$, extending out to $\sim 100 \, \rm kpc$ scales at surface brightness levels of $10^{-19} \, \rm erg \, arcsec^{-2} \, \rm cm^{-2}\, s^{-1}$ (Figure~\ref{fig_size_luminosity}).

Resonant scattering of Ly$\alpha$ photons from the broad line region all the way to halos scales of $\sim 100 \, \rm kpc$ had been thought to be difficult to achieve due to the high HI optical depths expected in the galactic nucleus. In such a scenario, Ly$\alpha$ escape would occur mainly via a frequency shift into the wings of the Ly$\alpha$ line. Scattering then relies on high optical HI depths at halo scales. This scenario applies exactly in our cosmological simulation performed without AGN feedback (Figure~\ref{fig_nebulae_noAGN}). In this case, most photons cease scattering at scales of $\sim 500 \, \rm pc$, well inside the quasar host galaxy. AGN feedback, which in our simulations operates mainly via radiation pressure on dust, changes this result dramatically. By blowing out the central gas reservoir, AGN feedback allows the Ly$\alpha$ flux to escape from the galactic nucleus more efficiently, allowing the photons to scatter off infalling gas out to scales of up to $\approx 100 \, \rm kpc$ (Figure~\ref{fig:intro}). Even if nebulae are powered by a combination of recombination radiation and collisional excitation, AGN feedback is still required to reduce dust absorption and the HI opacity in the central regions of the halo (Figure~\ref{fig_SBprofile_time}).

This paper thus reveals a close connection between AGN feedback and Ly$\alpha$ nebulae. 
Even though our simulations follow strong supernova feedback and stellar radiative feedback via photo-ionisation, photo-heating and radiation pressure, quasar feedback alone makes the difference between a Ly$\alpha$ nebula with properties in close agreement with observational constrains or, if at all, a much fainter nebula.
We, however, also find that energy deposited by AGN can ultimately cause the Ly$\alpha$ nebulae to shrink. If particularly strong, AGN feedback destroys so much HI gas, that photons produced in the central regions escape directly without having to scatter.  These results indicate that the observational properties of Ly$\alpha$ nebulae and in particular the slope of surface brightness profile may constrain the efficiency of AGN feedback.

A combination of dust absorption and scattering results in highly anisotropic Ly$\alpha$ escape (Figure~\ref{fig_escape}). The Ly$\alpha$ flux escapes preferentially along the rotation axis of the quasar host galaxy disc. Observer lines-of-sight closely aligned with the disc's polar axis thus detect the brightest Ly$\alpha$ nebulae. Lines-of-sight intercepting the quasar host disc edge-on can also display extended Ly$\alpha$ nebulae, but our simulations predict these to be both fainter and to display more asymmetric morphologies. These findings lead to the clear observational prediction (and test to our models) that asymmetric Ly$\alpha$ nebulae should, statistically, be associated to more edge-on central galaxy disc orientations, while rounder, brighter nebulae should surround face-on disc galaxies.

The agreement between observational constraints and our simulations, achieved without any parameter tuning, lend strong support to the theoretical prediction that quasars are hosted by rare, massive haloes, a condition which is required to explain the rapid growth of supermassive black holes at $z > 6$. Our simulations also predict the presence of extended Ly$\alpha$ nebulae with surface brightness levels of $10^{-18} \, \rm erg \, s^{-1} \, \rm cm^{-2} \, arcsec^{-2}$ at scales of $10 \-- 15 \, \rm kpc$ around $z \, = \, 7.5$ quasars (Figure~\ref{fig:SBmaps_z}). A future detection would point to the conclusion that AGN feedback shapes galaxy evolution from the very earliest stages of galaxy evolution.

\section*{Acknowledgements}
TC gratefully acknowledges L\'eo Michel-Dansac, J\'er\'emy Blaizot and collaborators for developing {\sc RASCAS}, making their code publicly available, for their interest and helpful advice. TC also thanks Martin Haehnelt, Chris Harrison and Volker Springel for invaluable comments on the manuscript. LCK was supported by the European Union's Horizon 2020 research and innovation programme under the Marie Sk\l{}odowska-Curie grant agreement No. 885990. TK was supported by the National Research Foundation of Korea (NRF) grant funded by the Korea government (MSIT) (No. 2020R1C1C1007079).

\section*{Data Availability}
The data underlying this article will be shared on reasonable request to the corresponding author.



\bibliographystyle{mnras}
\bibliography{lit} 


\appendix

\section{Scattering and asymmetric nebulae}
\label{appendix:a}

\begin{figure}
    \centering
    \includegraphics[width=0.475\textwidth]{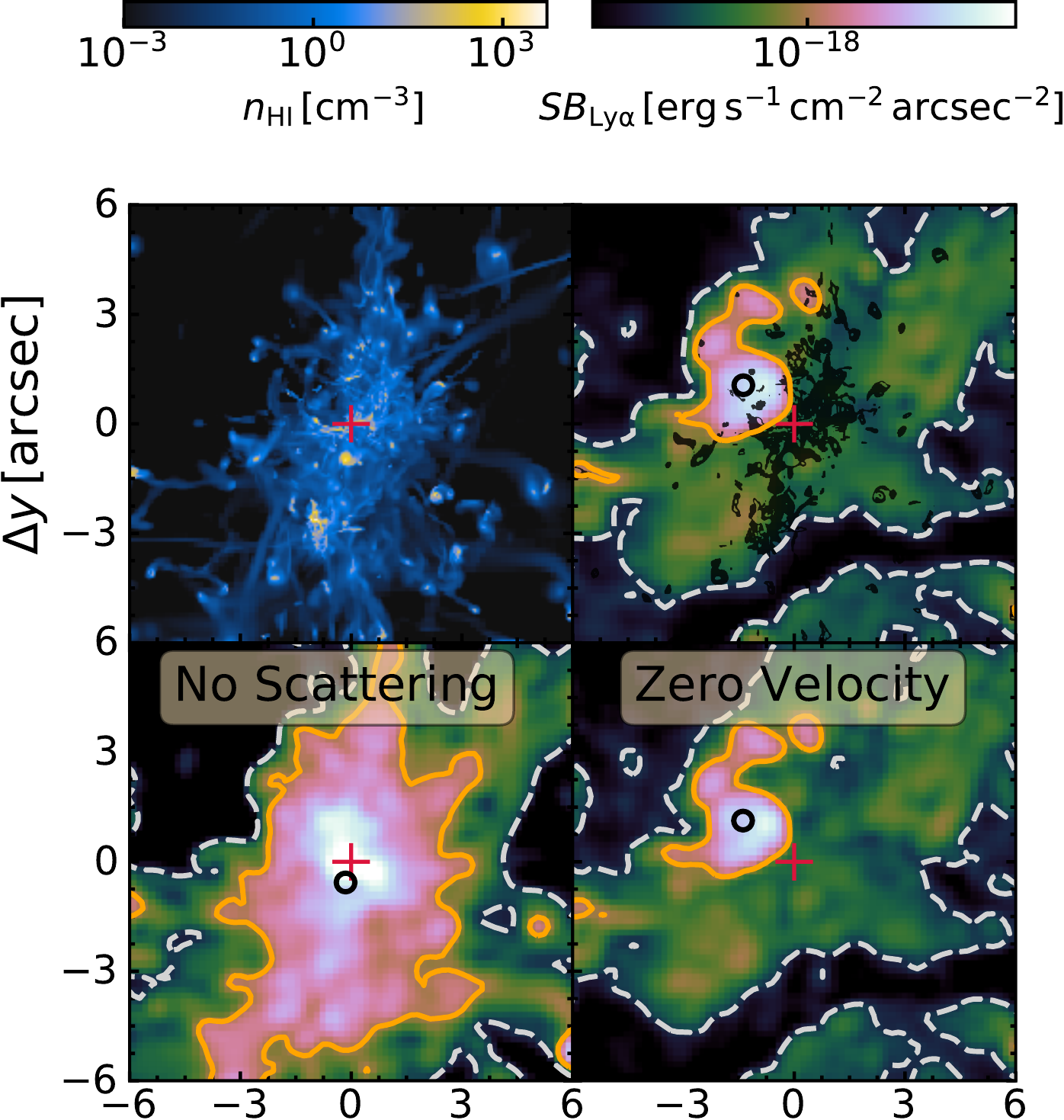}
    \caption{HI gas density at $z \, = \, 6.2$ (first panel) within the central regions with the position of the bright quasar marked by a red plus sign. The Ly$\alpha$ nebula associated to it is shown on the second panel, where a black circle marks its  . The nebula is spatially anti-correlated with the high HI density regions, marked for $n_{\rm HI} > 0.3 \, \rm cm^{-3}$ with a black shade. This spatial offset is caused by scattering, since it disappears if we neglect resonant scattering in our radiative transfer calculation (third panel). The fourth panel shows a test-run where scattering is enabled, but all velocities are set to zero. Neglecting the gas velocity field does not prevent the observed large spatial offsets. These are thus mostly caused by anisotropy in the HI distribution.}
    \label{fig_modelvariations}
\end{figure}

Figure~\ref{fig_modelvariations}) shows the HI density, together with Ly$\alpha$ maps, for the same line-of-sight as in panels 3 in Figure~\ref{fig_nebulae_los}, in the combined collisional excitation and recombination radiation scenario. The figure illustrates how the nebula morphology changes by neglecting resonant scattering or by setting the gas peculiar velocities to zero. The surface brightness distribution is unchanged if the velocity field is ignored in the Ly$\alpha$ radiative transfer computation. The morphological variation of the Ly$\alpha$ nebulae associated to our simulated halo thus appears to be set by the HI density distribution and not its dynamical state. For this line-of-sight, the HI gas phase (shaded in black in the top-right panel) is unequally distributed around the disc, with much lower HI densities above the quasar host galaxy (top left) than below (bottom right). Below the disc, the HI mass acts as a reflecting sheet for the Ly$\alpha$ photons emitted from the ISM of the host galaxy and from the BLR, focussing them towards the region above the host galaxy, creating a one-sided Ly$\alpha$ nebula.

\section{Moment maps}
\label{appendix:b}

\begin{figure*}
    \centering
    \includegraphics[width=0.9\textwidth]{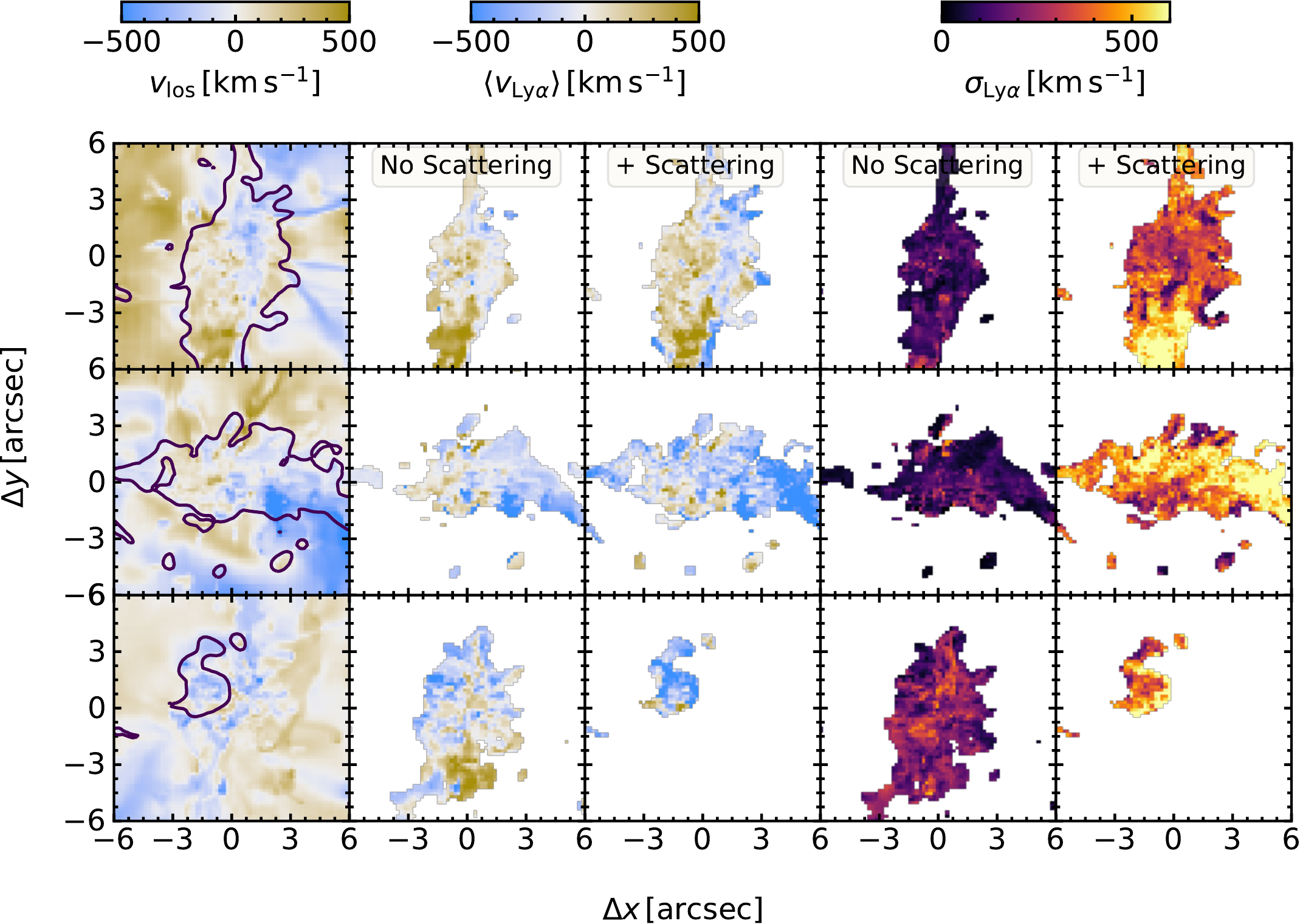}
    \caption{Density-weighted line-of-sight velocity as seen through three lines-of-sight (different rows) shown together with first moment maps (second and third columns) and second moment maps (fourth and fifth columns). The second and fourth columns show the moment maps obtained by neglecting resonant scattering in our Ly$\alpha$ radiative transfer calculations. The first moment traces the actual gas velocity reliably, even after considering resonant scattering. By broadening the Ly$\alpha$ line scattering, however, has a profound effect on the second moment.}
    \label{fig:moments}
\end{figure*}

We can associate a mean spectral velocity shift and a velocity dispersion to each pixel of our synthetic Ly$\alpha$ maps. It is interesting to then link the velocity information of the Ly$\alpha$ line with the underlying velocity field as output by the simulations, and assess the extent to which moment maps can constrain gas dynamics.

\begin{figure}
    \centering
    \includegraphics[width=0.475\textwidth]{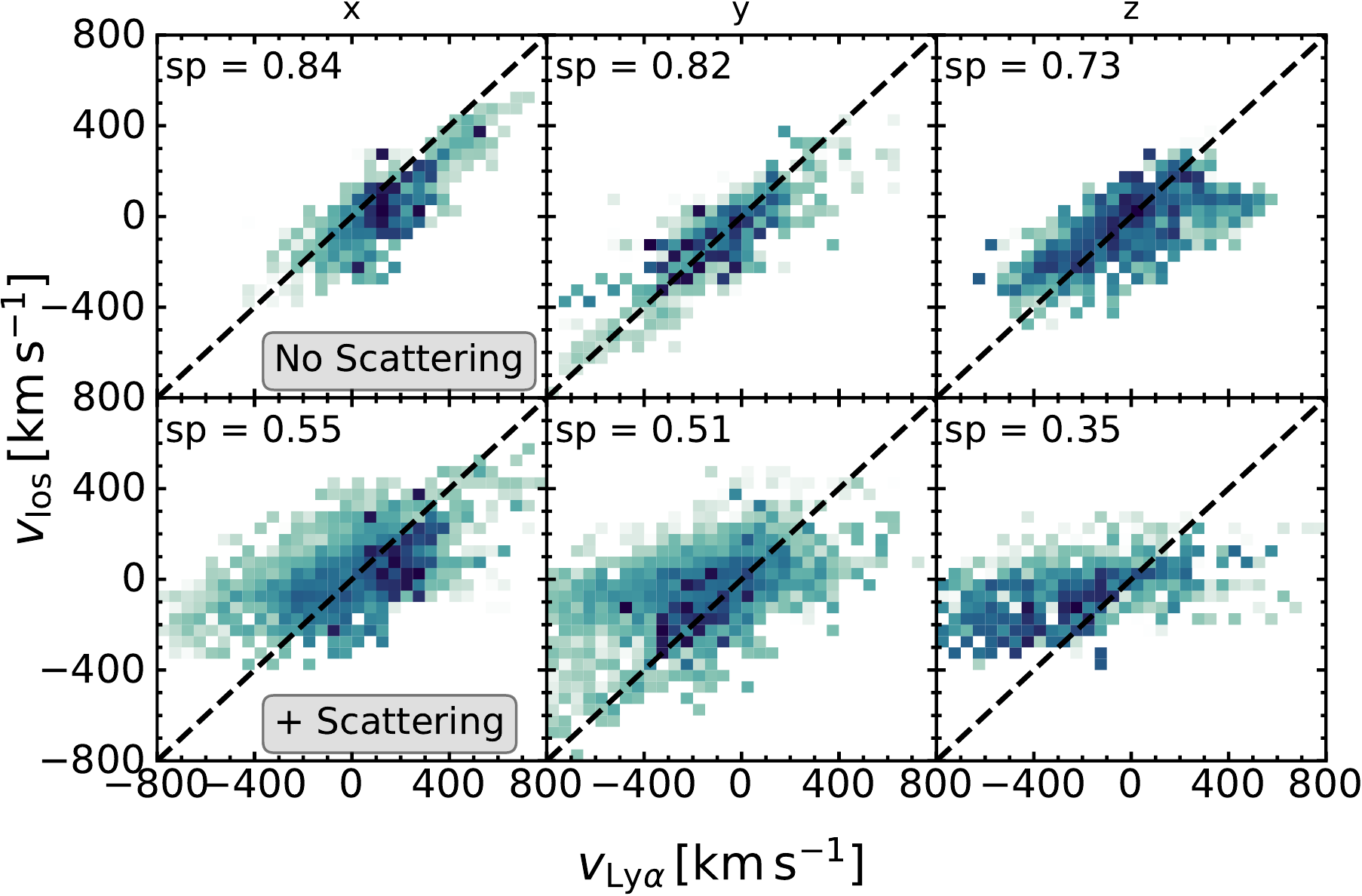}

    \caption{Relation between density-weighted line-of-sight velocity and the first moment of Ly$\alpha$ line measured pixel-by-pixel, for different orientations (different columns), ignoring and including the effect of resonant scattering (first and second rows, respectively). There is a good correspondence between density-weighted line-of-sight velocity and the first moment of Ly$\alpha$ even in the presence of scattering, as shown by high Spearman rank coefficients (given in each panel).}
    \label{fig:vlyman_vlos}
\end{figure}

Figure~\ref{fig:moments} shows moment maps for \texttt{Quasar-L3e47} for three different lines-of-sight and for a combination of all Ly$\alpha$ emission models (recombination radiation, collisional excitation and BLR scattering with $\rm FWHM_{\rm BLR} \, = \, 2400 \, \rm km \, s^{-1}$). The first column gives the density-weighted line-of-sight velocity as predicted by the simulation at $z \, = \, 6.2$. The second and third columns show the Ly$\alpha$ flux-weighted velocity shift, respectively neglecting and accounting for the effect of resonant scattering, considering only regions with $SB \geq 10^{-18} \, \rm erg \, s^{-1} \, cm^{-2} \, arcsec^{-2}$.
The velocity distribution we see in~Figure~\ref{fig:moments} is, as discussed in Section~\ref{sec:overview}, complex, though structures such as filaments (first row) and rotation patterns (third row) can be seen upon close look.
Comparing the first and second columns (which neglect scattering), we find that we can usually match the actual line-of-sight velocity with the first moment of the Ly$\alpha$ maps. Resonant scattering tends to change the shape of the Ly$\alpha$ nebula, as discussed in Section~\ref{sec:overview}, but does not appear to strongly modify its velocity structure. Figure~\ref{fig:vlyman_vlos} tests for a connection between the actual line-of-sight velocity and the flux-weighted Ly$\alpha$ velocity shift. We see that there is a strong correlation between these two quantities, as illustrated by Spearman correlation coefficients $> 0.7$, if scattering is ignored (first row). The bottom row of Figure~\ref{fig:vlyman_vlos} shows that scattering (i) broadens the Ly$\alpha$ velocity shift range, and (ii) increases the scatter in the relation, weakening the correlation between the two velocities. Nevertheless, Spearman correlation coefficients $\gtrsim 0.5$ still indicate a clear monotonic relation for some lines-of-sight. The weakest relations occurs for orientations where scattering is most important, e.g. the third column, which show configurations where the quasar host galaxy lies edge-on.

The most striking change introduced by resonant scattering is on the width of the spectral line. The last two columns of~Figure~\ref{fig:moments} give the standard deviation of the line velocity with and without scattering. If scattering is neglected, the second moments typically range from $10 \, \rm km \, s^{-1}$ to $\approx 300 \, \rm km \, s^{-1}$. Scattering raises the second moments to values $> 500 \, \rm km \, s^{-1}$ particularly in the outskirts of the nebulae, where scattering provides the most contribution.

Even through resonant scattering appears to be crucial in generating more spatially extended Ly$\alpha$ emission for some lines-of-sight, our simulations suggest that moment maps retain direct information about the dynamical state of gas. It reliably gives the sign of the gas velocity, though its magnitude can be strongly affected by scattering.

\bsp	
\label{lastpage}
\end{document}